\documentclass[longauth]{aa}
\usepackage{graphicx}
\usepackage{txfonts}
\usepackage{booktabs}
\usepackage{xcolor}

\begin{document} 

   \title{Potential scientific synergies in weak lensing studies between the CSST and \emph{Euclid} space probes}

   \titlerunning{CSST-{\it Euclid} synergies for weak lensing studies}

   \author{D.Z. Liu \inst{1}
          \and
          X.M. Meng \inst{2}
          \and
           X.Z. Er \inst{1} 
          \and 
          Z.H. Fan \inst{1}
          \and 
          M. Kilbinger \inst{3}
          \and
          G.L. Li \inst{4,5}
          \and
          R. Li \inst{6,7,8}
          \and
          T. Schrabback \inst{9,10}
          \and
          D. Scognamiglio \inst{9}
          \and
          H.Y. Shan \inst{11}
          \and
          C. Tao \inst{12}
          \and 
          Y.S. Ting \inst{13,14}
          \and 
          J. Zhang \inst{15}
          \and 
          S.H. Cheng \inst{16}
          \and
          S. Farrens \inst{3}
          \and 
          L.P. Fu \inst{17}
          \and 
          H. Hildebrandt \inst{18}
          \and 
          X. Kang \inst{4,5}
          \and 
          J.P. Kneib \inst{19}
          \and
          X.K. Liu \inst{1}
          \and 
          Y. Mellier \inst{20}
          \and 
          R. Nakajima \inst{9}
          \and
          P. Schneider \inst{9}
          \and 
          J.L. Starck \inst{3}
          \and 
          C.L. Wei \inst{4}
          \and
          A.H. Wright \inst{18}
          \and 
          H. Zhan \inst{2,21}
          }

  \institute{{South-Western Insitiute for Astronomy Research, Yunnan University, Kunming 650500, China}
             \email{zuhuifan@ynu.edu.cn}
        \and {Key Laboratory of Space Astronomy and Technology, National Astronomical Observatories, Chinese Academy of Sciences, Beijing 100101, China}
        \and {AIM, CEA, CNRS, Université Paris-Saclay, Université de Paris, F-91191 Gif-sur-Yvette, France}
             \email{martin.kilbinger@cea.fr}
        \and {Purple Mountain Observatory, Chinese Academy of Sciences, Nanjing 210023, China}
        \and {Zhejiang University-Purple Mountain Observatory Joint Research Center for Astronomy, Zhejiang University, Hangzhou 310027, China}
        \and {National Astronomical Observatories, Chinese Academy of Sciences, Beijing 100101, China}
        \and {School of Astronomy and Space Science, University of Chinese Academy of Sciences, Beijing 100049, China}
        \and {Institute for Frontiers in Astronomy and Astrophysics, Beijing Normal University,  Beijing 102206, China}
        \and {Argelander-Institut f\"ur Astronomie, Universit\"at Bonn, Auf dem Hügel 71, D-53121 Bonn, Germany}
        \and {Institut f\"ur Astro- und Teilchenphysik, Universit\"at Innsbruck, Technikerstr. 25/8, 6020 Innsbruck, Austria}
        \and {Shanghai Astronomical Observatory (SHAO), Nandan Road 80, Shanghai 200030, China; University of Chinese Academy of Sciences, Beijing 100049, China}
        \and {CPPM, Aix-Marseille Université, CNRS/IN2P3, CPPM UMR 7346, F13288 Marseille, France}
        \and {Research School of Astronomy \& Astrophysics, Australian National University, Cotter Rd., Weston, ACT 2611, Australia}
        \and {Research School of Computer Science, Australian National University, Acton ACT 2601, Australia}
        \and {Department of Astronomy, Shanghai Jiao Tong University, Shanghai 200240, China}
        \and {Department of Physics and Astronomy, The Johns Hopkins University, 3400 N Charles Street, Baltimore, MD 21218, USA}
        \and {Shanghai Key Lab for Astrophysics, Shanghai Normal University, Shanghai 200234, China}
        \and {Ruhr University Bochum, Faculty of Physics and Astronomy, Astronomical Institute (AIRUB), German Centre for Cosmological Lensing, 44780 Bochum, Germany}
        \and {Institute of Physics, Laboratory of Astrophysics, École Polytechnique Fédérale de Lausanne (EPFL), Observatoire de Sauverny, CH-1290 Versoix, Switzerland; Aix Marseille Univ, CNRS, CNES, LAM, F13388 Marseille, France}
        \and {CEA Saclay, DFR/IRFU, Service d’Astrophysique, Bat. 709, 91191 Gif-sur-Yvette, France; Institut d'Astrophysique de Paris, 98 bis, Boulevard Arago 75014 Paris, France}
        \and {Kavli Institute for Astronomy and Astrophysics, Peking University, Beijing 100871, China}
             }
 

   \date{Received ***; accepted ***}

 
  \abstract
   {}
   {With the next generation of large surveys coming to the stage of observational cosmology soon, it is important to explore their potential synergies and to maximise their scientific outcomes. In this study, we aim to investigate the complementarity of the two upcoming space missions {\it Euclid} and the China Space Station Telescope (CSST), focusing on weak gravitational lensing for cosmology. In particular, we analyse the photometric redshift (photo-$z$) measurements by combining NUV, $u, g, r, i, z, y$ bands from CSST with VIS, $Y, J, H$ bands from {\it Euclid} and other optical bands from the ground-based {\it Vera C. Rubin} Observatory Legacy Survey of Space and Time (LSST) and Dark Energy Survey. The advantages of combining the two space observational data in simplifying image deblending are also studied. For {\it Euclid}, weak lensing measurements use the broad optical wavelength range of 550-900 nm, for which chromatic point-spread function (PSF) effects are significant. For this, the CSST narrow-band data in the optical can provide valuable information for {\it Euclid} to obtain more accurate PSF measurements, and to calibrate the colour and colour-gradient biases for galaxy shear measurements.
   }
   {We create image simulations, using the Hubble Deep UV data as the input catalogue, for different surveys, and quantify the photo-$z$ performance  using \texttt{EAZY} template fitting. For blending analyses, we employ high-resolution HST-ACS CANDELS F606W and F814W data to synthesize mock simulated data for {\it Euclid}, CSST, and an LSST-like survey. We analyse the blending fraction for different cases, and the blending effects on galaxy photometric measurements. Furthermore, we demonstrate that CSST can provide a large enough number of high signal-to-noise ratio multi-band galaxy images to calibrate the colour-gradient biases for {\it Euclid}.
   }
   {The sky coverage of {\it Euclid} lies entirely within the CSST footprint. The combination of {\it Euclid} with CSST data can thus be done more uniformly than with the various ground-based data that are part of the {\it Euclid} survey. Our studies show that by combining {\it Euclid} and CSST, we can reach a photo-$z$ precision of $\sigma_{\rm NMAD} \approx 0.04$, and an outlier fraction of $\eta\approx 2.4\%$ at the nominal depth of the {\it Euclid} Wide Survey ($\mathrm{VIS}<24.5$ AB mag).
   For CSST, including the {\it Euclid} $Y, J, H$ bands reduces the overall photo-$z$ outlier fraction from $\sim 8.5\%$ to $2.4\%$. For $z>1$, the improvements are even more significant. Because of the similarly high resolutions, the data combination of {\it Euclid} and CSST can be relatively straightforward for photometry measurements. On the other hand, to include ground-based data, sophisticated deblending utilizing priors from high-resolution space observations is demanded. The multi-band data from CSST are very helpful in controlling the chromatic PSF effect for {\it Euclid} VIS shear measurements. The colour-gradient bias for {\it Euclid} galaxies with different bulge-to-total flux ratio at different redshifts can be well calibrated to the level of $0.1\%$ using galaxies from the CSST deep survey.  }
   {}


   \maketitle

\section{Introduction}
Rapid observational developments in the past few decades have changed cosmological studies dramatically from lacking of data to data driven. Consequently, our understanding about the Universe has increasingly deepened, leading to the emergence of the cosmic concordance model. According to it, the two dark components, namely dark matter and dark energy, play essential roles in shaping the present Universe. On the other hand, however, their nature is still to be understood, which is one of the greatest challenges in the research of fundamental physics \citep[e.g.][]{Weinberg2013}. 

The dark components affect both the global cosmic expansion and the formation and evolution of large-scale structures (LSS). Thus observables from both aspects are sought to as sensitive probes \citep{albrecht06}. Among others, the weak lensing (WL) effect has been widely recognized as one of the major tools in cosmological studies \citep{FuFan2014, Kilbinger2015, Mandelbaum2018}. Arising from the gravitational light deflection by foreground LSS, WL leads to tiny shape distortions and flux changes of background objects \citep{BS2001}. By observing those changes, we therefore can extract the underlying matter distribution in an unbiased way. Furthermore, WL depends on the distances between observer, foreground lenses, and background sources, and thus is sensitive to the global expansion of the Universe. On the other hand, however, WL cosmological studies are observationally challenging because of the weakness of the signals. The past twenty years witnessed great improvements in high-precision photometric surveys, which have made WL one of the fastest developing areas in cosmological observations. 

Represented by CFHTLenS \citep{Fu2008, Heymans2012, Kilbinger2013}, Stage-II \citep{albrecht06} surveys demonstrated fully the observational feasibility of WL cosmological studies. We are now in the Stage-III era with a number of surveys observing a few thousand square degrees of sky area, resulting in high quality shear samples containing $10^{7}$ to $10^{8}$ galaxies \citep{KiDS, DES, HSC}. These surveys have already provided valuable cosmological constraints, complementary to other probes such as galaxy clustering or cosmic microwave background (CMB) anisotropies by breaking the degeneracy of different cosmological parameters \citep{Heymans2021, KiDS1000, DESY3, HSC2019}. 
In particular, they reveal a potential $S_8$ discrepancy within the $\Lambda$CDM paradigm between the measurements from low-redshift large-scale structures including WL and galaxy clustering, and from the CMB observations \citep{Hildebrandt2017, Hildebrandt2020, Heymans2021, Lemos2021, Planck2020}. Here $S_8\equiv\sigma_8(\Omega_{\rm m}/0.3)^{0.5}$ and $\Omega_{\rm m}$ and $\sigma_8$ are the present dimensionless matter density and the root-mean-square of the linearly-extrapolated matter density fluctuations over a top-hat scale of $8h^{-1}\hbox{ Mpc}$ with $h$ being the dimensionless Hubble constant in units of $100\hbox{ km/s/Mpc}$. This has stimulated great attentions because it may imply new physics beyond our current understanding. On the other hand, however, from the current surveys, the $S_8$ difference is marginal with a low statistical significance of $<3\sigma$. It is still under debate if this is a real problem for the $\Lambda$CDM cosmology \citep[e.g.][]{DESY3}.

To further enhance the power of cosmological constraints, Stage-IV surveys will be in operation within the next five years or so. These include the ground-based {\it Vera C. Rubin} Observatory Legacy Survey of Space and Time \citep[LSST;][]{LSST2019}, the space-based {\it Euclid} telescope \citep{laureijs11}, the Roman Space Telescope \citep[Roman;][]{Roman2015}, and the China Space Station Telescope \citep[CSST;][]{Zhan2011, Zhan2018, Fan2018, Gong2019, Zhan2021}. These surveys target at high precision WL measurements for over a billion galaxies, an order of magnitude more than that of the current surveys. Consequently, the statistical power will be greatly increased. The $S_8$ problem, if it indeed exists, will show up unambiguously.  Utilising WL and galaxy clustering statistical analyses, the  constraints on the equation-of-state parameters $w_0$ and $w_a$ of dark energy can expectedly reach the precision of about $0.02$ and $0.1$, respectively, about an order of magnitude tighter than that from the current survey studies  \citep[e.g.][]{Euclid-prep-VII_ISTF_20,Gong2019}. Here the equation of state of dark energy is assumed to be $p_{\rm{DE}}=w(z)\rho_{\rm{DE}}$ and $w(z)=w_0+w_a[z/(1+z)]$ with  $p_{\rm{DE}}$ and $\rho_{\rm{DE}}$ the pressure and energy density , and $z$ the redshift. This will allow us to answer the very fundamental question with high statistical significance if the dark energy is in the form of cosmological constant or dynamical in nature.

The full realisation of the statistical capability of Stage IV surveys requires stringent controls on different systematics. For WL cosmological studies, the multiplicative shear measurement bias needs to be known to an accuracy of about $10^{-3}$. The photometric redshift (photo-$z$) bias and scatter in different redshift bins should be less than $10^{-3}(1+z)$ and $0.05(1+z)$, respectively. The catastrophic outlier fraction of photo-$z$s needs to be less than $10\%$ \citep[e.g.][]{laureijs11}. Different Stage-IV surveys have different designs. How to utilise their data to improve the control of systematics and thus to maximise the science outcome is an important task that has triggered significant studies \citep{Jain2015, Rhodes2017, Chary2020, Eifler2021, 2022zndo...5836022G}. 

The {\it Euclid} and the CSST are both space-based projects with common scientific objectives. Their high spatial resolutions allow accurate WL shear measurements for galaxies of small sizes unachievable from ground-based observations. The fraction of image blending, a serious issue for deep ground-based surveys, is much reduced in the two space surveys, which can benefit significantly the accurate shear and photo-$z$ measurements. The {\it Euclid} imaging survey has one broad optical VIS band from 550\,nm to 900\,nm for accurate WL shear measurements, and three near infrared (NIR) bands of $Y$, $J$ and $H$ for photo-$z$ estimates \citep{laureijs11}. To obtain photo-$z$s with the required accuracies, the {\it Euclid} needs external multi-band optical data \citep[e.g.][]{Rhodes2017,Euclidz2020,Guinot2022}. For shear measurements using the broad VIS band, the wavelength-dependent point-spread function (PSF) and the differences between the spectral energy distribution (SED) of stars and galaxies can lead to significant biases for galaxy shear measurements \citep[e.g.][]{eriksen18}. In addition, the spatial variation of colours across a galaxy can also induce a colour-gradient shear bias. They need to be carefully dealt with. The CSST has seven optical bands of NUV, $u, g, r, i, z$ and $y$ \citep{Zhan2021} for imaging surveys, very complementary to the {\it Euclid}. For photo-$z$ measurements, the CSST can provide external optical data necessary for the {\it Euclid}. On the other hand, the inclusion of the {\it Euclid} NIR data can significantly improve the CSST photo-$z$s at redshift $z>1$. The similar spatial resolutions of the two space missions make the data combinations less complicated than combining space- and ground-based data. The CSST multi-band data in optical can also be very helpful in calibrating the colour related shear measurement biases for the {\it Euclid}. 

Seeing the great complementarities, in this paper, we present the first study aiming to explore the potential synergies between the {\it Euclid} and the CSST with a focus on WL cosmology. This paper is organized as follows. Sect.~2 describes the characteristics of the CSST and the {\it Euclid}, respectively. In Sect.~3, we present the possible synergies for WL studies. We briefly mention a few other aspects of cosmological studies that can benefit from the joint analyses of the two missions in Sect.~4. We discuss our results in Sect.~5.  

\section{Introduction to the CSST and the \emph{Euclid}}

\begin{figure*}
\centering
\includegraphics[width=0.96\textwidth]{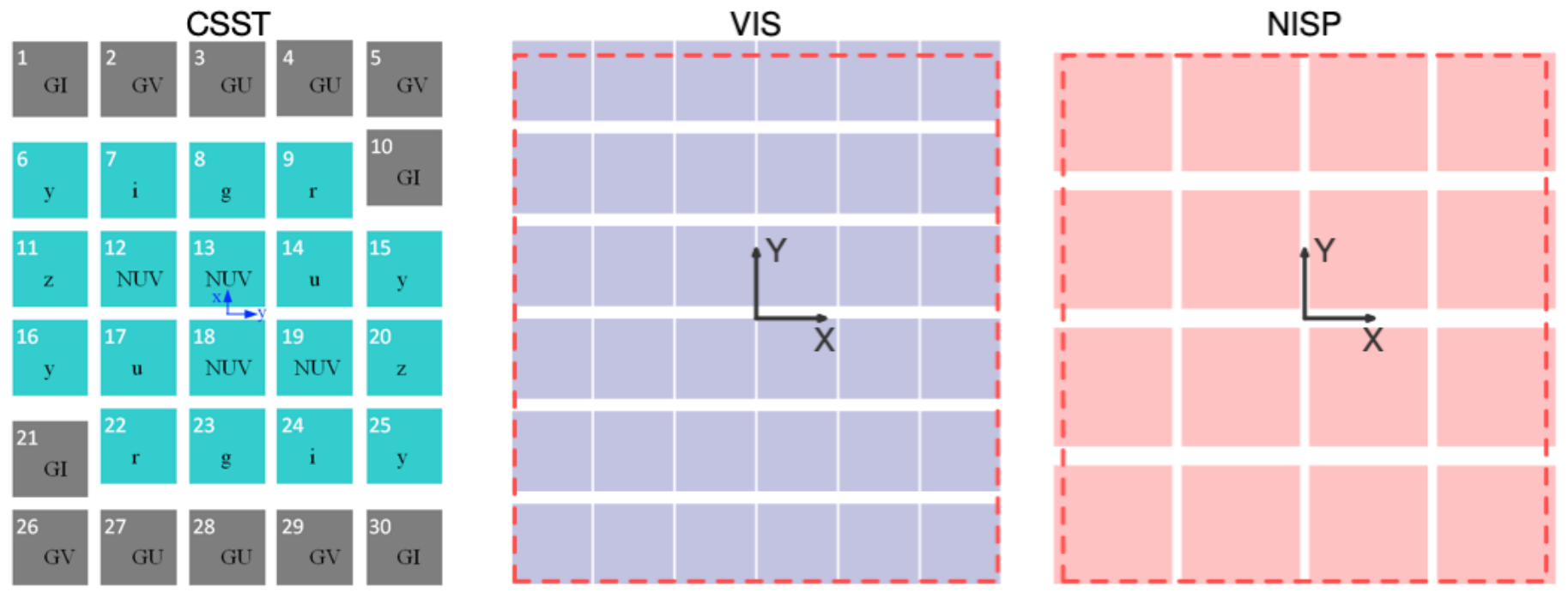}
\caption{{  Focal plane arrangements of the CSST wide-field survey camera ({\it left}), {\it Euclid} VIS ({\it middle}) and NISP ({\it right}). The X and Y axes in each camera define the reference frame in the focal plane. For CSST, the focal plane is segmented into 30 tiles with 18 for multi-band imaging (cyan tiles) and 12 for slitless spectroscopy (dark grey tiles). The layout of the filters and corresponding serial numbers for both imaging and spectroscopy are also shown. The 30 tiles, each of 9k$\times$9k pixels, are arranged into a $5\times6$ array with the FOV of $1\fdg1\times 1\fdg0$. For {\it Euclid} VIS and NISP, the schematic focal planes are taken from \citet{Scaramella2021}. The VIS imager comprises 36 4k$\times$4k CCDs with a pixel scale of $0\farcs1$, while the NISP includes 16 arrays of 2k$\times$2k HgCdTe detectors with a pixel scale of $0\farcs3$. The joint FOV of the two instruments, which is about $0.54$ deg$^2$, is shown in the red dashed frames.}}\label{fig:focalP}
\end{figure*}

\subsection{CSST}
CSST is a major science project of the China Manned Space Program and is planned to be launched around 2024 \citep{Zhan2011, Gong2019}. The telescope is a 2m $f/14$ Cook-type off-axis three-mirror anastigmat (TMA) system 
with five first-generation instruments: a survey camera, a terahertz receiver, a multichannel imager, an integral field spectrograph, and a cool planet imaging coronagraph \citep{Zhan2021}. A fold mirror is placed near the exit pupil to direct light to different instruments. The pointing accuracy of CSST is specified to be $\leq 5\arcsec/10\arcsec$ with stability of $\leq 0\farcs05/0\farcs85$ ($3\sigma$) over 300\,s with/without guide stars. 
The roll angle accuracy is $\leq 10 \arcsec$ with $\leq 1\farcs5$ stability ($3\sigma$) over 300\,s. The jitter requirement is $\leq 0\farcs01$ ($3\sigma$).

The wide-field survey camera will be used to carry out a large-area high-resolution multi-band imaging and slitless spectroscopic survey. The imaging survey data will enable precision cosmological studies with multiple probes, including weak lensing, strong lensing, and galaxy clustering. The spectroscopic survey will map the galaxy redshift distribution for extracting cosmic signals of the baryon acoustic oscillation and the redshift-space distortion at different redshifts.
The high-quality survey data will also enable a wide range of astrophysical studies, from galaxy formation and evolution to solar system objects. 


The primary focal plane of the survey camera is shown in the left panel of Fig.~\ref{fig:focalP}. It is segmented into $30$ tiles with $18$ for multi-band imaging, and $12$ for slitless spectroscopic observations. They are arranged into a $5\times 6$ array. The field-of-view (FOV) is $1\fdg1\times 1\fdg0$.
Covering the wavelength range from $255$ to $1000$\,nm, the imaging bands consist of NUV, {\it u}, {\it g}, {\it r}, {\it i}, {\it z} and {\it y}, and the pixel scale is $0\farcs074$. The slitless spectroscopy bands include GU (255--400nm), GV (400--620nm), and GI (620--1000nm) with the same pixel scale.
These different bands are distributed in the focal plane, with each  occupying 2 or 4 segments.
The targeted angular resolution for imaging observations is about $ 0\farcs15$  
in terms of the radius at 80\% encircled energy ($r{\rm EE80}$). The resolution has a slight band variation. The angular resolution for the slitless spectroscopy is $0\farcs3$/pixel,
and the average spectral resolution is $R\geq$ 200. 

The planned surveys for both imaging and spectroscopy include, but are not limited to, a wide-field survey over about 17\,500 deg$^2$ sky area and a deep-field survey of about 400 deg$^2$. The wide-field imaging survey will reach an average limiting magnitude better than 25.5\, AB mag at 5$\sigma$ for point sources. For the spectroscopic survey, it is $\sim$\,23.7\,mag.  They are about 1.2 mag deeper for the deep-field survey. For WL studies from the wide-field survey, the expected galaxy number density can reach $n_{\textrm{g}} \sim 20 \hbox{ arcmin}^{-2}$, and the median redshift is about $0.8$. 


Around the primary focal plane, a flux calibration unit and a NIR unit will be equipped.
The flux calibration unit is designed to assist flux calibration in {\it r}-band by observing stars that will be too bright for the main survey.
This will be achieved by placing the detector off the focus, allowing a saturation limit 4\,mag brighter than that of the main survey. 
The NIR unit has 8 HgCdTe detectors in two groups, imaging and slitless spectroscopy. 
Each group has four detectors, two in the $J^{\prime}$-band (0.9--1.3$\mu$m) and two in $H^{\prime}$-band (1.3--1.7$\mu$m).
Each detector has a FOV of $\sim1\farcm1\times 0\farcm9$ ($\sim 1/100$ of the detectors in the primary focal plane). 
The limiting magnitude for NIR imaging is $\sim 23.4$ mag in $J^{\prime}$  and $\sim 22.8$ mag in $H^{\prime}$ (150\,s exposure, point source 5$\sigma$). The spectral resolution for slitless spectroscopy is $R\geq 100$, and point sources brighter than $20.5$ mag are detectable.

\subsection{{\it Euclid}}

{\it Euclid} \citep{laureijs11} is an ESA M-class astrophysics and cosmology mission with a currently expected launch early 2023. For cosmology it will employ two primary probes, weak lensing and galaxy clustering, aiming to
constrain dark energy properties with  a figure of merit \mbox{$\mathrm{FoM}>400$} \citep[ e.g.][]{albrecht06}, the exponent $\gamma$ of the growth factor with a \mbox{$1\sigma$} precision of \mbox{$<0.02$}, the sum of the neutrino masses with a \mbox{$1\sigma$} precision of \mbox{$<0.03$} eV, and initial conditions of cosmic structure formation parameterised by \mbox{$f_\mathrm{NL}$}
to a \mbox{$1\sigma$} precision of \mbox{$\sim2$} in combination with {\it Planck} CMB observations \citep{Planck2020, Euclid-prep-VII_ISTF_20}.

{\it Euclid}'s payload includes a 1.2m Korsch  telescope equipped with two instruments, which simultaneously probe a common FOV of \mbox{$\sim 0.54$} deg$^2$. The optical VIS imager comprises 36 4k$\times$4k CCDs with $0\farcs1$/ pixel. The Near-Infrared Spectrograph and Photometer (NISP) includes 16 arrays of 2k$\times$2k NIR-sensitive HgCdTe detectors with a pixel scale of $0\farcs3$. {  The middle and right panels of Fig.~\ref{fig:focalP} show the focal planes of the two instruments.} {\it Euclid} will survey 15\,000 deg$^2$ of the extragalactic sky avoiding the ecliptic plane due to increased zodiacal background in its Wide Survey and use step and stare mode with four dither pointings per step \citep{Scaramella2021}, plus 40  deg$^2$ split over three deep fields\footnote{\url{https://www.cosmos.esa.int/web/euclid/euclid-survey}} with a depth increase by 2 mag.

NISP will obtain NIR imaging in $Y$, $J$, and $H$ bands for photometric redshifts reaching 24 AB mag at 5$\sigma$ for point sources in the Wide Survey, plus slitless spectroscopy for galaxy clustering studies with a flux limit of \mbox{$3\times 10^{-16}$} erg cm$^{-2}$ s$^{-1}$ at $3.5\sigma$ for a $1\arcsec$ diameter source.

VIS images will be taken in a broad \mbox{$550$--$900$} nm optical bandpass, reaching \mbox{$\sim 24.5$} AB mag at 10$\sigma$ for galaxies with a size larger than 1.25 times the full width at half-maximum (FWHM) of the PSF. 
These data will be used to obtain galaxy shape measurements for weak lensing. 
In order to optimise the thermal stability and minimise temporal PSF variations for WL measurements, {\it Euclid} will observe from an
orbit  around the Sun-Earth Lagrange point L2. In addition, the variation of space-craft orientations with respect to the Sun will be limited to maximise PSF stability.

For the WL analysis, galaxies need to be split into tomographic bins using photometric redshift. 
For this, the {\it Euclid} data need to be combined with depth-matched multi-band optical observations, such as, from ground-based wide-field imaging surveys. These optical colour measurements are also needed to account for the colour-dependence of the diffraction-limited VIS PSF when modelling the PSF and measuring galaxy shapes \citep{cypriano10,eriksen18}.
Shape measurements based on VIS will also require at least statistical corrections for the impact of galaxy colour gradients, which cannot be constrained individually from seeing-limited colour measurements \citep{voigt12,semboloni13,er18}.

\section{Synergies for weak lensing science}

In this section, we present the aspects in WL studies that can benefit from the joint analyses of the two missions.   


\subsection{Photometric redshifts}
\label{sec:photoz}

Accurate photo-$z$ measurements have become an indispensable part of weak lensing cosmology \citep{Hildebrandt2010, Newman2015,Tanaka2018, Hildebrandt2021,Myles2021}. The accuracy of photo-$z$ depends critically on the wavelength coverage and the filter bands employed in observations \citep{2017AJ....153...53L,Salvato2019}. While optical bands are the core for photo-$z$ measurements, the inclusion of NIR bands is important to improve photo-$z$ for $z>1$. Additionally, observations of $u$ and NUV bands at shorter wavelength can reduce the low-redshift outlier fraction by correctly disentangling the $4000\, \mathring {A}$ break and the redshifted Lyman break and Lyman line features \citep{Rafelski2015}. 

 CSST has seven filters consisting of NUV, $u, g, r, i, z, y$. For {\it Euclid}, there is a single broad VIS band in the optical from 550-900\,nm 
 mainly for galaxy weak lensing shear measurements, and three NIR bands $Y, J, H$ for photo-$z$ determination. The optimal utilization of the data from the two surveys can improve the photo-$z$ accuracy profoundly. In addition, the two space-based surveys have similar resolutions in optical bands, and thus the complexity in image deblending can be much less than that of combining space- and ground-based data. 
 
 Below we present our investigations on the photo-$z$ accuracy by utilizing CSST and {\it Euclid} observations. For comparison, we also study combinations of {\it Euclid} and ground-based observations that match the characteristics of the LSST \citep{LSST2019} and the Stage-III Dark Energy Survey \citep[DES;][]{DES}. Table \ref{tab:bandsinfo} shows the band information and the designed performance parameters of each survey used in our analyses. {  In our study here, four setups are considered for the photo-$z$ analyses: 1) CSST-only setup containing 7 optical bands, 2) {\it Euclid}+CSST setup with 11 optical and NIR bands, 3) {\it Euclid}+LSST-like setup having 10 optical and NIR bands, and 4) {\it Euclid}+DES-like setup consisting of 9 optical and NIR bands. As stated in \citet{Euclidz2020}, the {\it Euclid} WL shear sample is defined by a set of galaxies with VIS-band signal-to-noise ratio higher than 10, corresponding to the magnitude depth of $m_\mathrm{VIS}<24.5$\,mag, and being not flagged as having bad photometry or active galactic nuclei (AGNs). In addition, concerning  the reliability of photo-$z$ estimates, the {\it Euclid} shear sample is set to be in the range of $0.2<z_{\rm {phot}}<2.6$. In this work, we apply the same criteria to analyze the photo-$z$ accuracy. Detailed image simulation and photo-$z$ measurements are presented as follows.}
 

\subsubsection{Method and simulations}
\label{sec:methods_and_sims}


{For the photo-$z$ analyses here, we generate simulated mocks based on the Hubble Deep UV (HDUV) Legacy Survey GOODS-N/CANDELS-Deep data catalogue \citep{Oesch2018}. The sky area is $56.5\hbox{ arcmin}^2$ centred on (RA, DEC)=($12^{\rm h}37^{\rm m}$, $62^{\circ }14\arcmin$). The source detection of the HDUV catalogue is based on the coadded images of the three HST WFC3 bands $F125W$, $F140W$ and $F160W$ with an average PSF ${\mathrm{FWHM}}\sim 0\farcs19$ \citep{Skelton2014}. This is comparable to the resolution of the {\it Euclid} VIS band and of the CSST bands, and thus this catalogue is very suitable for our studies. In contrast, for the COSMOS2015 sample \citep{laigle16} that has been extensively used in photo-$z$ studies \citep[e.g.][]{Rhodes2017, cao18}, the source detection is primarily based on UltraVISTA NIR bands \citep{McCracken2012} and Subaru $z^{++}$ band \citep{Taniguchi2015}. These images have PSFs with a typical value of $\mathrm{FWHM}\sim 0\farcs6$. This lower resolution resulted in a lower number density of galaxies in the COSMOS2015 catalogue than that of HDUV. For our studies here concerning the space-based {\it Euclid} and CSST, photo-$z$ analyses based on the high-resolution HDUV catalogue are optimal.} 

{  With the HDUV catalogue as the input, we use \texttt{GALSIM}, a flexible framework to simulate galaxy images \citep{2015A&C....10..121R}, to perform image simulations for different considered bands at the pixel level.}  The procedures are as follows.

\begin{table*}
\centering
\caption{Designed performance of CSST, {\it Euclid}, LSST, and DES. }
\label{tab:bandsinfo}
\begin{tabular}{@{\extracolsep{1em}}ccccccl}
\toprule
Telescope/              & Band     & $\lambda_\mathrm{eff}$ & $\Delta \lambda (\mathrm{FWHM})$ & Detection Limit & Pixel Scale & PSF Size\\
Project                 &          &  (\AA)                 &    (\AA)                &   (mag)         &    (arcsec) & (arcsec)\\
\midrule
CSST\tablefoottext{a}   & NUV      &   2880                 &    694                  &    25.4         &             & 0.135\\
                        & {\it u}  &   3726                 &    866                  &    25.4         &             & 0.135\\
                        & {\it g}  &   4734                 &    1455                 &    26.3         &             & 0.135\\
                        & {\it r}  &   6107                 &    1417                 &    26.0         & 0.074       & 0.135\\
                        & {\it i}  &   7548                 &    1465                 &    25.9         &             & 0.145\\
                        & {\it z}  &   8975                 &    1082                 &    25.2         &             & 0.165\\
                        & {\it y}  &   9606                 &    542                  &    24.4         &             & 0.165\\
\midrule
{\it Euclid}\tablefoottext{b} & VIS      &   6726                 &    3699                 &    24.5         & 0.1         &  0.18\\
                        & $Y$        &   10\,678                &    2665                 &    24.0         & 0.3         &  0.62\\
                        & $J$        &   13\,333                &    4052                 &    24.0         & 0.3         &  0.63\\
                        & $H$        &   17\,328                &    5023                 &    24.0         & 0.3         &  0.70\\
\midrule
LSST\tablefoottext{c}   & {\it u}  &   3734                 &    623                  &    26.1         &             & 0.81\\
                        & {\it g}  &   4731                 &    1427                 &    27.4         &             & 0.77\\
                        & {\it r}  &   6139                 &    1359                 &    27.5         &             & 0.73\\
                        & {\it i}  &   7487                 &    1247                 &    26.8         & 0.2         & 0.69\\
                        & {\it z}  &   8671                 &    1022                 &    26.1         &             & 0.68\\
                        & {\it y}  &   9677                 &    855                  &    24.9         &             & 0.71\\
\midrule
DES\tablefoottext{d}    & {\it g}  &   4734                 &    1295                 &    24.7         &             & 1.11\\
                        & {\it r}  &   6342                 &    1485                 &    24.4         &             & 0.95\\
                        & {\it i}  &   7748                 &    1480                 &    23.8         & 0.263       &  0.88 \\
                        & {\it z}  &   9139                 &    1475                 &    23.1         &             & 0.83 \\
                        & {\it y}  &   9880                 &    660                  &    21.7         &             & 0.90 \\
\bottomrule
\end{tabular}
\tablefoot{\tablefoottext{a}{{  The detection limits are the 5$\sigma$ limiting magnitudes of point sources  and the PSF sizes are the radii at 80\% encircled energy \citep{Zhan2021}}}; 
\tablefoottext{b}{For {\it Euclid} VIS band, the detection limit is the 10$\sigma$ limiting magnitude of extended sources and the PSF size is defined as the FWHM, while for {\it Euclid} NIR bands, the limits are the 5$\sigma$ limiting magnitudes of point sources and the PSF sizes are the radii at 80\% encircled energy \citep{laureijs11}};
\tablefoottext{c}{The detection limits for LSST bands are the 5$\sigma$ depth of point sources after ten year observations and the PSF sizes are the FWHM \citep{LSST2019}};
\tablefoottext{d}{The detection limits for DES bands are the 10$\sigma$ depth of point sources and the PSF sizes are the FWHM \citep{2021arXiv210105765D}.}}
\end{table*}

\vspace{2mm}
\noindent (1) {We select galaxies in the HDUV sample by setting the parameter \texttt{use\_phot==1}, which combines a number of criteria described in \citet{Skelton2014}, such as not being a star, not being close to a bright star, etc.. {  In real observations, bright stars can affect the photometry of nearby galaxies and may lead to catastrophic photo-$z$ measurements. Therefore, in WL studies  bright stars need to be masked with specified radii to reduce their effects  \citep[e.g.][]{KiDS}. Here we rely on the flag of \texttt{use\_phot==1} in the original HDUV catalogue that already excludes galaxies close to bright stars. Therefore the photometry of the selected galaxies should not be affected significantly \citep{Skelton2014}.  In addition, in our simulation analyses, we only use the position and the SED information of the galaxies, and their images are generated with \texttt{GALSIM}. We therefore expect negligible impacts from bright stars on our studies. For real {\it Euclid} and CSST data processing, however, this issue needs to be carefully dealt with, which is beyond the scope of the current paper. After applying the cut, the resulting total number of galaxies is $13\,790$.} For each galaxy, there are data from 24 bands covering a wavelength range from $\sim 0.25-8\,\mu m$ \citep{Oesch2018}. The size and orientation parameters of these galaxies are the 3D-HST GOODS-N catalogue \citep{Skelton2014}, which contains the relevant information measured from the coadded images of the three HST WFC3 bands $F125W$, $F140W$ and $F160W$.}


\noindent (2) {We estimate the SED of each individual galaxy in the above sample by fitting the multi-band data from HDUV using the \texttt{EAZY} template fitting code \citep{Brammer2008}. We note that the original photo-$z$s of the HDUV galaxies are derived using the same code but with additional template optimisation and zero-point corrections \citep{Skelton2014, Oesch2018}. Because of the unavailability of the optimised templates, here we do not use the photo-$z$ information given in the catalogue in obtaining the best-fit SED for a galaxy. Instead, we fit its photo-$z$ and the SED simultaneously from its 24-band data using the fiducial set of \texttt{EAZY} templates, and regard them as the ground truth of the galaxy. Specifically, we take the peak value of the photo-$z$ distribution of a galaxy, $z_{\rm peak}$, from \texttt{EAZY} as its true redshift, denoted as $z_{\rm input}$. {  We compare the derived $z_{\rm peak}$ with those in the original catalogue, and in general they are in good agreements with a scatter $\sigma_\mathrm{NMAD}=0.018$ and the outlier fraction $\eta=1.87\%$ (see the definitions in Sect.~\ref{subsec:photozRes}).} Although our photo-$z$s may not be as accurate as those given in the catalogue due to the lack of SED optimisation, for the purpose of the studies here, our approach is self-consistent.}

\noindent (3) With the best-fitted SED for a galaxy from (2), \texttt{EAZY} calculates its flux in each band of the surveys considered in this study. These fluxes are adopted to generate the idealized images of the galaxy. Its intrinsic size $\sigma$ is estimated by $\sigma=\sqrt{\sigma^2_o-\sigma^2_{\rm psf}}$, where $\sigma_o$ is the half-light radius from the 3D-HST catalogue mentioned in step (1), 
and $\sigma_{\rm psf}$ is the PSF half-light radius calculated from \mbox{$\mathrm{FWHM}=0\farcs19$} assuming Gaussianity \citep{Skelton2014}. We then employ the bulge+disk model for galaxies and adopt the de Vaucouleurs profile for the bulge and the Sersic profile with $n=1$ for the disk, and set their intrinsic half-light radii both to be $\sigma$. In reality, bulge and disk can have different size distributions. But our simplified treatment here should not have significant impacts on our photo-$z$ studies because the overall size of a galaxy is consistent with that of the 3D-HST measurement, and the effect of colour gradient is not considered in our analyses here. {For the bulge to total flux ratio ${\rm {B/T}}$, the input distribution in our simulations is taken to be consistent with that used in \citet{Miller2013}. The bulge-dominated elliptical galaxies have ${\rm {B/T}}=1$ and account for $10\%$ of galaxies. For the rest of disk-dominated galaxies, it is taken to be a truncated Gaussian distribution in the range of $0\le {\rm {B/T}}<1$ with the peak at ${\rm {B/T}}=0$ and the dispersion of $0.1$.} As discussed in \citet{Miller2013}, this distribution is a reasonable representation of the observed bulge fraction found at low and high redshifts  \citep[e.g.][]{1996ApJ...464...79S,2002ApJS..142....1S}. With the position, size and ${\rm {B/T}}$ information of a galaxy, we first simulate a circular image, and then stretch it to an elliptical shape with an axis ratio and orientation consistent with the 3D-HST measurements \citep{Skelton2014}.



\noindent (4) {  For the model images from (3), circular Gaussian PSFs are applied to generate PSF-smeared images\footnote{For the circular Gaussian PSF, the conversion between the FWHM and the radius $r_{c}$ at a given percent (defined as 100c\%) encircled energy is $\mathrm{FWHM} = 2r_{c}\times \sqrt{\frac{\ln(2)}{\ln(1/(1-c))}}$.}.  The PSF sizes of different bands of different surveys are listed in  Table~\ref{tab:bandsinfo}. We then use \texttt{GALSIM} to simulate galaxy multi-band images on pixels with the corresponding pixel scales also shown in Table~\ref{tab:bandsinfo}. The photon shooting method implemented in \texttt{GALSIM} enables to include Poisson noise in the simulated galaxy images. The resulting flux counts between any two pixels are uncorrelated.  The galaxy positions are taken to be the same as that given in the HDUV catalogue.} 


\noindent (5) {{  Finally, different random Gaussian background noises
are added to the images of different bands of different surveys to match their limiting magnitudes. Unlike real observations that normally need to stack multiple exposures to achieve the limiting magnitudes and thus introduce pixel-by-pixel noise correlations \citep[e.g.][]{2000AJ....120.2747C,2007ApJS..172..219L,2013ApJS..207...24G,Skelton2014}, the random Gaussian background noise added here for each band is uncorrelated between adjacent pixels. Because of the simplicity in generating the simulated images, we do not generate separate weight maps that are typically needed in real observations to characterise the noise properties.} 

The input sample size from HDUV GOODS-N is relatively small, especially after applying a cut with {\it Euclid} VIS-band ${\rm {mag}}<24.5$. To control the impact of statistical fluctuations on the analysis results, for each survey, we generate 20 sets of images with different Gaussian noise realizations using different random seeds. We then regard these 20 sets as independent mock images of the survey for photo-$z$ analyses. This approach effectively suppresses the statistical uncertainties from noise. On the other hand, however, the cosmic variance intrinsic to the input catalogue still exists.} 

\vspace{2mm}

\begin{figure}
\centering
\includegraphics[width=0.48\textwidth]{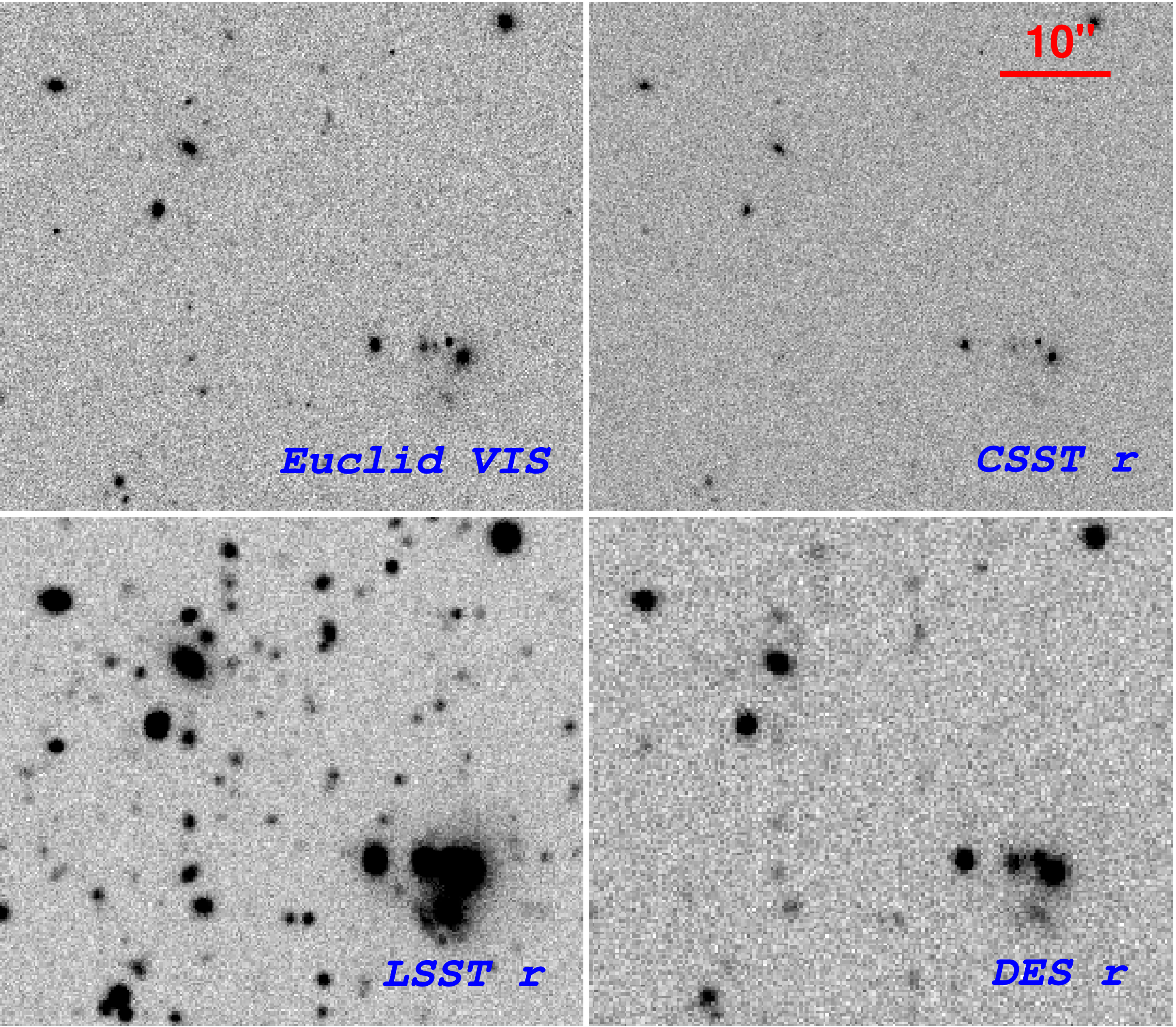}
\caption{An example of simulated images for {\it Euclid} VIS band (upper left), CSST $r$ band (upper right), LSST-like $r$ band (lower left) and DES-like $r$ band (lower right), respectively. }\label{fig:imgSim}
\end{figure}

\begin{figure*}
\centering
\includegraphics[width=0.96\textwidth]{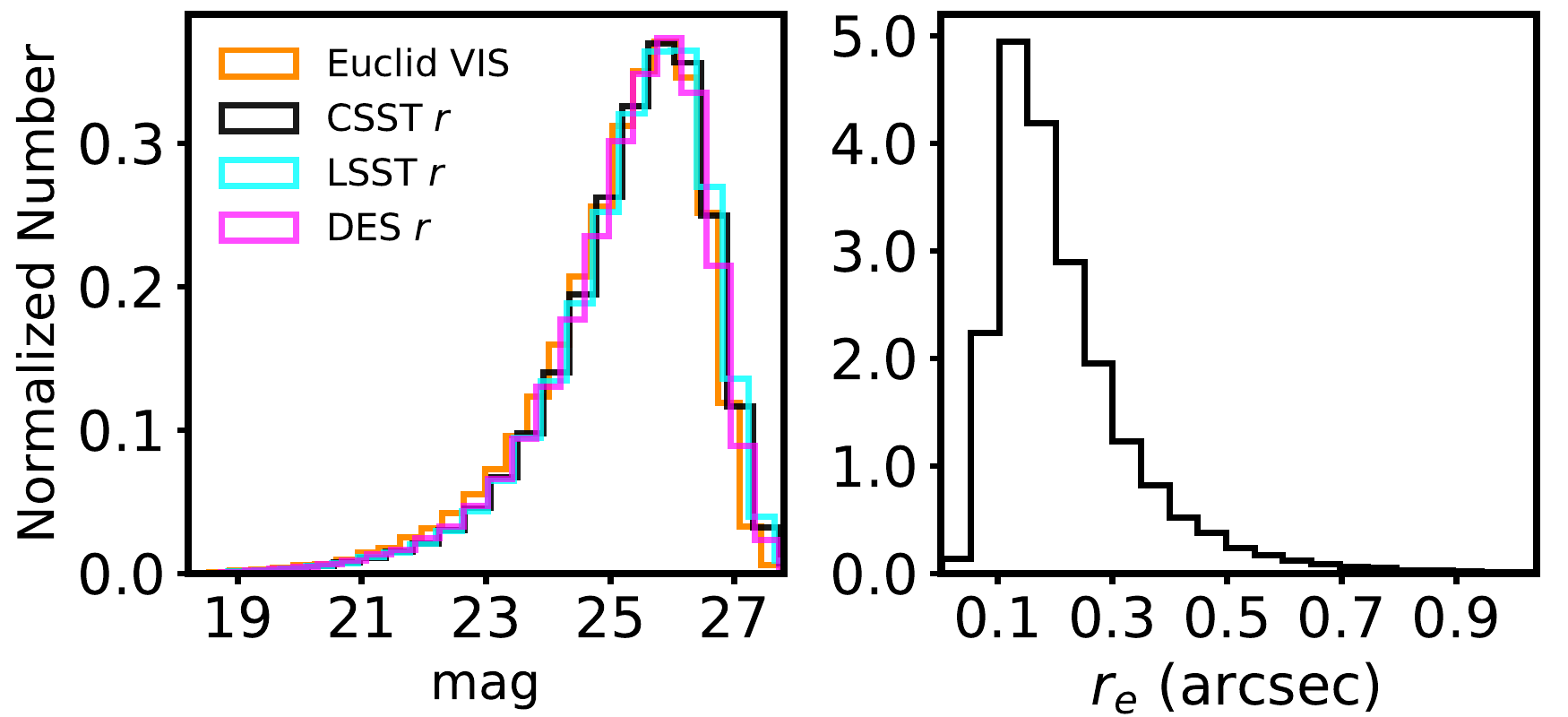}
\caption{The input magnitude distributions in the {\it Euclid} VIS band and the $r$ bands of the CSST, LSST-like and DES-like setups (left)  and the size distribution (right) of the simulated galaxies.}\label{fig:magsize}
\end{figure*}

\begin{figure}
\centering
\includegraphics[width=0.48\textwidth]{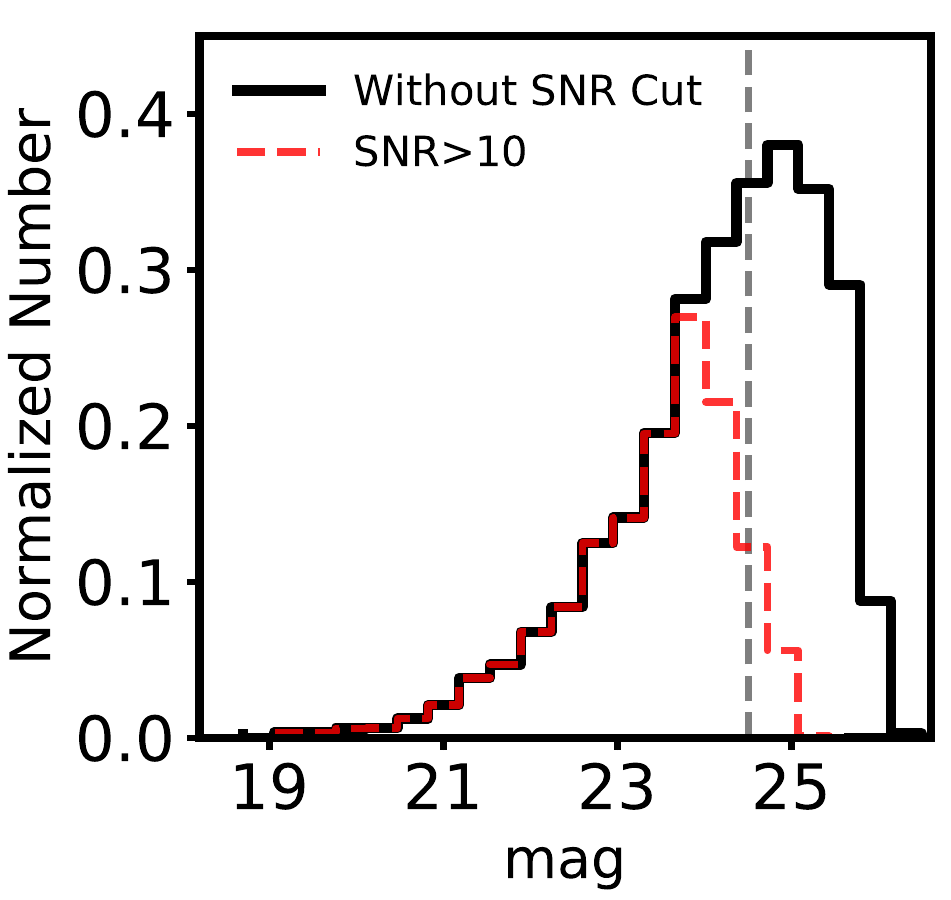}
\caption{The magnitude distribution of the detected galaxies from {\it Euclid} VIS-band images (black) and the ones with ${\rm {S/N}}>10$ (red). The vertical dashed line is for $\rm {mag}=24.5$.}\label{fig:magEuclid}
\end{figure}

In total, we generate 22 sets of mock simulations corresponding to the 22 different bands, 7 for CSST, 4 for {\it Euclid}, 6 for LSST-like {  setup} and 5 for DES-like setup. {For each band, we have 20 sets of images with different noise realizations.}
Each image contains $13\,790$ input galaxies covering an area of $56.5\hbox{ arcmin}^2$ from HDUV GOODS-N. 
Fig. \ref{fig:imgSim} shows examples of the simulated images for the {\it Euclid} VIS band, CSST $r$ band, LSST-like $r$ band, and DES-like $r$ band, respectively. Different survey characteristics are clearly seen. In particular, galaxies that are severely blended in ground-based observations can be better resolved in the two space-based surveys. In Fig.~\ref{fig:magsize}, we present the synthetic magnitude distributions (left) of the four surveys and the distribution of the intrinsic half-light radius $\sigma$ (right) of the simulated galaxies from step (3). 

{  With the mock images, we perform galaxy detection based on the  {\it Euclid} VIS images employing \texttt{SExtractor}, and determine the 2-D elliptical aperture, known as the Kron radius \citep{1980ApJS...43..305K}, in unit of arcsec for each galaxy. The parameters of the scaling factor $k$ and the minimum radius $R_\mathrm{min}$ of a Kron ellipse in  \texttt{PHOT\_AUTOPARAMS} are set to be the default values of 2.5 and 3.5, respectively \citep{BA1996}. The corresponding flux and magnitude enclosed within the elliptical aperture are defined as \texttt{FLUX\_AUTO} and \texttt{MAG\_AUTO} in \texttt{SExtractor}, respectively.} Fig. \ref{fig:magEuclid} shows the \texttt{SExtractor} \texttt{MAG\_AUTO} magnitude distribution of the detected galaxies in the solid histogram, while the dashed histogram corresponds to the distribution of galaxies with signal-to-noise (S/N) ratio above 10. {  The S/N is defined by the inverse of the magnitude error $\texttt{MAGERR\_AUTO}$, that is S/N=1.0857/\texttt{MAGERR\_AUTO}, which is a robust estimate here because the noise correlations do not presented in our simulated images \citep{1991PASP..103..122N,BA1996}.} This figure shows that substantial incompleteness occurs for \texttt{MAG\_AUTO} $\gtrsim24$, approximately matching the designed limit for {\it Euclid}. {From the 20 sets of {\it Euclid} VIS images, we obtain the average number of detected galaxies with \texttt{MAG\_AUTO}$<24.5$ being $1852$ in the area of $56.5\hbox{ arcmin}^2$. The corresponding number density is $\sim 32.8 \,\hbox{arcmin}^{-2}$, well in line with the {\it Euclid} specification. For comparison, we also perform parallel mock simulations using the COSMOS2015 catalogue as the input. The corresponding number density of {\it Euclid} VIS-detected galaxies is $17.5\,\hbox{arcmin}^{-2}$, much lower than that from our fiducial simulations based on the HDUV catalogue.} 

With the aperture determined from the VIS band, we perform forced photometry for each galaxy at the other bands. {  To ensure that the flux measurements from different bands contain the same fraction of the total flux, we perform PSF homogenization for all the bands. In our analyses, we assume the PSFs of different bands follow the two-dimensional circular Gaussian function, and thus the homogenization kernels are also Gaussian. Therefore for a certain band $j$, the kernel size $\sigma_\mathrm{kernel}$ can be derived as $\sigma_\mathrm{kernel} = \sqrt{\sigma_{0}^2 - \sigma_{j}^2}$, where $\sigma_{j}$ is the PSF size of the band $j$, and $\sigma_{0}$ is the size of the target PSF (denoted as $\mathrm{PSF_{0}}$). We then convolve the image of band $j$ with the Gaussian kernel to get the PSF homogenized image. This is performed through the \texttt{convolve\_fft} function implemented in \texttt{Astropy} which is a community Python package for astronomy \citep{2013A&A...558A..33A}. To validate the quality of the convolution, we calculate the fractional error  $(\mathrm{PSF_{j}^{conv}} - \mathrm{PSF_{0}})/\mathrm{PSF_{0}}$, where $\mathrm{PSF_{j}^{conv}}$ is the convolved PSF of band $j$. It turns out that the fractional error is less than $10^{-4}$, indicating that the PSF homogenization is accurate.  We note that for real observations, the PSF profiles can be more complicated. Thus the PSF homogenization needs to be done with sophisticated care \citep{2012MNRAS.421.2355H,Skelton2014,2019A&A...632A..34W}.

The specific PSF homogenizations for different data sets are as follows. For  the CSST-only setup, the homogenization is based on the CSST $y$-band PSF which has the largest size with \mbox{$\mathrm{FWHM}=0\farcs22$}. For the {\it Euclid}+DES-like setup, the DES-like $g$-band PSF is used for homogenization. For the {\it Euclid}+LSST-like setup, we use the {\it Euclid} $H$-band PSF for the homogenization. For the {\it Euclid}+CSST, because the large difference between the PSF sizes in the optical and NIR bands, we adopt a similar strategy as in \citet{Skelton2014} to perform homogeneous photometry. For the CSST and {\it Euclid} optical bands, we homogenize the images based on the CSST $y$-band PSF,  while for the {\it Euclid} NIR images, the homogenization is based on the $H$-band PSF.} 

Because of the different treatment in the case of {\it Euclid}+CSST, we need to convert the measured aperture flux to the total flux of a galaxy.  This is done using
\begin{equation}
f_j=f_{\rm VIS}/\tilde f_{\rm VIS}^{j}\times \tilde f_j\; ,
\label{flux}
\end{equation}
where the scaling factor $f_{\rm VIS}/\tilde f_{\rm VIS}^{j}$ is the ratio of the total galaxy flux measured from the mock {\it Euclid} VIS images without PSF homogenization and the aperture flux of the corresponding PSF-homogenized {\it Euclid} VIS images,  and $\tilde f_j$ is the aperture flux measured from the homogenized images of band $j$. {  The \texttt{FLUX\_AUTO} measured by \texttt{SExtractor} is used as the representation of the total flux $f_{\rm VIS}$  of the galaxy. The aperture flux is measured by a circular aperture of radius $r=b/\sqrt{2}$, where $b$ is the semi-minor axis of the Kron ellipse which is used to measure the total flux $f_{\rm VIS}$. Since the Kron ellipses are different for different galaxies, the circular apertures are adaptive for different galaxies.  We test this by applying different circular aperture settings, with $r=n\times b/\sqrt{2}$ where n ranges from 0.6 to 2.0 with a step of 0.2, and find that $n=1$ gives the best photo-$z$ results which are presented in Sect.~\ref{subsec:photozRes}. The photo-$z$ analyses with fixed apertures for all galaxies are also  discussed in the next section.} After the PSF homogenization the PSFs of the {\it Euclid} VIS images and the band $j$ images are expected to be identical. {  For the CSST and {\it Euclid} optical bands, $\tilde f^j_{\rm VIS}$ is the aperture flux measured from the VIS-band images after being homogenized to the CSST $y$-band PSF, 
while for the {\it Euclid} NIR bands, $\tilde f^j_{\rm VIS}$ is measured from the VIS-band images after being homogenized to the $H$-band PSF.} For the other three 
setups, we also apply the conversion with the ratio being the same for all bands because a uniform homogenization is applied.

{  Differently from the original simulated images, it is noted that the PSF homogenization procedure can result in  correlated noise between adjacent pixels. The flux errors measured by \texttt{SExtractor}  are therefore underestimated since it does not consider the correlated noise \citep[e.g.][]{2007ApJS..172..219L}. To estimate the flux errors with the presence of correlated noise, we follow the method as in \citet{Skelton2014} which is also adopted in many other studies \citep[e.g.][]{2006ApJS..162....1G, 2017AJ....153...53L}. For a specific aperture, the flux errors of galaxies in the PSF homogenized images of band $j$ can be estimated by 
\begin{equation}\label{eq:cnoise}
    \sigma_{\tilde f_{j}} = \sqrt{\tilde f_j + A^{2\beta}\sigma_\mathrm{bkg}^2},
\end{equation}
where $A$ is the pixel area in a given photometric aperture, $\sigma_\mathrm{bkg}$ is the standard deviation of the background noise, and $\beta$ is the correlation parameter to be determined. In the case of no correlation presented, $\beta=0.5$, which is the formula used in \texttt{SExtractor}. On the other hand, if the adjacent pixels are perfectly correlated, $\beta=1.0$. Generally, the value of $\beta$ is within the range of $0.5<\beta<1.0$. To determine $\beta$ of band $j$, we randomly generate a set of $>$2000 positions on the image that do not overlap with the detected objects. The fluxes are then measured for each position with different apertures. For a given aperture, a Gaussian function is applied to fit the distribution of the measured fluxes to derive the noise dispersion $\sigma$. It turns out that this dispersion $\sigma$ gets larger with the increase of aperture size. We then use the power-law equation $\sigma \sim A^{\beta}$ to fit the measured noise dispersion to find the best-fit value of $\beta$. The derived $\beta$ values, typically ranging from 0.5 to 0.6 for different bands, are applied to  Eq.~(\ref{eq:cnoise}) to calculate the flux errors. Finally, the flux errors are corrected to total $\sigma_{f_j}$ in the same way as for the fluxes (see Eq.~(\ref{flux}) above).
}

For photo-$z$ measurements with \texttt{EAZY}, we use the $f_j$ and corresponding $\sigma_{f_j}$ as input, and denote the obtained photo-$z$ as $z_{\rm output}$. {  Because the purpose in this work is to compare the photo-$z$s under different survey setups, we run \texttt{EAZY} with the same input configuration used to generate the SED templates. The configuration here ignores the template uncertainties in photo-$z$ estimates which is different from real cases that the optimizations of SED templates are necessary to improve the photo-$z$ quality as the real galaxy SEDs are not exactly the same as the input templates \citep{Skelton2014}.}




\subsubsection{Results}\label{subsec:photozRes}

\begin{figure}
\centering
\includegraphics[width=0.48\textwidth]{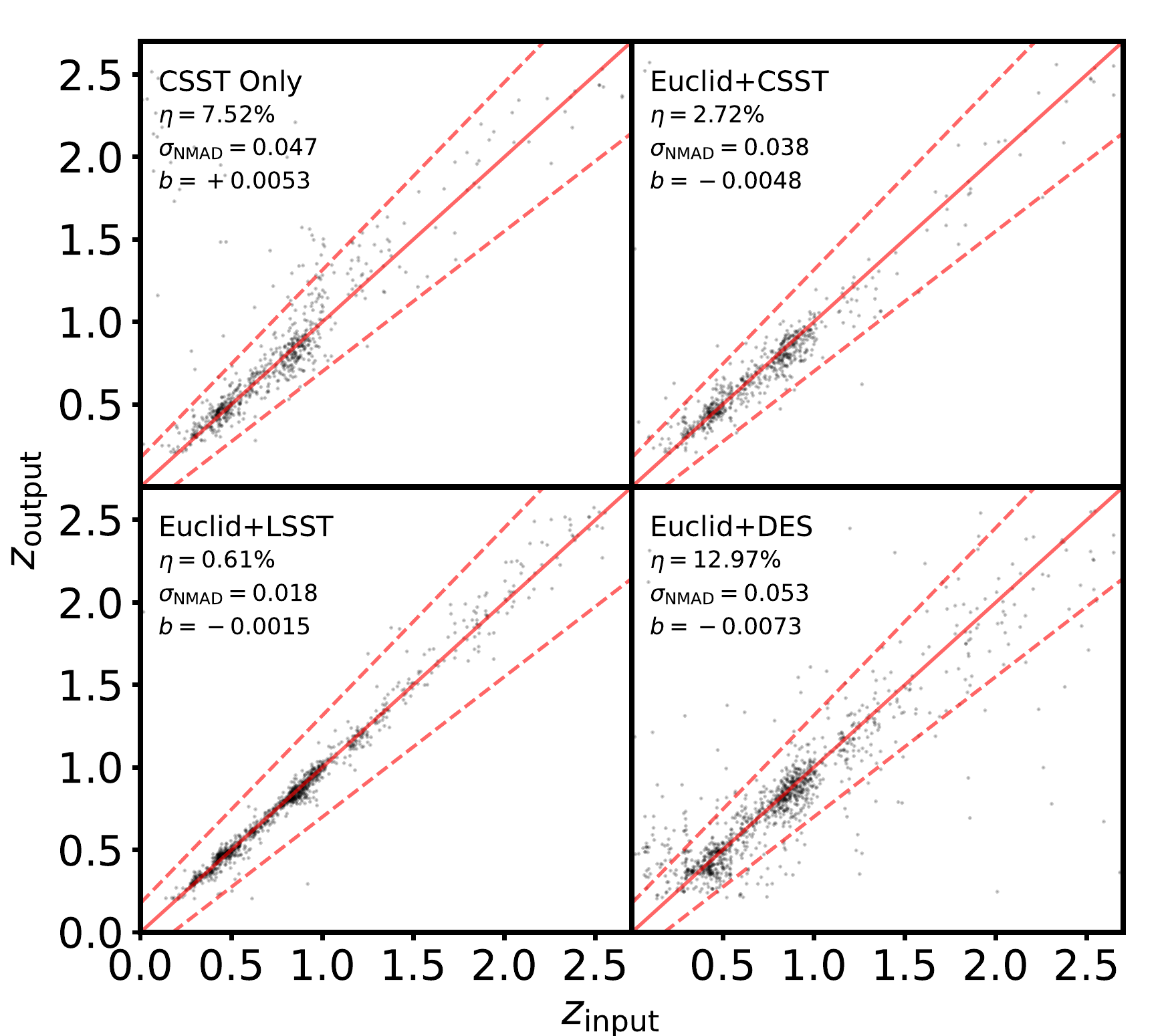}
\includegraphics[width=0.48\textwidth]{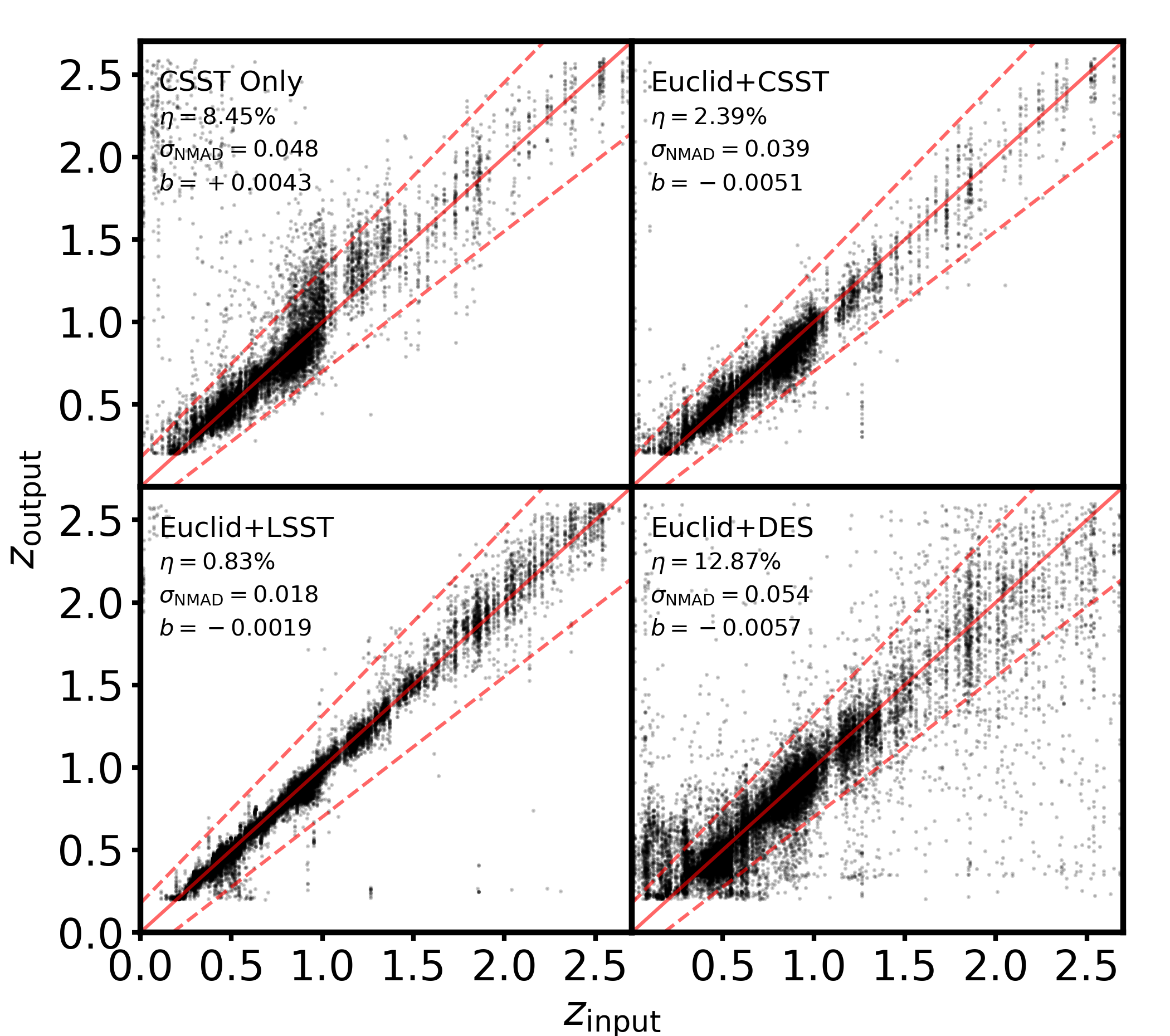}
\caption{{  Top: photo-$z$ results from data with only one noise realization. Each panel shows the $z_{\rm output}$ vs. $z_{\rm input}$ for CSST-only (upper left), {\it Euclid}+CSST (upper right), {\it Euclid}+LSST-like setup (lower left) and {\it Euclid}+DES-like setup (lower right), respectively. The red solid and dashed lines indicate $z_{\rm photo}=z_{\rm input}$, and $|\delta{z}|/(1+z_{\rm input})=0.15$, respectively. Bottom: photo-$z$ results from 20 sets of data with  different noise realizations.}}
\label{fig:zphCom}
\end{figure}

{We show our photo-$z$ results in Fig.~\ref{fig:zphCom}, where the upper left and right panels are for CSST-only and {\it Euclid}+CSST, and the lower left and right ones are for {\it Euclid}+LSST-like and {\it Euclid}+DES-like setups, respectively. {  Here the top plots show the results from data with a specific noise realization, while the bottom plots present the scatters from 20 sets of data with different noise realizations as described in the step (5) in Sect.~\ref{sec:methods_and_sims}. The vertical stripes seen in the bottom plots reflect the statistical fluctuations of different noise realizations.}} The photo-$z$ statistics for galaxies with VIS $< 24.5$ and $0.2<z_{\rm output}<2.6$ are shown in Table \ref{tab:photozstat}. Here we use the conventional definitions with the bias given as 
\begin{equation}
b = \left\langle \frac{\delta{z}}{1+z_\mathrm{input}} \right\rangle\; ,
\label{zpbias}
\end{equation}
with $\delta z= z_{\rm output}-z_{\rm input}$, 
the normalized median absolute deviation of the residuals $\sigma_{\rm NMAD}$ calculated by 
\begin{equation}
    {\sigma_{\rm NMAD}}=1.48\times {\rm median}\hbox{ }\bigg[\frac{|\delta z-{\rm median}\hbox{ }(\delta z)|}{1+z_{\rm input}}\bigg ]\; ,
\label{zpRMS}
\end{equation}
and the outlier fraction $\eta$ being the fraction of galaxies with 
\begin{equation}
    \frac{|\delta{z}|}{1+z_{\rm input}}>0.15\;. 
\label{zpoutlier}
\end{equation}
%

Because of the deeper limiting magnitudes of the LSST-like survey after 10 years of observations, the {\it Euclid}+LSST-like case gives the best photo-$z$ accuracy with $\sigma_{\rm NMAD}\approx 0.018$ and  $\eta\approx 0.83\%$ considering galaxies with S/N$>10$ in at least one of the LSST $g$ or $i$ bands. For the {\it Euclid}+DES-like setup, the numbers are $0.054$ and $12.87\%$. For the combination of the two space missions, {\it Euclid}+CSST, we have ${\sigma_{\rm NMAD}}\approx 0.04$, and $\eta \approx 2.39\%$. For the CSST-only setup, ${\sigma_{\rm NMAD}}\approx 0.048$ and $\eta \approx 8.45\%$. It is seen that for CSST, adding the {\it Euclid} NIR bands can improve the photo-$z$ measurements significantly, reducing the outlier fraction by more than a factor of $3$. 

\begin{figure}
\centering
\includegraphics[width=0.48\textwidth]{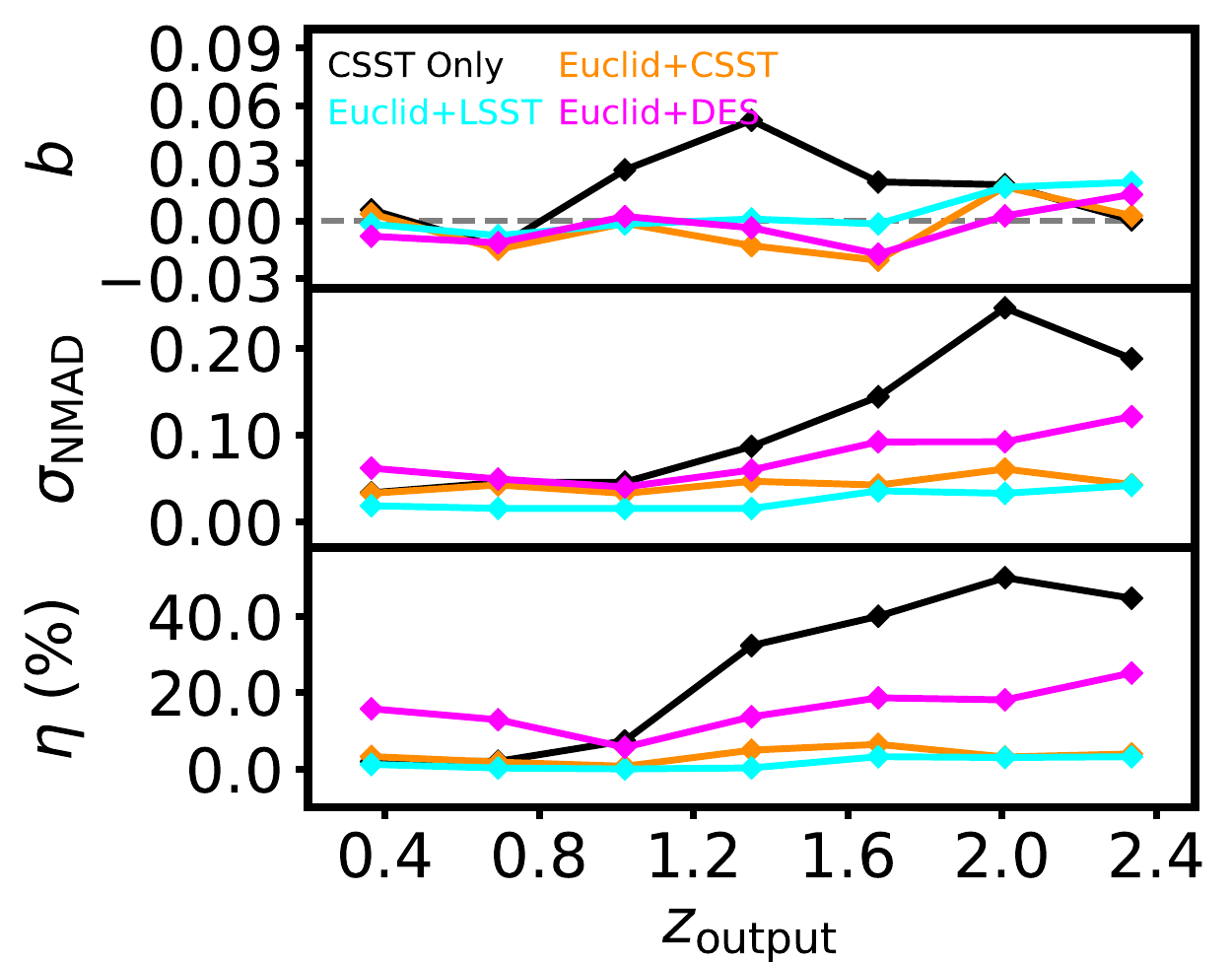}
\caption{The redshift dependence of the bias, $\sigma_{\rm NMAD}$ and the outlier fraction $\eta$ for different combinations of surveys. The horizontal dashed line in the top panel indicates $b=0$.}
\label{fig:zphAnaz}
\end{figure}

{In Fig.~\ref{fig:zphAnaz}, we show $b$, $\sigma_{\rm NMAD}$ and $\eta$ at {different redshifts $z_{\rm output}$}. By comparing the results of {\it Euclid}+CSST with that of CSST-only, all the three photo-$z$ statistics are improved significantly for $z_{\rm output}>1$ by including the NIR bands. 
As discussed extensively in the literature  \citep[e.g.][]{2008MNRAS.387.1215S,Salvato2019,2022ARA&A..60..363N}, the absence of ultraviolet or NIR data can result in difficulties in correctly identifying the Balmer and Lyman breaks in a galaxy SED. This will introduce large photo-$z$ errors. In our study, this problem is more significant for the photo-$z$ result of CSST-only setup because of the lack of NIR bands.} This leads to larger $b$, $\sigma_{\rm NMAD}$ and $\eta$ at $z>1$ than the other cases as shown in Fig.~\ref{fig:zphCom} and Fig.~\ref{fig:zphAnaz}. 

\begin{figure}
\centering
\includegraphics[width=0.48\textwidth]{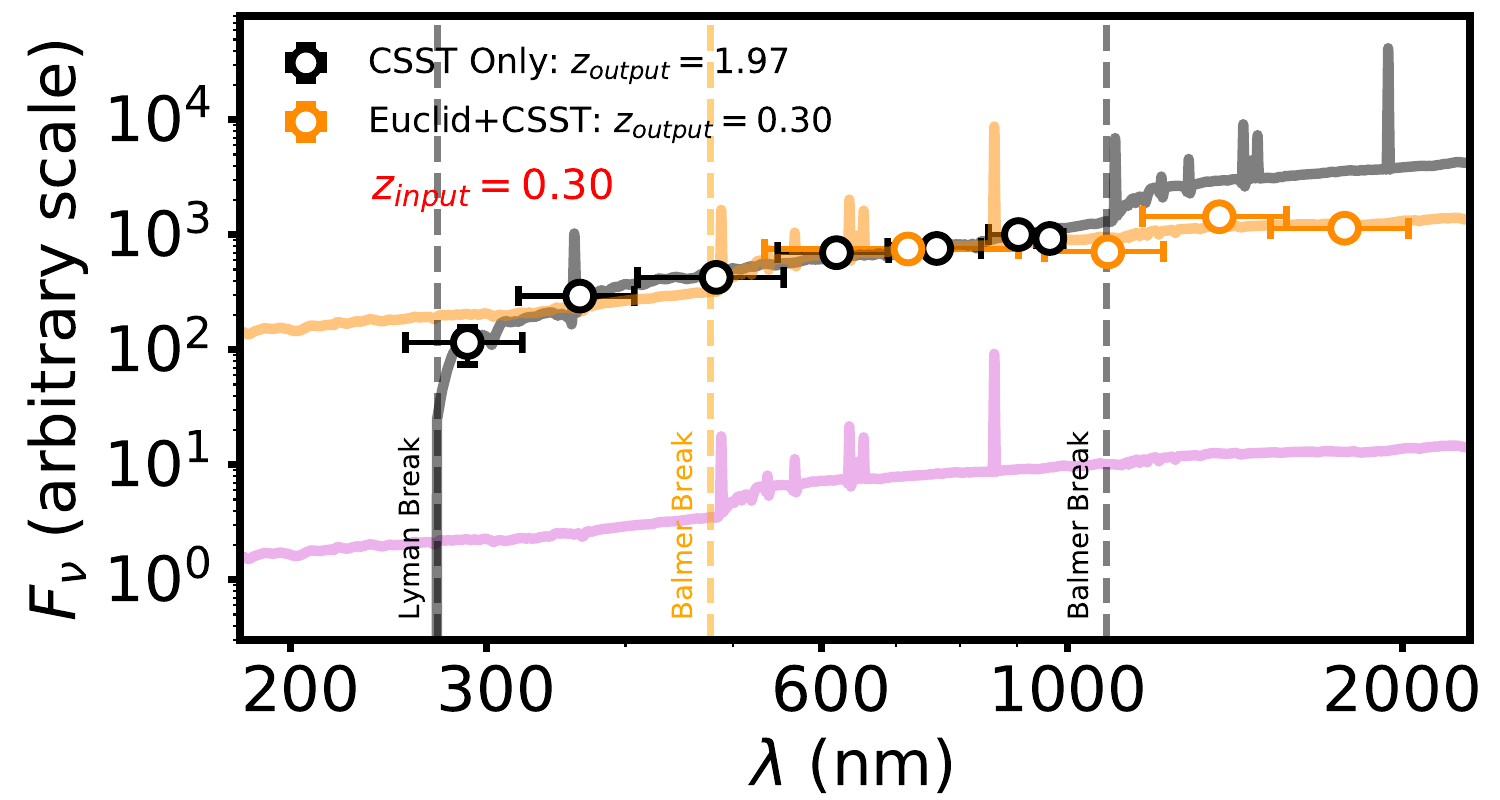}
\includegraphics[width=0.48\textwidth]{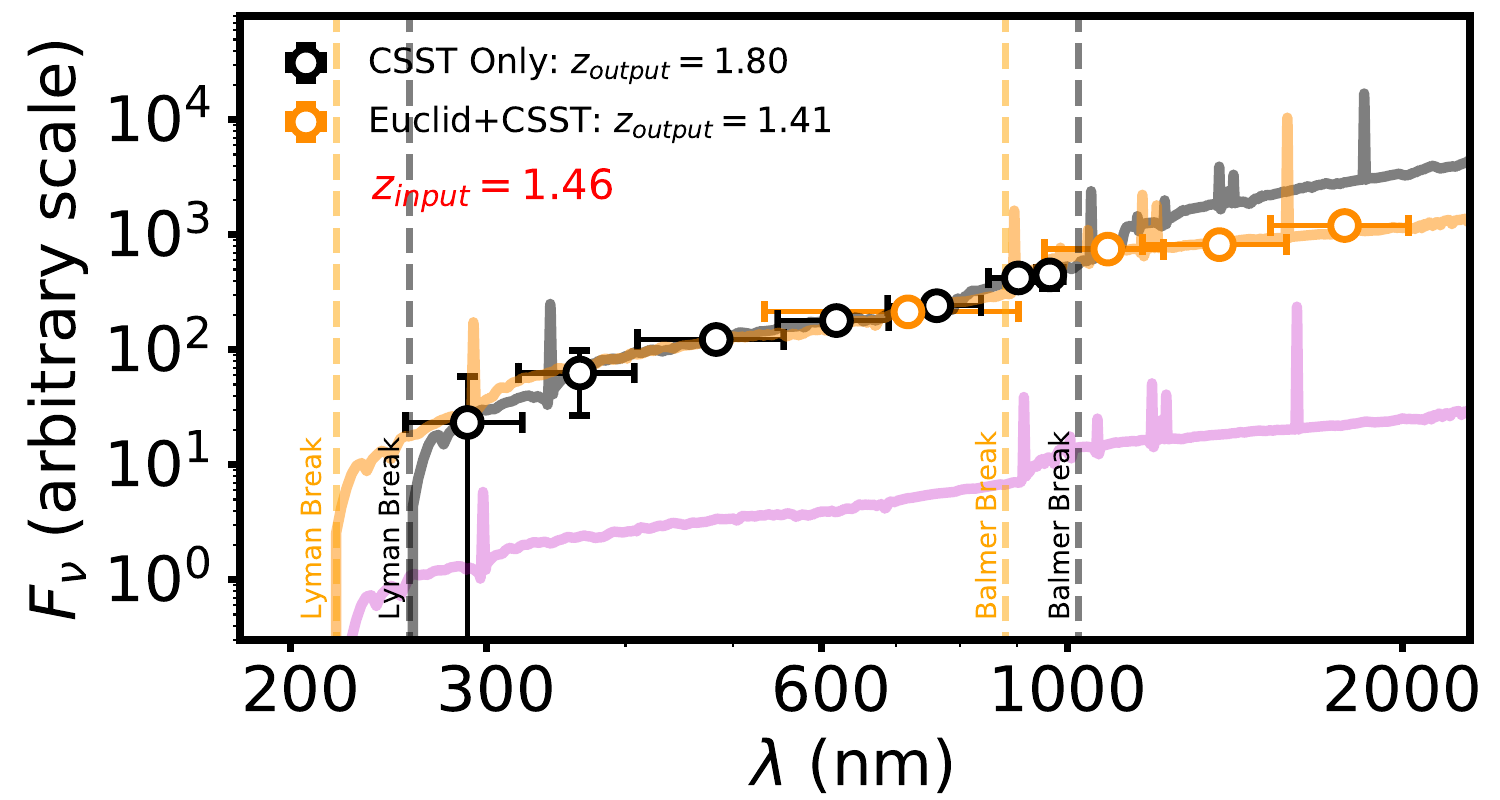}
\includegraphics[width=0.48\textwidth]{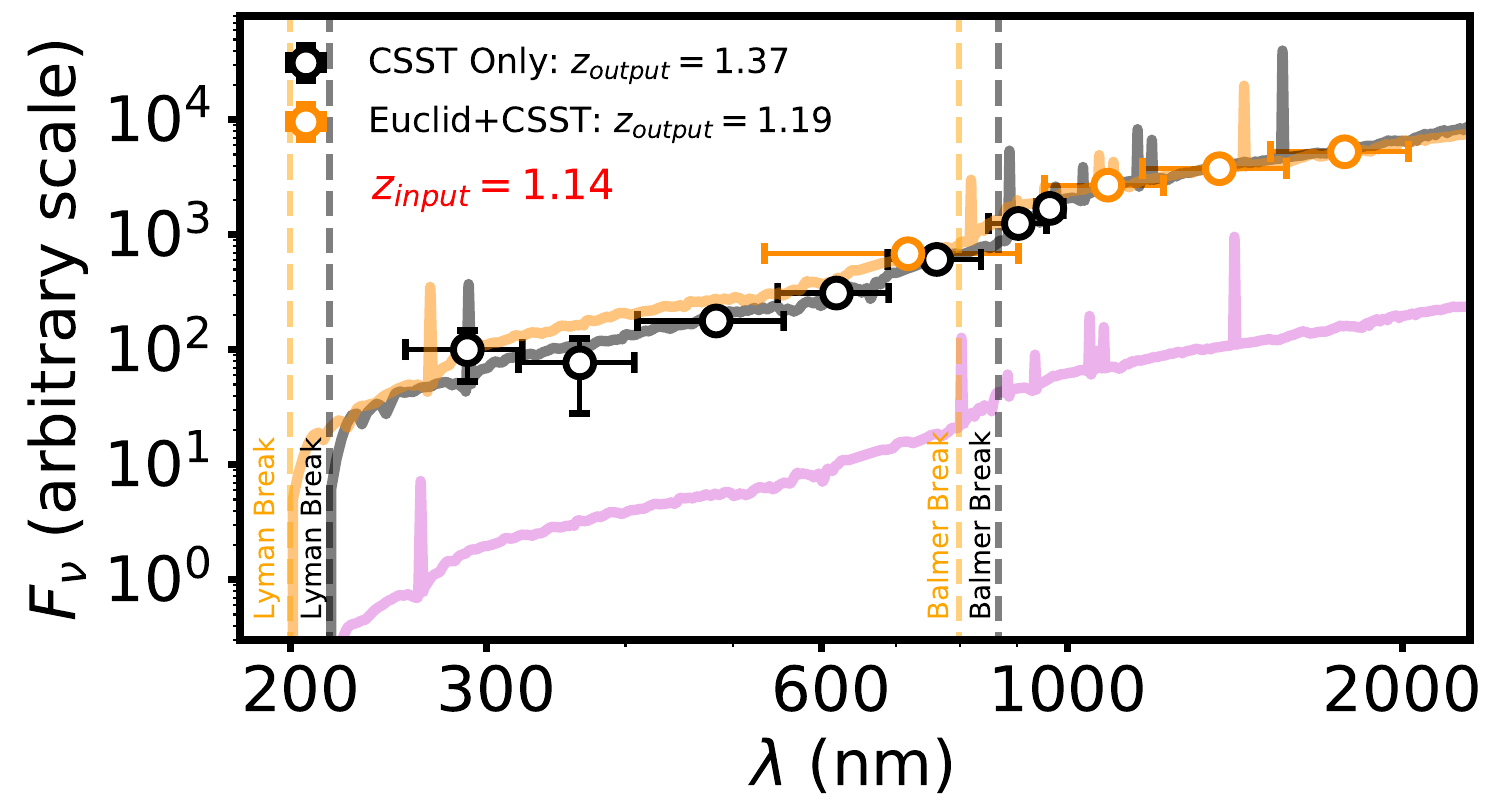}
\caption{Three examples illustrating the best-fit SEDs and photo-$z$s in the case of CSST-only (black) and that in {\it Euclid}+CSST (orange). The original SEDs of the galaxies (purple) are shifted down for clarity. In each panel, the vertical black and orange dashed lines from left to right indicate the Lyman break and Balmer break of the best-fit SED from the CSST-only and the {\it Euclid}+CSST setups, respectively. The black and orange dots represent the fluxes of CSST and {\it Euclid} bands, respectively. The errorbar of each dot along the horizontal axis represents the FWHM of that band, while the errorbar along vertical axis represents the photometric error.}
\label{fig:zphDemo}
\end{figure}

We check the large outlier part in the upper left corner of the CSST-only case in Fig.~\ref{fig:zphCom}, and also particularly analyse multiple cases in the region of  $1<z_{\rm{input}}<1.5$ where the CSST-only photo-$z$s show relatively large positive biases $b$.  It is noted that for $1<z_{\rm{input}}<1.5$, the Balmer break and most of the emission lines except Ly$\alpha$ are redshifted to the wavelength longer than $i$ band, and thus it is difficult to identify them accurately without NIR data. Furthermore, the true Lyman break is still in the blue side of NUV, and thus hard to be located correctly. The trend with relatively large photo-$z$ biases in this redshift range with optical data only is also shown in, for instance, \citet{Salvato2019} and \citet{cao18}. In Fig.~\ref{fig:zphDemo}, a few examples are presented to illustrate clearly the reasons for the bias in the CSST-only case and the effects of NIR data. There the best-fit SEDs in the CSST-only (black) and {\it Euclid}+CSST (orange) cases, as well as the input SEDs (purple),  are shown. 

The top panel of Fig.~\ref{fig:zphDemo} is for a case with the CSST-only photo-$z$ result drastically wrong by assigning a high $z_{\rm{output}}=1.97$ to a low-redshift galaxy with $z_{\rm{input}}=0.3$. The middle panel shows a case with $z_{\rm{input}}=1.46$ and the CSST-only $z_{\rm{output}}=1.80$. For these two cases, the SED behaviours are similar. With the CSST-only data, the  Balmer break feature cannot be located well, and the best-fit redshifted SED gives rise to relatively higher NIR fluxes than the input SED. The Lyman break position (left black vertical line) is determined to be closer to CSST NUV than the true one, leading to higher $z_{\rm{output}}$ than $z_{\rm{input}}$. Adding {\it Euclid} NIR data, the NIR part of the SED is better fitted and the Balmer break as well as the Lyman break positions (vertical orange lines) can be correctly identified. As a result, the best-fit SED and consequently the photo-$z$ from {\it Euclid}+CSST are in better agreement with the input ones. The bottom panel of Fig.~\ref{fig:zphDemo} shows another case with $z_{\rm{input}}=1.14$ where the CSST-only best-fit SED has about the correct shape in NIR comparing to the input one, but the Lyman break position is determined to be at a somewhat longer wavelength, and thus a higher $z_{\rm{output}}=1.37$. In this case, the inclusion of {\it Euclid} NIR adds weights to adjust the overall SED including the two break positions, delivering a better photo-$z$ of $z_{\rm{output}}=1.19$.

For the combinations of {\it Euclid} with LSST-like, CSST, and DES-like setups, respectively, over the entire redshift range, both $\sigma_{\rm NMAD}$ and $\eta$ are the largest for the case of {\it Euclid}+DES-like.
It should be noted that in our analyses, the same set of templates are used in generating the input fluxes in different bands and in the photo-$z$ fitting. Thus, the overall small bias here is largely due to photometric errors. In real situations, if the true SEDs of some galaxies are not included in the templates, their photo-$z$ bias from template fitting can be significantly larger than the numbers shown here. 

\begin{table}
\centering
\caption{Photo-$z$ statistics in different cases. }
\label{tab:photozstat}
\begin{tabular}{@{\extracolsep{1em}}cccc}
\toprule
Combination      &  bias        & $\sigma_{\rm NMAD}$     & $\eta$ \\
\midrule
CSST-only        &  $+$0.0043     & 0.048                   &    8.45\%   \\
\midrule
{\it Euclid}+CSST      &  $-$0.0051     & 0.039                   &   2.39\%    \\
\midrule
{\it Euclid}+LSST-like &  $-$0.0019     & 0.018                   &   0.83\%   \\
\midrule
{\it Euclid}+DES-like  &  $-$0.0057     & 0.054                   &   12.87\%   \\
\bottomrule
\end{tabular}
\end{table}

We emphasize that our purpose here is to compare different data combinations, and thus the relative differences in photo-$z$ accuracy between different cases are more meaningful to us. While the absolute photo-$z$ accuracy can be sensitive to the adopted methods and codes \citep{Hildebrandt2010, Euclidz2020, Zhou2021}, the relative accuracy should depend mainly on the data used. We therefore only use the \texttt{EAZY} template fitting code to do the photo-$z$ analyses here and leave the thorough investigations on photo-$z$ measurement methods to future studies. 

{It should also be noted that in our analyses, {  we do not consider the effect of AGNs. Studies have shown that galaxies  hosting AGNs generally have worse photo-$z$ quality \citep[e.g.][]{Salvato2019}. At high redshifts, faint AGNs cannot be easily identified \citep[e.g.][]{2013A&A...557A..16M,2021MNRAS.503.4136G}. Thus their flux contaminations to their host galaxies can lead to errors in the photo-$z$ estimation. \citet{2010PASP..122..485M} studied the photo-$z$ bias induced by AGN contamination, and concluded that this bias can potentially corrupt the derived cosmological parameters, especially for future WL surveys reaching deeper limiting magnitudes. 
As described in Sect.~\ref{sec:photoz}, {\it Euclid} selects its shear sample with one of the criteria by requiring that the galaxies therein should not be flagged as AGNs \citep{Euclidz2020}. This can mitigate the AGN-induced photo-$z$ bias to some extent. However, for unidentified AGNs, their effects on photo-$z$ measurements need to be addressed in future studies.} 

In addition, we also do not perform careful deblending for ground-based images. On the other hand, to suppress the blending effects on ground-based photometry, we employ the positions and apertures measured from high-resolution {\it Euclid} VIS simulated data for flux measurements on LSST-like or DES-like images. A potential problem for such an approach is that for some galaxies, the {\it Euclid} VIS apertures can be too small for the two ground-based-like images, and thus their measured $\rm {S/N}$ are sub-optimal}. 

{To see the impact of photometric apertures, we measure galaxy fluxes using different apertures taken to be $N\,D$, where $N$ is a factor and $D$ is the FWHM of the PSF after homogenization. Specifically, $D=0\farcs92, 1\farcs11, 0\farcs92$ and $0\farcs22$ for the case of {\it Euclid}+LSST-like, {\it Euclid}+DES-like, {\it Euclid}+CSST, and CSST-only, respectively. As explained previously, for {\it Euclid}+CSST, high-resolution images from CSST and the {\it Euclid} VIS-band as well as the PSF homogenized {\it Euclid} $Y, J, H$-band images are used in the photo-$z$ analyses. Thus, we take $D=0\farcs92$ here, which is the FWHM of the {\it Euclid} $H$-band PSF used in the $Y, J, H$ homogenization. The aperture flux in each band is scaled to the total flux with Eq.\,(\ref{flux}).} 

\begin{figure}
\centering
\includegraphics[width=0.48\textwidth]{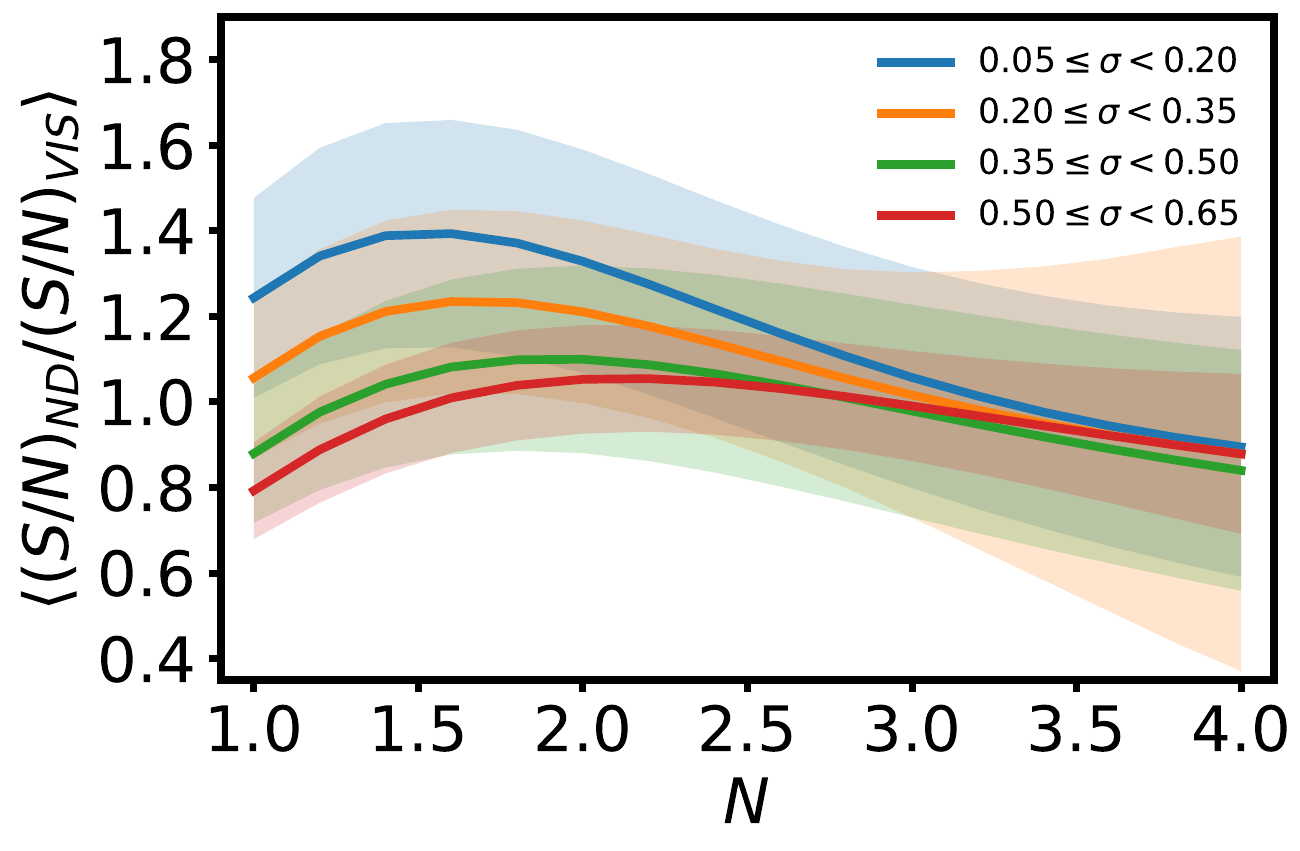}
\caption{The S/N variations with the change of the photometric aperture. The lines are the average and the shaded regions are the corresponding dispersion within the bins.}\label{fig:SNRwithAper}
\end{figure}

{We calculate the $\rm {S/N}$ variations with the change of $N$ for LSST-like galaxies. Different from point sources, galaxies have various intrinsic sizes, and the dependence of their $\rm {S/N}$ on the aperture size can be different. We therefore bin galaxies based on their intrinsic size $\sigma$ defined in the simulation step (3), and calculate the average $\rm {S/N}$ within a bin. We further note that galaxies of similar size can have a range of intrinsic luminosity. To clearly illustrate the aperture effect on $\rm {S/N}$, for each LSST-like galaxy, we calculate the ratio of the signal-to-noise ratio measured within an aperture diameter $N\, D$ to that measured within the {\it Euclid} VIS aperture, that is $({\rm {S/N}})_{ND}/({\rm {S/N}})_{\rm VIS}$. The results are shown in Fig.~\ref{fig:SNRwithAper}, where the lines are the average and the shaded regions are the corresponding dispersion. It is seen that for galaxies with $\sigma >0\farcs35$, the optimal aperture is about $2D$ where $({\rm {S/N}})_{ND}/({\rm {S/N}})_{\rm VIS}\approx 1$. In other words, for these relatively large galaxies, using {\it Euclid} VIS aperture is a good choice. For smaller galaxies, $\rm {S/N}$ reaches the maximum at $\sim 1.5D$ and $({\rm {S/N}})_{ND}/({\rm {S/N}})_{\rm VIS}\sim 1.2$--1.4, that is {\it Euclid} VIS apertures are sub-optimal for LSST-like images homogenized to a PSF with FWHM $D=0\farcs92$.}

\begin{figure}
\includegraphics[width=0.48\textwidth]{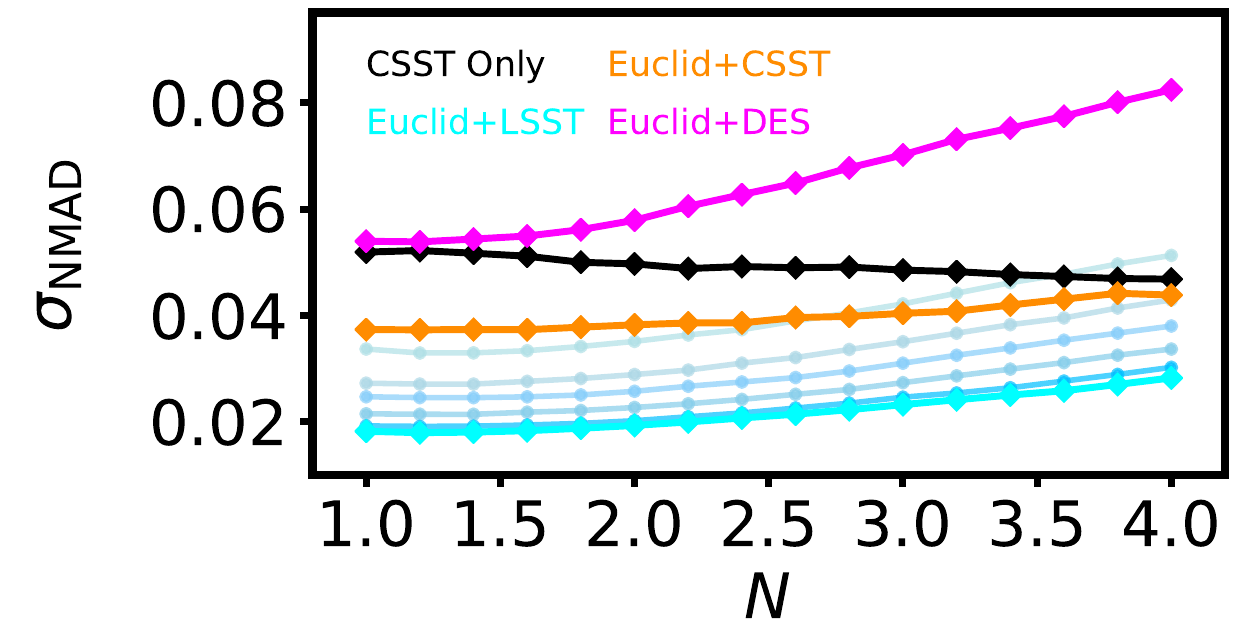}
\includegraphics[width=0.48\textwidth]{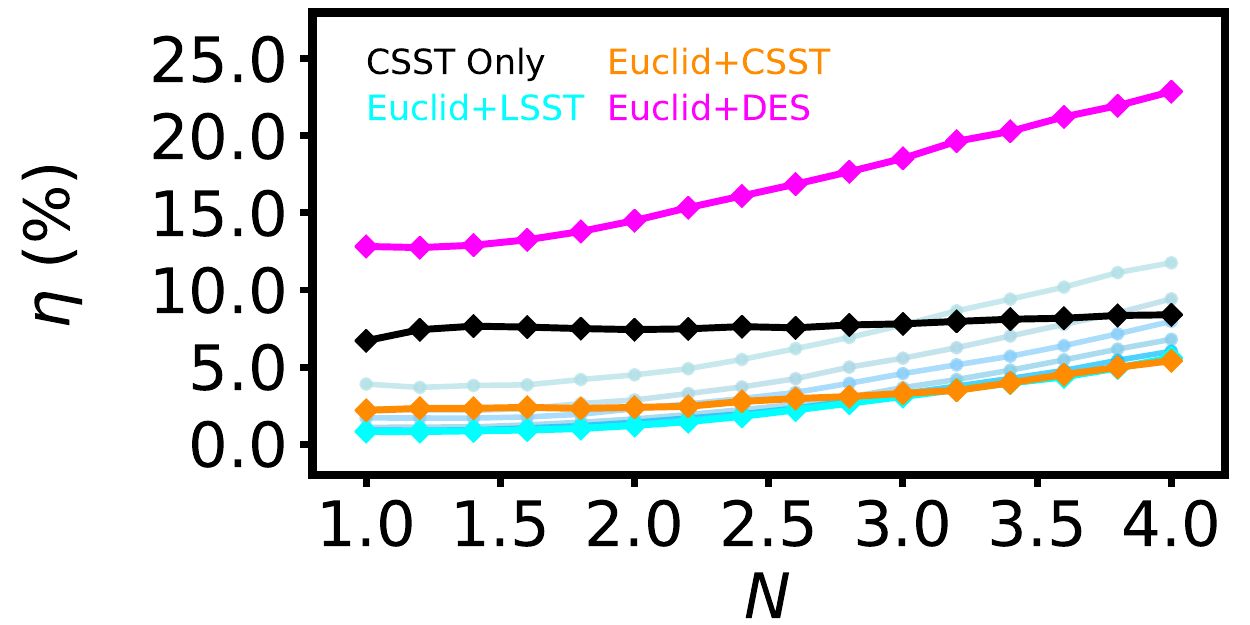}
\caption{The effects of the size of the photometric aperture on $\sigma_{\rm NMAD}$ and $\eta$. For the {\it Euclid}+LSST-like setup, lines with similar colours from top to bottom show the results from LSST-like mock data corresponding to Year01 to Year10.}\label{fig:AperD}
\end{figure}

{To illustrate the impact of the aperture on the photo-$z$, we show the dependence of $\sigma_\mathrm{NMAD}$ and $\eta$ on the aperture parameter $N$ in Fig.~\ref{fig:AperD} for different cases. 
There we also show the results of {\it Euclid}+LSST-like setup, using LSST-like data of different depths, which correspond to different years of accumulation. We see that for both {\it Euclid}+LSST-like and {\it Euclid}+DES-like, $\sigma_{\rm NMAD}$ and $\eta$ are nearly flat for $N\le 2$, and their values are very close to the ones shown in Table \ref{tab:photozstat} using {\it Euclid} VIS apertures. This seems to indicate that although {\it Euclid} VIS apertures are not optimal for small galaxies, their effects on photo-$z$ are not very significant. Increasing the aperture factors from $N=2$ to $N=4$, the outlier rate $\eta$ increases considerably, for the case of {\it Euclid}+LSST-like by factors of $\sim 3$ to $5$ for LSST-like Year01 to Year10. The $\sigma_{\rm NMAD}$ also increases by about a factor of $1.5$. {These trends are largely attributed to the light blending of nearby galaxies plus the reduced ${\rm {S/N}}$ at large apertures as shown in Fig.~\ref{fig:SNRwithAper}. Similar behaviours are seen for the results of the {\it Euclid}+DES-like case.} 
For {\it Euclid}+CSST, the changes of $\eta$ and $\sigma_{\rm NMAD}$ are much flatter over the whole range from $N=1$ to $N=4$. This is because the high resolution optical images from CSST and {\it Euclid} VIS are much more compact than those of LSST-like or DES-like setups, and therefore the light blending effect is much weaker even with similar photometric apertures. In terms of the outlier rate $\eta$, {\it Euclid}+CSST can reach approximately the same accuracy as that of {\it Euclid}+LSST-like YEAR02 data when using an aperture diameter less than $\sim 2D$.} 

{We need to point out that accurate multi-band photometric measurements are challenging and should be carefully investigated based on the specific data characteristics. Our analyses here, either using {\it Euclid} VIS apertures in our fiducial analyses or the tests using different fixed apertures, are somewhat sub-optimal in the sense that we do not perform detailed aperture optimisation band by band and galaxy by galaxy. With more sophisticated optimisations \citep[e.g.][]{Kuijken2008, KiDS, Wright2016, Bosch2018, Huang2018, Sevilla-Noarbe2021}, the photo-$z$ statistics may be further improved compared to the results shown here. Although we do not expect qualitative changes of our conclusions, quantitative studies and comparisons with different photometric measurements are desired in future investigations.  }

{We also emphasize again that we do not perform careful deblending for ground-based-like images. The performance of photo-$z$ in the case of {\it Euclid}+LSST-like (DES-like) can be improved by employing sophisticated approaches for accurate photometric measurement. On the other hand, for the data combination of {\it Euclid} and CSST, the blending in optical bands is minimal, and thus the photometric measurements can be relatively straightforward.}

\subsection{Blended galaxies}


As we discussed in the previous section, with the increase of observational depth blended galaxies for ground-based observations are becoming increasingly problematic for accurate photometry. For HSC, the estimated fraction of blended galaxies is about 58\% (Bosch et al. 2018). For LSST, the fraction will be even higher. To achieve high-precision cosmological results from galaxy imaging surveys, sophisticated deblending algorithms are critically needed.

In \texttt{SExtractor} the deblending is done on single-band images by using a tree structure to find sub-objects with the intensity above a certain fraction $\delta_\textrm{c}$ of the total intensity of the composite object. The intensity of overlapping pixels are assigned to different sub-objects according to bivariate Gaussian fits to the sub-objects \citep{BA1996}.  In recent years, more advanced methodologies have been developed. For example, \texttt{MuSCADeT} \citep{MuSCADeT} and \texttt{SCARLET} \citep{SCARLET} utilize multi-band pixel information for deblending, and thus galaxy SEDs are taken into account. However, similarly to \texttt{SExtractor}, these methods still make certain assumptions about galaxy properties. Another line of study is to use machine learning for deblending, which on the one hand can avoid overly simplifying assumptions about galaxies. On the other hand, appropriate training sets are crucial for machine learning \citep{Reiman2019, Arcelin2021, Farrens2022}. 

In this regard, space observations with high resolution are clearly advantageous in reducing the fraction of blended galaxy images, and thus simplifying the photometric measurements.
To explore the advantages of utilizing data from the two space missions to control blending, we create a set of dedicated simulations, which differ from the simulations presented in Sect.~\ref{sec:methods_and_sims}. 
We start here from HST/CANDELS GOODS-N images of F606W and F814W \citep{Giavalisco2004, Grogin2011, Koekemoer2011}. The effective area is about 158 arcmin$^{2}$. These image stacks have been created using 0\farcs06 pixels. We synthesize {\it Euclid} VIS and CSST $r$-band images from them. For comparison, we also generate LSST $r$-band images. 

Specifically, to generate {\it Euclid} VIS images, we first combine CANDELS F606W and F814W pixel data using linear interpolation as follows,
\begin{equation}
I^{\rm VIS}_0= w\Bigg[\frac{\lambda-\lambda_1}{\lambda_2-\lambda_1}\frac{\mu_2f_2}{w_2}+\frac{\lambda_2-\lambda}{\lambda_2-\lambda_1}\frac{\mu_1f_1}{w_1}\bigg],
\label{VIS}
\end{equation}
where $\lambda_{1}=606$ nm, $\lambda_2=814$ nm, and $\lambda=725$ nm are the central wavelengths of F606W, F814W, and the {\it Euclid} VIS band, respectively. The zero-points related to the rescaling of F606W and F814W are $\mu_1$ and $\mu_2$. We take $\mu_1=1$ and thus $\mu_2=10^{-0.4({\rm zp}_2-{\rm zp}_1)}$ with zp$_i$ the zero points for the two CANDELS bands. The corresponding widths of the three bands are $w_1$, $w_2$ and $w$. For CSST $r$- and LSST $r$-bands, we simply use the CANDELS F606W images, ignoring the slight band difference. 

We then convolve these idealized images, labeled as $I_0$, with the corresponding PSF, and resample them according to the pixel scales of {\it Euclid} VIS, CSST $r$- and LSST $r$-bands, respectively, with the parameters listed in Table \ref{tab:bandsinfo}. We denote these PSF- and pixel-scale-matched images as $I_1$. In the last step, we add Gaussian noises to $I_ 1$ according to the three survey depths to generate the final $I_2$ images. In this simulation setting, we assume that the HST CANDELS images are ideal without considering their own PSF and noise. For the purpose of analysing blending effects of different surveys, this simplification should not affect the results significantly.  

\begin{figure}
\centering
\includegraphics[width=0.48\textwidth]{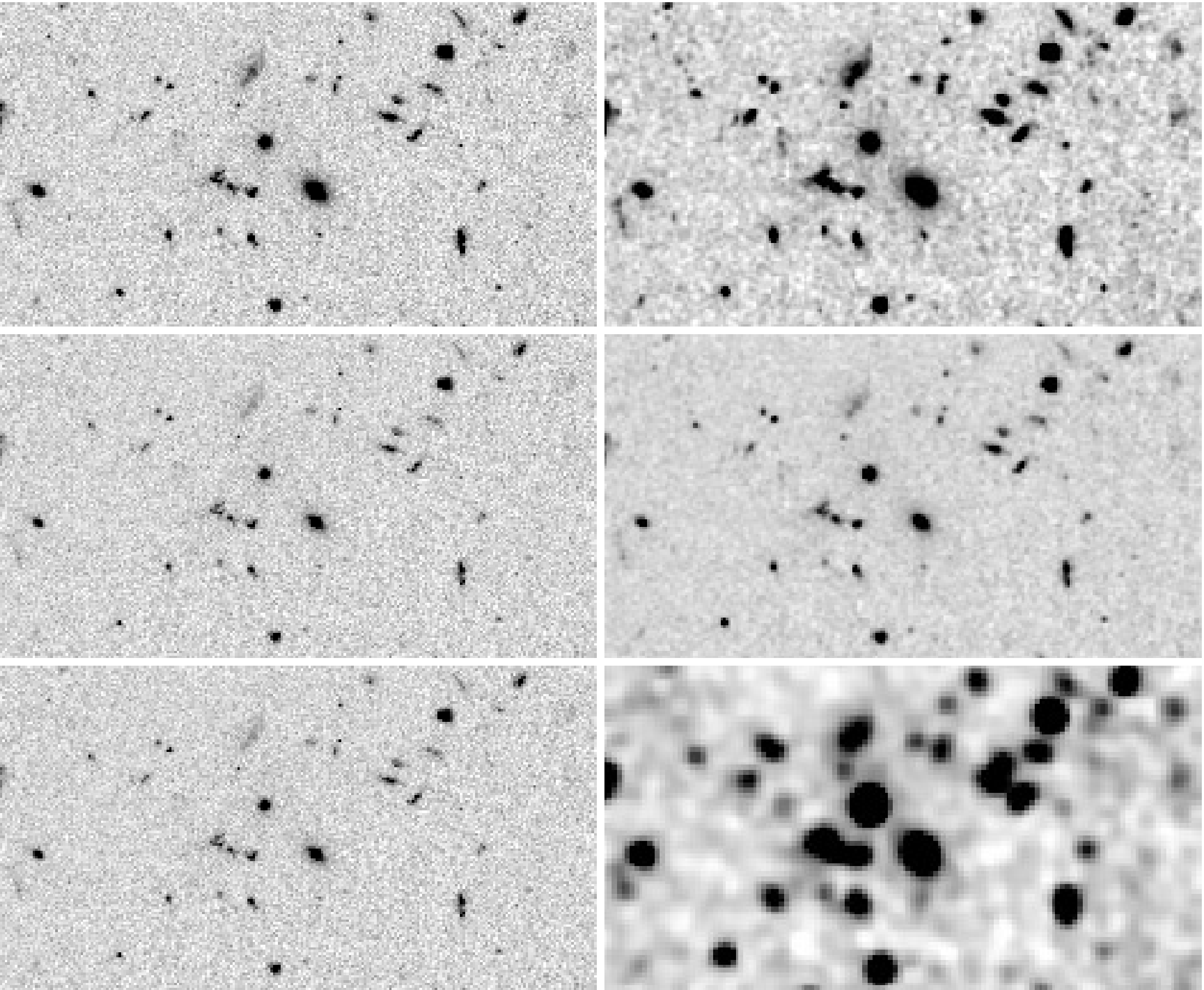}
\caption{Examples of $I_0$ (left column) and $I_1$ (right column) images for {\it Euclid} VIS (top row), CSST $r$ (middle row) and LSST $r$ (bottom), respectively. }
\label{fig:I0I1images}
\end{figure}

In Fig. \ref{fig:I0I1images}, we show examples of the simulated images, where the left panels are $I_0$ images and the right ones are the corresponding $I_1$ images. It is seen clearly that both {\it Euclid} VIS and CSST $r$ band observations can resolve galaxies similarly to the HST CANDELS data. For ground-based LSST-like data, however, the blending effect is much more severe. 

\begin{figure}
\centering
\includegraphics[width=0.48\textwidth]{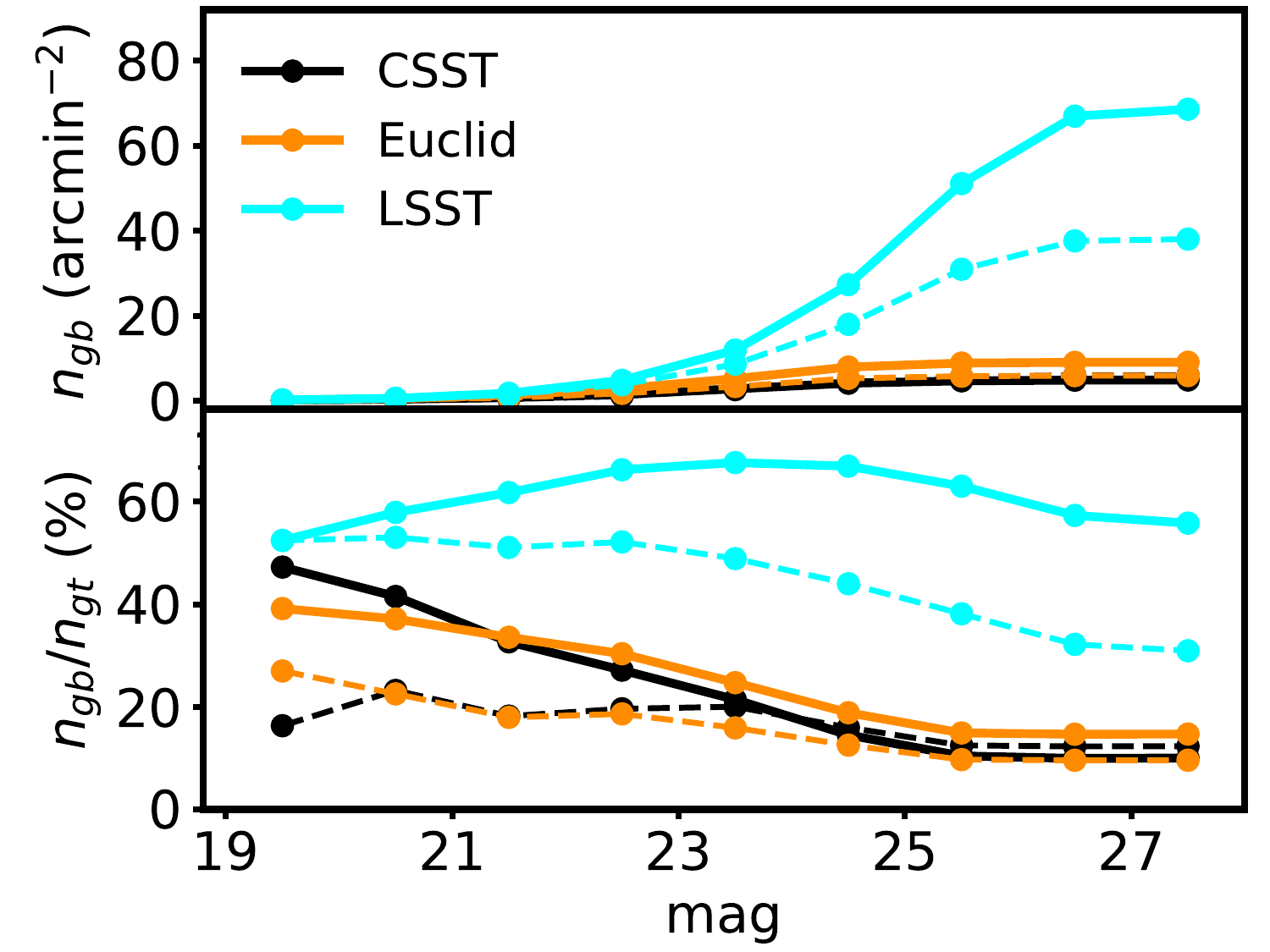}
\caption{{ The cumulative number density of blended galaxies $n_{gb}$ as a function of magnitude for different cases (upper panel) and the corresponding fractions relative to the total detected number densities of galaxies $n_{gt}$ at the same magnitude cut as $n_{gb}$ (lower panel).} The dashed and solid lines are the results using \texttt{SExtractor} flags and the overlapping of 2.5 times expanded Kron ellipses, respectively.}
\label{fig:blendcomp}
\end{figure}

{Fig. \ref{fig:blendcomp} shows quantitatively the blending statistics. Here we perform source detection using \texttt{SExtractor} on $I_2$ images for the three cases separately. 
The dashed lines show the results with the blended galaxies defined using the \texttt{SExtractor} \texttt{FLAGS}=1 and 2, and the values containing the contributions from these two flags. Here, \texttt{FLAGS}=1 indicates a likely contamination of the aperture photometry by neighbouring sources, and \texttt{FLAGS}=2 shows that the object has been deblended \citep{BA1996}. The solid lines present the blending statistics defined as the occurrence of overlapping between 2.5 times expanded Kron ellipses with neighbouring galaxies. The upper and lower panels show the dependence of the cumulative number density of blended galaxies on the magnitude, and the corresponding fractions. Note that the horizontal axis shows the magnitude of the galaxies we are concerned with. To judge if they are blended, however, we are not limited to this magnitude, but use all detected galaxies instead. We can see that for LSST-like images, the number density of blended galaxies can reach about 30--50 arcmin$^{-2}$ at ${\rm mag}\sim 26$ depending on the definition of blending. For {\it Euclid} and CSST, the corresponding number density is about 5 arcmin$^{-2}$. The blended fraction is computed as the ratio of the numbers shown in the upper panel and the corresponding total detected number densities of galaxies at the same magnitude cut.
It is $\sim 10\%$ at ${\mathrm {mag}}<26$ for both {\it Euclid} and CSST, while for the LSST-like data, it is in the range of $40\%$ to $60\%$.} 


\begin{figure}
\centering
\includegraphics[width=0.48\textwidth]{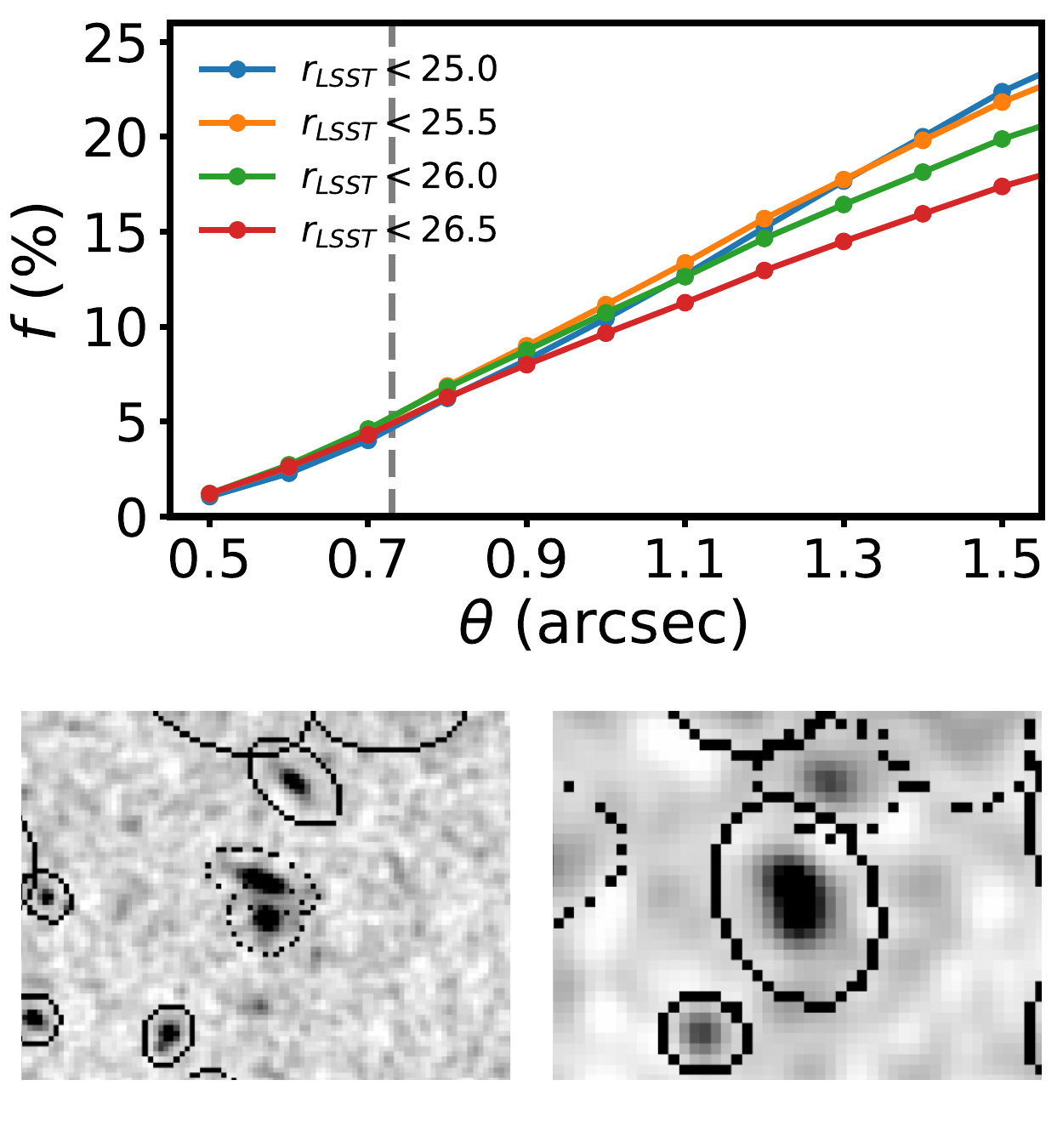}
\caption{Top: The fraction of total blends in LSST-like $I_1$ images estimated by matching to {\it Euclid} VIS galaxies. The horizontal axis is the matching radius and the vertical dashed line indicates $\theta=0.73\arcsec$, approximately the PSF FWHM of LSST-like data. Bottom: An example of total blends with left and right being the simulated {\it Euclid} VIS and LSST-like $r$ band images, respectively. {  \texttt{SExtractor} Kron ellipses are overlaid for the detected objects.}}
\label{fig:totalblend}
\end{figure}

{In the above analyses, the source detection and the blending identification are done independently for each of the three sets of mock survey simulations. There can be cases that two or more close-by galaxies in a high-resolution image are identified as a single galaxy in a low-resolution LSST-like image. We refer to these as total blends, and they cannot be correctly counted in Fig. \ref{fig:blendcomp}. To calculate the fraction of total blends for the LSST-like case, we perform analyses by matching the {\it Euclid} VIS $I_1$ detected galaxies with those detected from LSST-like $I_1$ images. Specifically, we first perform a double-match with a given matching radius between {\it Euclid} VIS and LSST-like galaxies, and then remove those matched ones from the {\it Euclid} VIS galaxy catalogue. For the remaining galaxies, we perform a second double-match with the full LSST-like galaxy catalogue. If two {\it Euclid} galaxies are matched to a single LSST-like galaxy, we regard the latter as a total blend. In Fig. \ref{fig:totalblend}, the bottom panels show an example of total blends with the left and right panels being {\it Euclid} VIS and LSST-like $r$-band images, respectively. The top panel shows the fraction of total blends in the LSST-like case with respect to the matching radius $\theta$. For $\theta\sim 0\farcs73$, the size of the LSST-like PSF FWHM, the fraction is about $5\%$, and it increases to $\sim 10\%$ for $\theta=1\arcsec$.}

\begin{figure}
\centering
\includegraphics[width=0.48\textwidth]{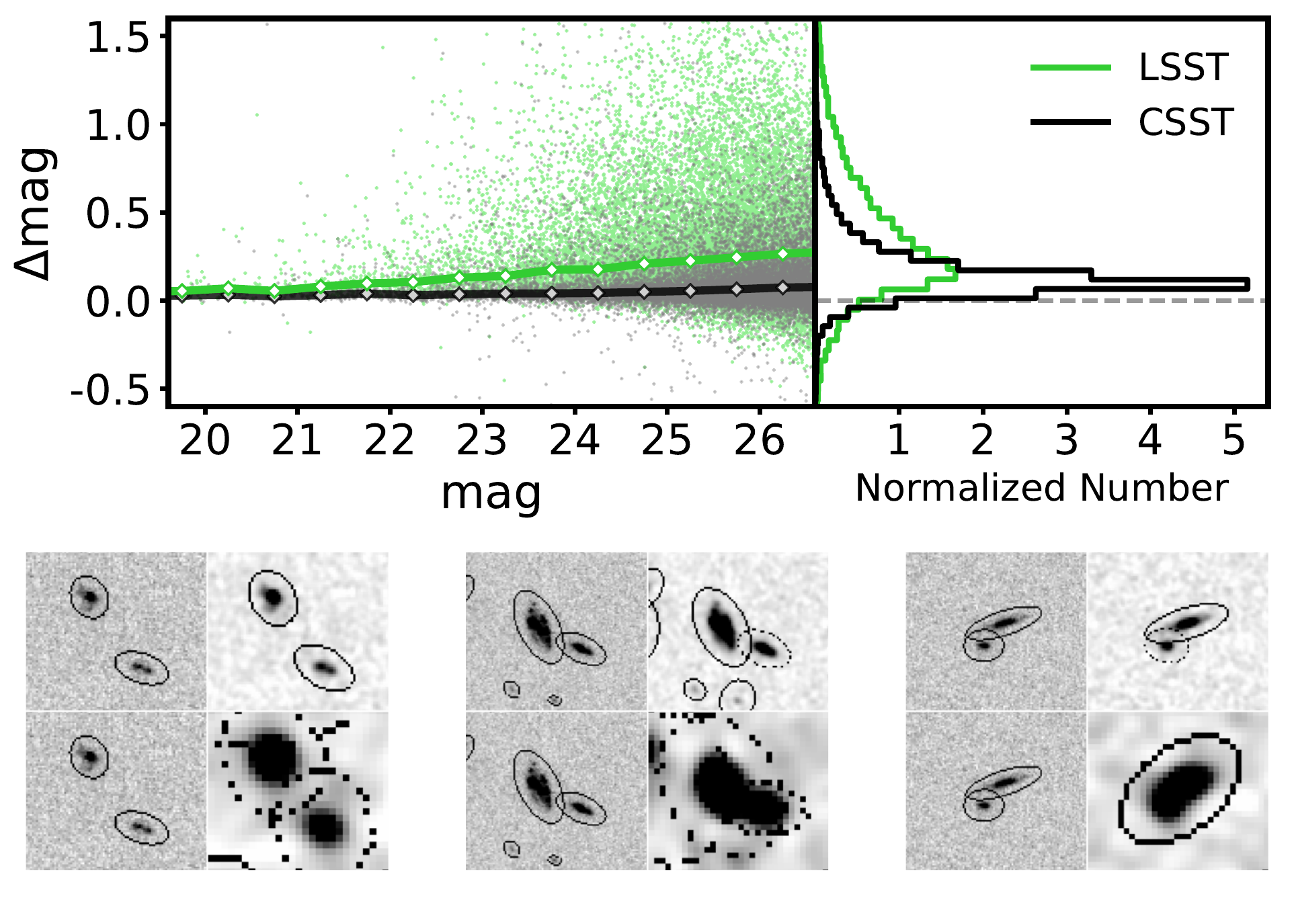}
\caption{Top left: The difference of the \texttt{MAG\_AUTO} measurements from the $I_0$ and $I_1$ images, $\Delta \textrm{mag} = \rm{mag}_{\rm auto}^{I_0}-\rm{mag}_{\rm auto}^{I_1}$ vs. $\rm{mag}_\mathrm{auto}^{I_0}$. The lines show the median of $\Delta {\rm mag}$. Top right: The distribution of $\Delta {\rm mag}$. Bottom: Three examples of galaxy images illustrating the impact of blending with the top for CSST $I_0$ (left) and $I_1$ (right), and the bottom for the corresponding LSST-like case. {  \texttt{SExtractor} Kron ellipses are overlaid for the detected objects.}}
\label{fig:magauto}
\end{figure}

To study the impacts of blending on photometry, we first measure the \texttt{MAG\_AUTO} magnitude of each detected object using \texttt{SExtractor} on $I_0$ images of the LSST-like $r$ band and the CSST $r$ band, respectively. They are at high resolution without noise. We then carry out the photometry measurements on the corresponding $I_1$ images, which are PSF-convolved and pixel scale-matched images but contain no noise. The comparison of these two sets of photometry can reveal cleanly the blending effects. For the {\it Euclid} VIS band, the results are similar to those of CSST because of the similar PSF, and thus they are not shown here. The top left panel of Fig. \ref{fig:magauto} shows the scatter plot of the two sets of measurements, where the horizontal axis shows the $I_0$ magnitude and the vertical axis is the difference of the magnitudes measured in the $I_0$ and $I_1$ images. The lines are the median bias of the corresponding cases. It is seen clearly that in the LSST-like case, the severe blending effects lead to magnitude estimates that are significantly biased low, that is brighter, with the median bias at ${\rm mag}=23$, $24$, and $25$ being 0.135, 0.175 and 0.215 ${\rm mag}$, respectively. For CSST, the corresponding bias is about 0.038, 0.042, and 0.052 mag. The distributions of the $\Delta {\rm mag}$ are shown in the top right. 

{  From the distributions, it is noted that the $\Delta\mathrm{mag}$ is systematically positive, meaning that the \texttt{MAG\_AUTO} magnitudes measured in $I_1$ images are systematically brighter than that of $I_0$ images. Our tests show that this is attributed to the larger Kron ellipses (in unit of arcsec) used in $I_{1}$ images than that of in $I_{0}$ images due to both the PSF effect and the pixel downsampling from $I_{0}$ to $I_{1}$. Firstly, the PSF smoothing operation tends to smear the surface brightness distributions of the galaxies and reduce the random background noise. Secondly, the downsampling operation of the pixel scale from $I_0$ to $I_{1}$ images can also reduce the random background noise and thus increase the S/N of the galaxies in $I_{1}$. 
Both result in a smaller \texttt{MAG\_AUTO} measured from $I_{1}$ and thus the distributions shown in the top panel of  Fig. \ref{fig:magauto} being peaked at a positive $\Delta\mathrm{mag}$. On the other hand, it is the blending effects that lead to a significantly extended tail at the positive side of the distribution, especially in the case of LSST-like.} The bottom panels of Fig. \ref{fig:magauto} present three examples of galaxy images. In each of these panels, the left sub-panels are the CSST (top) and LSST-like (bottom) $I_0$ images while the right sub-panels show the corresponding $I_1$ images. They demonstrate how the blending effect impacts the ground-based photometry at different levels. {  Here again, the larger Kron ellipses due to PSF smoothing and image downsampling can be seen on the $I_{1}$ images.} Note that \texttt{SExtractor} provides a deblending procedure. However, this is clearly not good enough. More sophisticated deblending techniques are needed. On the other hand, for the two space missions, the blending effect is much lower, and the photometric measurements can be significantly simplified.

\begin{figure}
\centering
\includegraphics[width=0.48\textwidth]{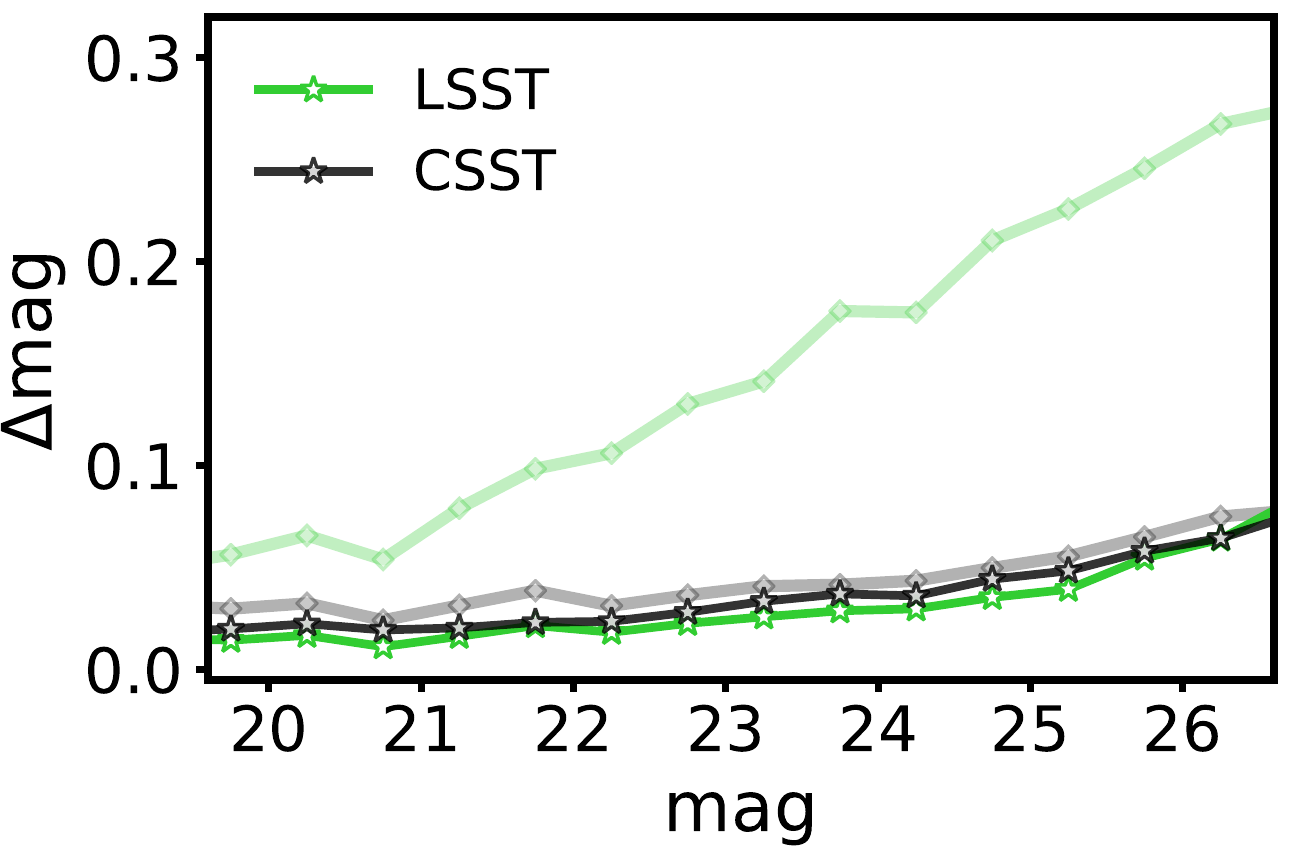}
\caption{The median of the magnitude difference measured from $I_0$ and $I_1$ images using {\it Euclid} VIS apertures and rescaled to the total magnitude following Eq.(\ref{flux}) (dark coloured lines). For comparison, the results of Fig.~\ref{fig:magauto} are shown in light coloured lines.}
\label{fig:euclidaperture}
\end{figure}

The photometry results shown in Fig. \ref{fig:magauto} correspond to \texttt{SExtractor} \texttt{MAG\_AUTO} magnitudes, which are intended to measure the total magnitude of galaxies directly from images. In other words, if considering joint analyses of different surveys, such measurements can be used for data combinations at the catalogue level. {  Because of the PSF smoothing and image downsampling, the aperture used to compute \texttt{MAG\_AUTO} is significantly larger in the case of LSST-like simulations compared to the simulated space-based data, causing severe blending effects and biases in the photometry.} Using space observations as priors, the photometry can be improved significantly \citep[e.g.][]{Chary2020, SCARLET}. This demands that the data combination must be done at the pixel level. In fact, in our photo-$z$ analyses in Sect.~\ref{sec:photoz}, we adopt the {\it Euclid} VIS detection to provide a position prior, and the photometric aperture for a galaxy is also from {\it Euclid} VIS. We then apply them to other cases considering the scaling relation of Eq.~(\ref{flux}). We perform such measurements here to $I_0$ and $I_1$ images of different cases, and the results are presented in Fig. \ref{fig:euclidaperture}. By comparing with the results shown in Fig. \ref{fig:magauto}, we can see that the bias in the LSST-like case is indeed reduced significantly. For CSST, the results are nearly the same. This reveals to us that for the two space missions, the data combination at the catalogue level may be sufficient. By including ground-based data, however, careful treatments at the pixel level are necessary.

\begin{figure*}
\centering
\includegraphics[width=0.9\textwidth]{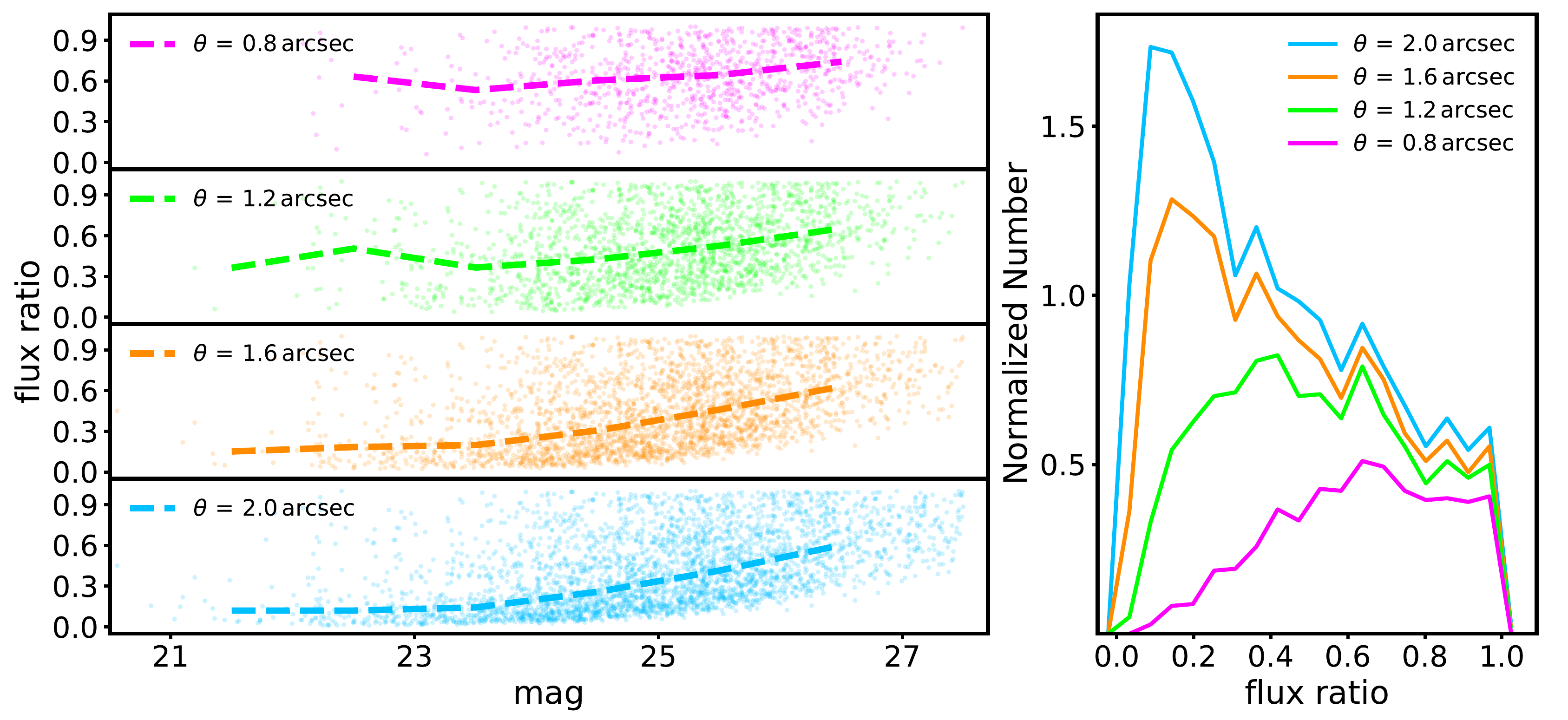}
\caption{{  The flux ratio of the contributing  {\it Euclid} VIS $I_1$ galaxies in the LSST-like total blends. The sub-panels in the left panel show the scatters and the median values of the flux ratios with different matching radii with the horizontal axis showing the LSST-like $r$-band magnitude of the total blends. The right panel is the corresponding results of the probability distributions normalized using the number of total blends for a matching radius of $2\arcsec$.}}
\label{fig:fluxratio}
\end{figure*}

{For the total blends in the LSST-like case shown in Fig. \ref{fig:totalblend}, we calculate the flux ratio of the blended galaxies from high-resolution {\it Euclid} VIS $I_1$ images. The results are presented in Fig. \ref{fig:fluxratio}, where the left panel shows the flux ratio vs. the LSST-like $r$-band magnitude of the total blends. {  Different sub-panels are for the total blends using different matching radius $\theta$, and the dashed lines are the corresponding medians.} The distributions of the flux ratio for the cases shown in the left are presented in the right panel where they are all divided by the number of total blends in the case of $\theta=2\arcsec$. Thus the areas under the curves reflect the relative probabilities with respect to the case of $\theta=2\arcsec$. The median flux ratio increases with the decrease of the matching radius, but the probability of total blends with smaller matching radius is lower. For $\theta=2\arcsec$, the median flux ratio is $\sim 0.1$, $\sim 0.2$ and $0.3$ at the LSST-like ${\rm mag}=23$, $24$ and $25$, respectively.} 

\begin{figure}
\includegraphics[width=0.48\textwidth]{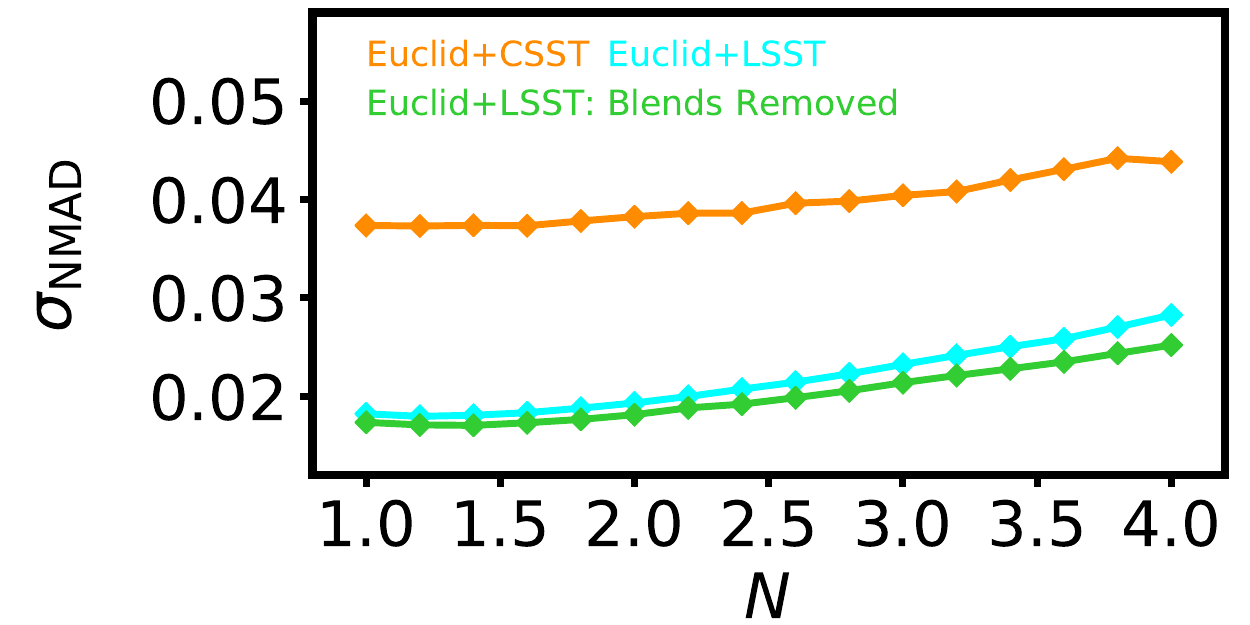}
\includegraphics[width=0.48\textwidth]{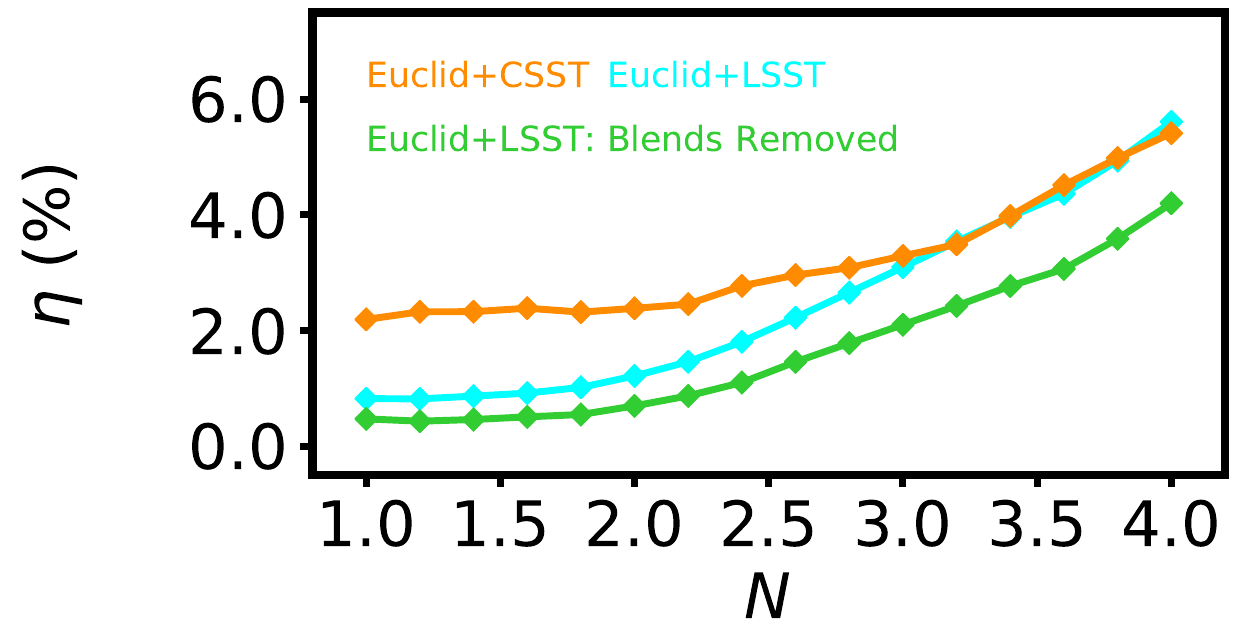}
\caption{The aperture dependence of $\sigma_{\rm NMAD}$ (top) and $\eta$ (bottom) for {\it Euclid}+LSST-like Year10 data (blue), {\it Euclid}+LSST-like Year10 with galaxies having neighbours within $2\arcsec$ removed (green), and {\it Euclid}+CSST (orange).}
\label{fig:EAZYD}
\end{figure}

{To show the blending effects on photo-$z$s, 
in Fig.~\ref{fig:EAZYD}, we re-plot the results of {\it Euclid}+LSST-like Year10 data and {\it Euclid}+CSST with different photometric apertures specified in Sect.~\ref{sec:photoz}. 
In addition, we also show the results for the {\it Euclid}+LSST-like case but discarding the galaxies with neighbours within $2\arcsec$. It is seen that by removing those potential blended galaxies, the outlier rate is reduced significantly for the aperture size larger than $2D\sim 2\arcsec$. This shows clearly the blending effect on the photo-$z$ measurements for ground-based observations, and careful deblending, which is not studied here, is needed for relatively large aperture sizes. For {\it Euclid}+CSST, because of the high-resolution images in optical bands, the effect of the aperture size resulting from the blending on photo-$z$ accuracy is much weaker, showing the advantage of the space-space data combination of {\it Euclid} and CSST with similarly high resolutions in optical bands. }

\subsection{PSF chromaticity}

The wide {\it Euclid} VIS band is designed to increase the signal-to-noise ratio of galaxies for accurate shear measurements. However, this leads to a significant chromaticity of the PSF. The effective PSF depends on the SED of each object. It can differ systematically for stars and galaxies. This can lead to significant biases in the shear measurement if not carefully accounted for \citep[e.g.][]{cypriano10,eriksen18,2015ApJ...807..182M,2018MNRAS.479.1491C}.

CSST has seven bands from NUV to $y$, and can provide valuable SED information for stars and galaxies. Such information can be used for {\it Euclid} to obtain accurate PSF measurements from stars. The PSF interpolation to galaxies can also be improved by including galaxy SEDs at high angular resolution, and thus reducing the PSF chromaticity bias in shear measurements. In addition, CSST is equipped with slitless spectrometers for galaxy redshift surveys. They cover the wavelength range from 255 nm to 1000 nm, divided into the three bands GU, GV, and GI. The spectral resolution is $\sim 200$. For the wide survey, the cumulative magnitude limit within each of the three bands is about ${\rm mag}\sim 23$ (5$\sigma$ for point source). For the deep survey, it is one magnitude deeper. The spectral data from CSST can then provide high-resolution stellar SEDs, which can further improve the {\it Euclid} PSF construction compared to the use of photometry-based SED estimates.

For extended galaxies, besides the PSF chromaticity bias described above, there is also a colour gradient (CG) bias in the shear measurement resulting from SED differences within a galaxy. The amplitude of the CG bias depends on several factors: the SED of the galaxy, the relative size of the galaxy compared to the PSF, and the width of the filter. The wide bandpass of the {\it Euclid} VIS filter can lead to a non-negligible CG bias. 
It has been demonstrated that the {\it Euclid} VIS CG bias can be calibrated using high spatial resolution data from two narrower bands covering the VIS bandpass \citep{semboloni13,er18}. The analyses show that for each type of galaxies based on, e.g. galaxy size, bulge-to-total flux ratio ${\rm {B/T}}$ and SED, their average multiplicative CG bias can be calibrated down to the 0.1\% level using 50 images with ${\rm {S/N}}\ge 50$ in the two narrower bands \citep{er18}.

CSST has a similar spatial resolution and a smaller pixel scale of $0\farcs074$ than that of {\it Euclid} VIS. The $g, i, r, z$ bands from CSST can thus provide narrower-band data for {\it Euclid} CG bias calibration. In order to see if CSST surveys can provide a sufficient number of high-${\rm {S/N}}$ galaxy images, we calculate the statistics of the ${\rm {S/N}}$ of CSST galaxies from our mock CSST simulations. 

{As noted in Sect.~4.1, the HDUV catalogue has a relatively small size. Although we generate 20 sets of mock images by adding different realizations of noise, for each galaxy, its input parameters, including position, redshift, intrinsic size $\sigma$ and ${\rm {B/T}}$ are the same in all the 20 sets. Thus the statistics in terms of these parameters concerned in the CG bias calibration are limited by the input HDUV catalogue. In comparison, the COSMOS2015 catalogue covers a much larger sky area and thus contains many more galaxies \citep{laigle16}. We therefore use the parallel CSST mock simulations based on the COSMOS2015 catalogue for the statistical analyses here.} 

{Specifically, we consider a portion of the COSMOS field covering an area of $0.6\times 0.6 \deg^2$ centered at (RA, DEC) = ($150\fdg0, 2\fdg0$). The selected field provides a large enough galaxy sample while being less affected by saturated star masks. Meanwhile, to avoid bad photometric measurements, we select galaxies by constraining the parameter \texttt{FLAGS} in $i^+$ band to be smaller than 3, meaning that galaxies suffering from the impact of saturation or truncation are excluded, but those with close neighbours are kept. The redshift of a galaxy is taken to be the median value of its photo-$z$ likelihood distribution (\texttt{PHOTOZ==ZPDF}) given in the COSMOS2015 catalogue. To acquire the size and orientation parameters of these galaxies, we match them with the HST/ACS F814W-band catalogue \citep{2007ApJS..172..219L} using a matching radius of $1 \arcsec$. Finally, we have 99\,264 galaxies as input for the image simulations. We follow the simulation steps described in Sect.~\ref{sec:methods_and_sims} to generate mock multi-band images for the CSST wide survey and also for the CSST deep survey which is about one magnitude deeper than the wide one.}
In accord with the studies in \citet{er18}, we consider calibrations
using two narrower bands. We therefore stack the images of $g, r$ bands and $i, z$ bands separately to build the calibration data sets.

\begin{figure}
\centering
\includegraphics[width=8cm]{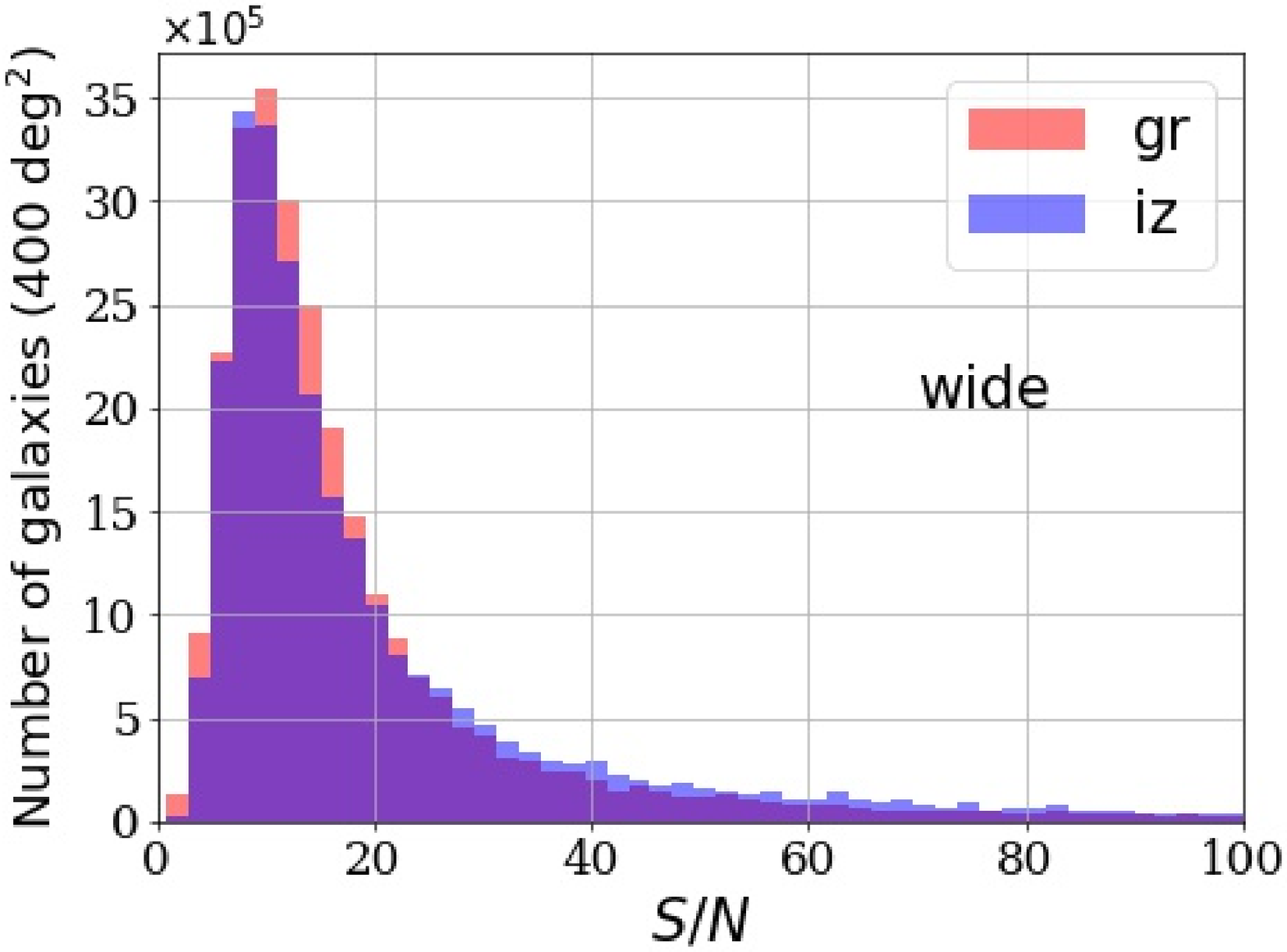}\\
\includegraphics[width=8cm]{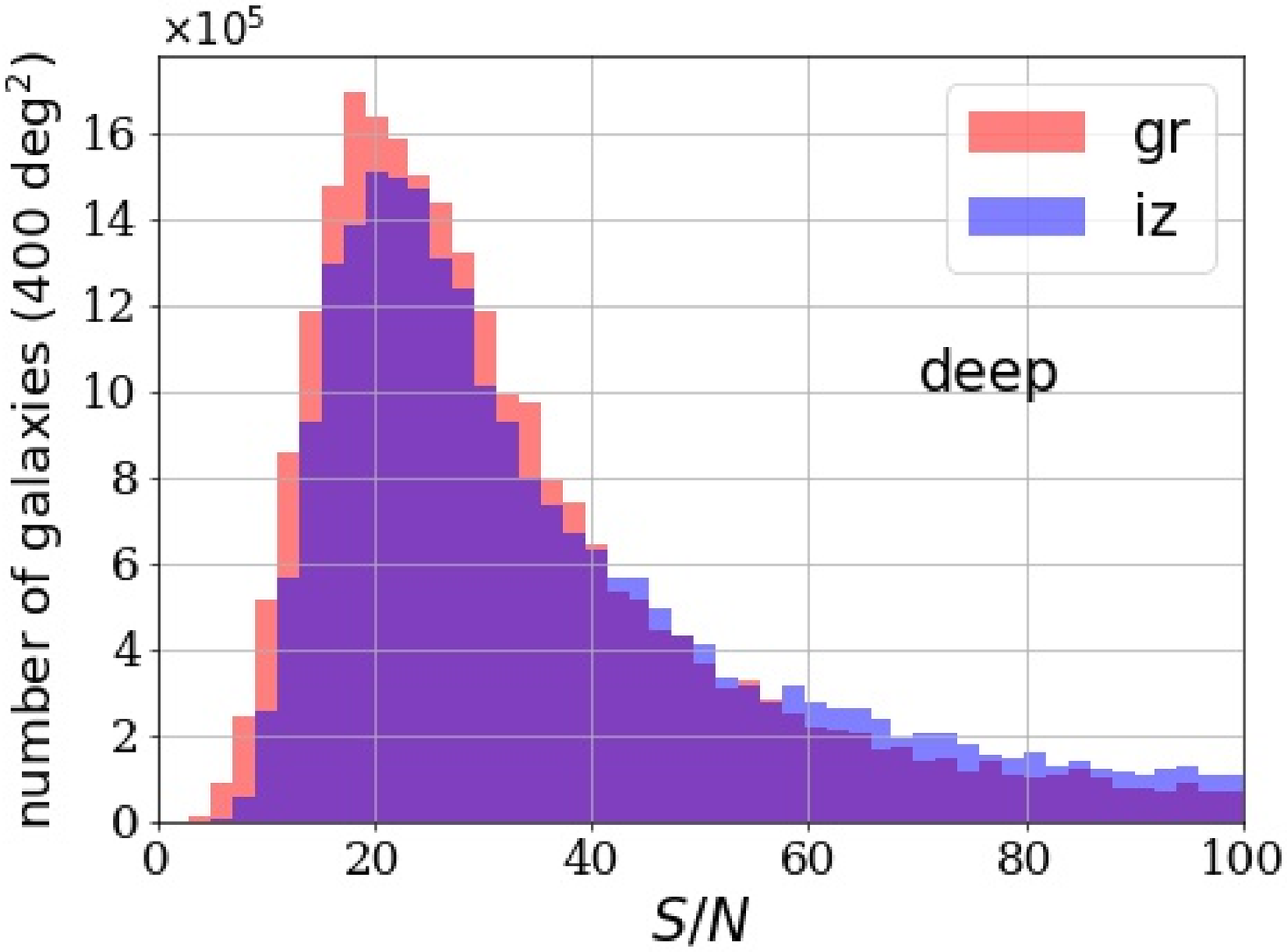}
\caption{The distribution of galaxy ${\rm {S/N}}$ from CSST simulations. Top and bottom panels are for wide and deep surveys, respectively. Only galaxies with {\it Euclid} VIS mag$<$ 24.5 are included. }
\label{fig:s2n_csst_wide_deep}
\end{figure}

\begin{figure}
\centering
\includegraphics[width=8cm]{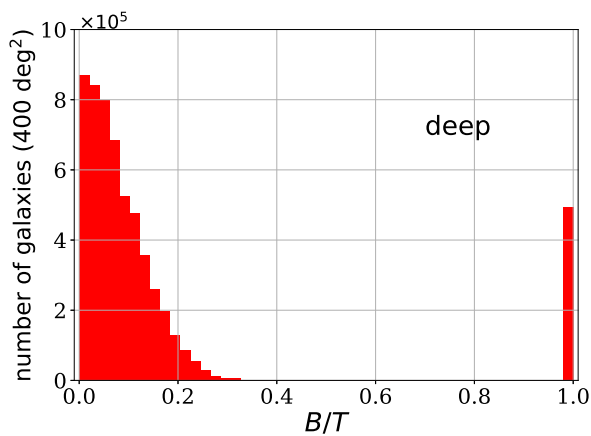}
\caption{The number distribution of galaxies with ${\rm {S/N}}>50$ as a function of ${\rm {B/T}}$ expected from CSST deep survey of $400$ deg$^2$. }
\label{fig:b2t_distr_deep}
\end{figure}

In Fig.~\ref{fig:s2n_csst_wide_deep}, we show the ${\rm {S/N}}$ distributions of galaxies from CSST wide (top) and deep (bottom) simulations. Here we only consider galaxies with {\it Euclid} VIS ${\rm {mag}}<24.5$. For comparison, we scale the galaxy numbers here to a sky area of $ 400 {\hbox{ deg}}^2$, the planned coverage of the CSST deep survey. The corresponding number of galaxies with ${\rm {S/N}}\ge 50$ is $2.07\times 10^6$ and $5.83\times 10^6$ for the wide and deep cases, respectively. We further consider galaxies with different properties. It is known that the bulge and disk components of galaxies typically have different SEDs. Thus galaxies with different ${\rm {B/T}}$ can have different CG biases, and need to be calibrated separately. We therefore bin galaxies based on ${\rm {B/T}}$ and calculate the ${\rm {S/N}}$ for galaxies in each bin. Fig.~\ref{fig:b2t_distr_deep} shows the number distribution of galaxies with ${\rm {S/N}}>50$ with respect to ${\rm {B/T}}$ from CSST deep simulations. {  It is noted that the absence of galaxies in the range $0.3\lesssim {\rm {B/T}}< 1.0$ is attributed to the input ${\rm {B/T}}$ distribution as described in Sect.~\ref{sec:methods_and_sims}}. We can see that the $400\hbox{ deg}^2$ CSST deep survey can provide sufficient numbers of galaxies in nearly all the ${\rm {B/T}}$ bins for {\it Euclid} to calibrate the CG bias. 

\begin{figure}
\centering
\includegraphics[width=8cm]{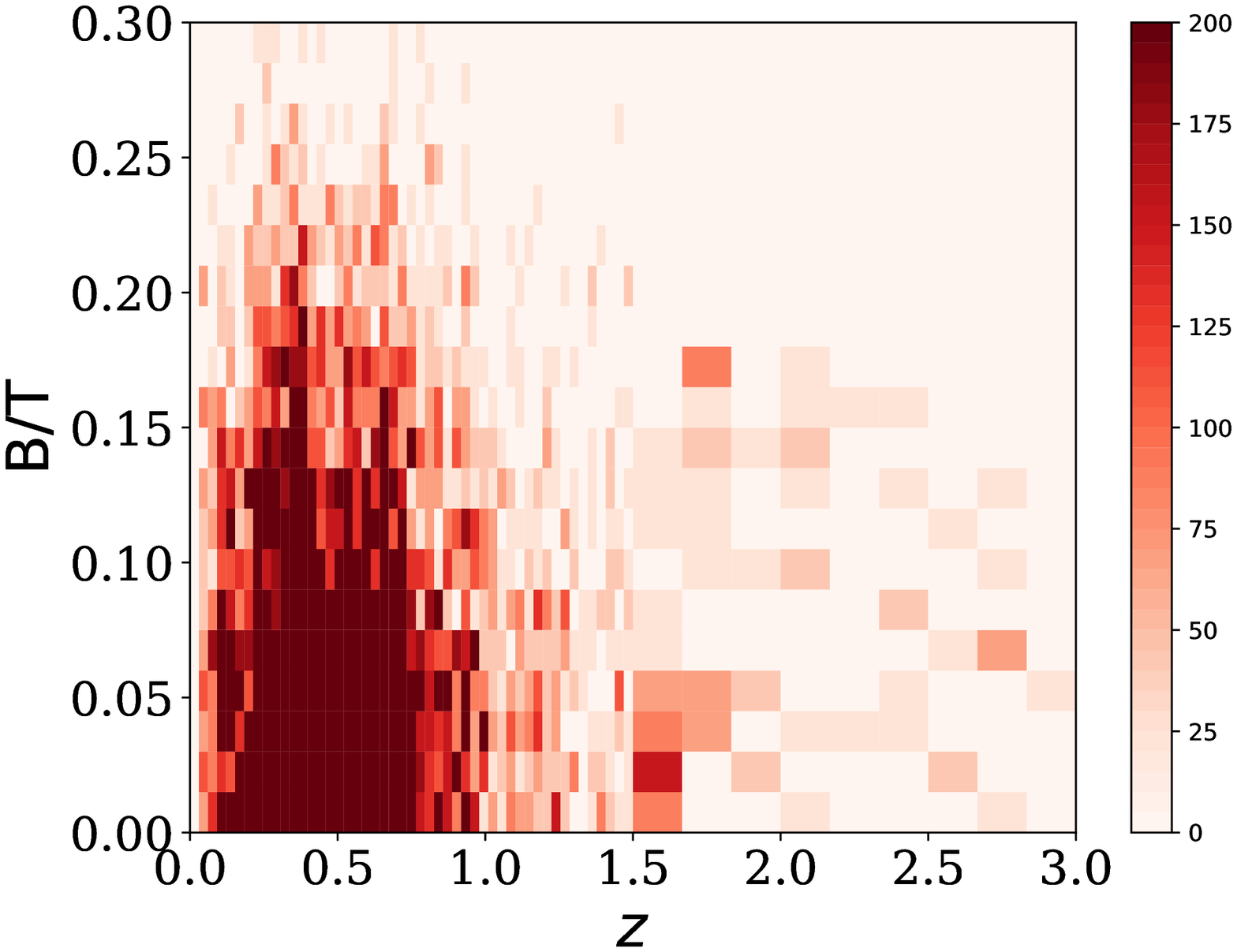}\\
\includegraphics[width=8cm]{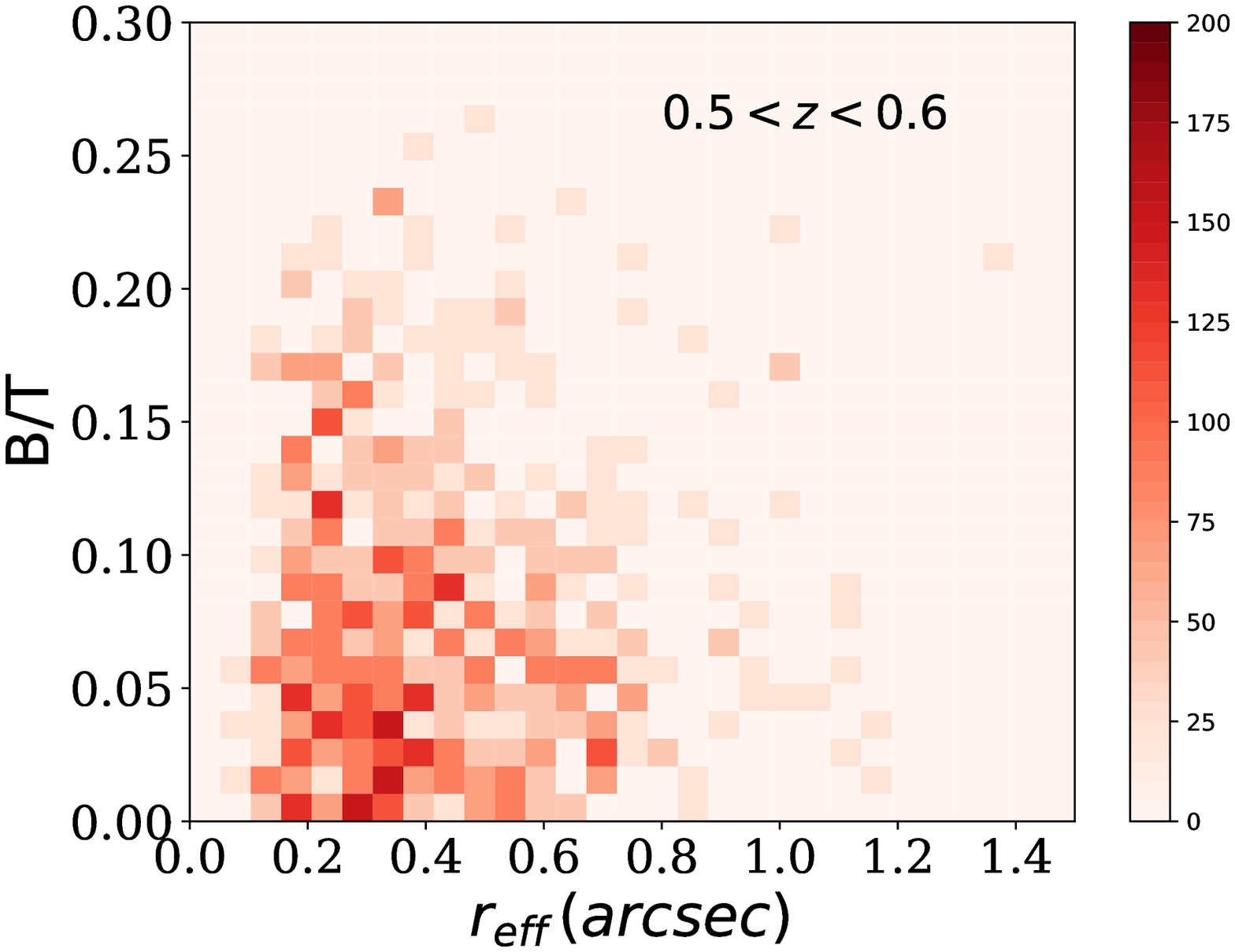}
\caption{The number distribution of galaxies with ${\rm {S/N}}>50$ in 2-D grids of ${\rm {B/T}}$ vs. redshift $z$ (top panel) and of ${\rm {B/T}}$ vs. the intrinsic size $\sigma$ (bottom panel), expected from the full 400 deg$^2$ of CSST deep survey. The colour bar represents the number of galaxies in unit of 50, and truncated at 200 for clear visibility. In the top panel, the bin size of ${\rm {B/T}}$ is 0.015, and that for $z$ is 0.03 for redshifts within the range of $[0,1.5]$ and 0.15 within $(1.5,3]$. In the bottom panel, the galaxies are selected with redshifts within $[0.5,0.6]$. }
\label{fig:zdis_snr50}
\end{figure}

We further analyse the feasibility of calibration in finer grids of galaxy parameters. In Fig.~\ref{fig:zdis_snr50}, the top panel shows the number distribution of galaxies with ${\rm {S/N}}\ge 50$ expected from the full CSST deep survey in 2-D grids of ${\rm {B/T}}$ vs. redshift $z$. Here the bin sizes are $0.015$ for ${\rm {B/T}}$, and $0.03$ ($0.15$) for $z$ at $z < 1.5$ ($z \ge 1.5$). {The bottom panel is for the number distribution in ${\rm {B/T}}$ vs. the intrinsic size of galaxies $\sigma$ defined in Sect.~4.1 for galaxies in the redshift range of $0.5\le z\le 0.6$.} 
One can see that the deep survey shall be able to provide a calibration sample up to  $z\sim 3$. At $z<1$, we can perform analyses for different types of galaxies separately, leading to more accurate calibrations for the CG bias for galaxies with different ${\rm {B/T}}$ and size.

{As we discussed, because of the ground-based source detection, the COSMOS2015 catalogue has a lower number density of galaxies than that of the HDUV catalogue with the source detection from the high-resolution HST images. Thus the statistics shown here are conservative. We expect that the CSST deep survey can provide a larger number of high-${\rm {S/N}}$ galaxies than that seen in Fig.~\ref{fig:s2n_csst_wide_deep} to Fig.~\ref{fig:zdis_snr50} for {\it Euclid} to calibrate the CG bias. }
 
Besides CG bias calibration for different samples of galaxies, it may even be possible to correct for the bias for individual galaxies with CSST high $S/N$ multi-band data. We leave this for future studies. 

%

\section{Further science}


Apart from WL analyses, the complementarity of the data from {\it Euclid} and CSST can benefit many more areas of research. Here we briefly mention a few topics that are outside of WL but relate to cosmology. They especially make use of the multi-band imaging data and the resulting photometric measurements.

The high resolution and large sky coverage of the two surveys allow us to find $\sim10^5$ strong-lensing systems. Their identification requires deep-learning-based machinery \citep[e.g.][]{Metcalf2019}. The colour information from the joint multi-band imaging data can improve the efficiency and accuracy of the strong-lensing system detection considerably \citep[e.g.][]{Petrillo2019,Pearson2019}.
Furthermore, {\it Euclid}+CSST can provide accurate photo-$z$s for both lenses and sources, and thus additional information on the geometrical distances of lens systems. This can tighten the lens modelling significantly.   
One of the major science cases of strong lensing in both CSST and {\it Euclid} is to detect small perturbations to Einstein rings/giant arcs, which can put constraints on the nature of dark matter \citep[e.g.][]{Li2016,Li2017}. This requires very accurate modelling of the lens mass distribution. Data from independent surveys can provide unique discriminating tests of the model accuracy. In addition, because of the wavelength-dependence of the light distribution for a source galaxy, its lensed image positions in the optical and NIR can be different. Therefore, the joint lensing analyses from CSST optical and {\it Euclid} NIR observations provide more sample points to constrain the lens potential, helping to break degeneracies in the lens modelling. 

The photo-$z$ measurements from {\it Euclid}+CSST can also improve the detection of clusters of galaxies. In particular, the NIR bands from {\it Euclid} can lead to more efficient identifications of high-redshift clusters comparing to that with CSST 7 bands only \citep{Wen2021}. The improved photo-$z$ accuracy is also important to allow us to probe tomographic galaxy clustering using photometric galaxy samples, which typically have a much larger number of galaxies extending to higher redshifts than that of spectroscopic galaxy samples. 

{  Additionally, the data from NUV to $H$ from {\it Euclid}+CSST will form a great data pool for legacy studies, such as galaxy formation and evolution, near-field cosmology, etc.. As an example, the high-resolution multi-band data from the two space missions will allow robust SED modeling, with, e.g.  CIGALE \citep{2019A&A...622A.103B,2020MNRAS.491..740Y}, to derive the physical properties, such as the stellar mass, star formation rate for a large sample of galaxies. Carefully decomposing light from AGNs and host galaxies can also provide information about the AGN luminosity function to the faint end. Combining with the spectroscopic data from the two surveys, it is expected that the synergy will give new insights into the formation and evolution of galaxies and AGNs as well as the large-scale structures of the Universe. Because both projects have different working groups investigating on different subjects, detailed explorations of the synergy in areas other than WL cosmology is beyond the scope of our studies here.}

\section{Discussion and conclusions}

The high-quality multi-band data from upcoming large surveys will allow us to probe the Universe with an unprecedented precision. How to efficiently utilize data from different surveys to enhance the science gains is becoming an increasingly important topic. 

In this study, we focus on the two space missions, {\it Euclid} and CSST, and investigate their potential synergies emphasizing the aspects related to the weak lensing studies. While having a similar high spatial resolution and largely overlapping sky coverage, the two missions cover different wavebands. Their combination can produce multi-band data from NUV to $H$ for over a billion galaxies. We analyse the photo-$z$ performance from {\it Euclid}+CSST. For galaxies with {\it Euclid} VIS ${\rm {mag}}<24.5$ and $0.2<z_{\mathrm{output}}<2.6$, the dispersion is $\sigma_{\rm NMAD}\approx 0.04$, and the outlier fraction is $\eta \approx 2.4\%$. The precision is comparable to that from the combination of the {\it Euclid}+LSST-like 2 -- 3 year data. Because of the similar sky coverage, the utilization of the data from the two space missions can be done in a uniform way, and thus the systematic errors in the data processing can be well controlled. This is advantageous over the combination of {\it Euclid} with several different ground-based surveys with different performance characteristics. Compared with the photo-$z$ results from CSST seven bands only, the inclusion of the {\it Euclid} NIR bands leads to a significant reduction of the outlier fraction from $\sim 8.5\%$ to $\sim 2.4\%$. 

We also carry out simulations to investigate the image blending effects. For {\it Euclid} and CSST, because of their high resolutions, the blending effects are minimal, and thus a catalogue-level data combination from the two space missions is feasible. For combining with ground-based data, however, pixel-level analyses with careful deblending are necessary. While there are still blended galaxies for the simulated space-based observations, their fraction is much lower compared to ground-based observations. Their deblending can be done efficiently by employing multi-band pixel data from CSST and {\it Euclid}. This will be investigated in future studies. 

Analyses of {\it Euclid} VIS data are significantly affected by the PSF chromaticity. Accordingly, the addition of multi-band imaging and slitless spectroscopic observations from CSST can improve the {\it Euclid} VIS PSF modelling, and provide calibrations for chromaticity bias and the CG bias in shear measurements. Our studies show that the CSST 400 deg$^2$ deep multi-band photometric survey can provide a large sample of galaxies with high ${\rm {S/N}}>50$. Therefore the multiplicative CG bias can be well calibrated to the level of $0.1\%$ for {\it Euclid} galaxies with different ${\rm {B/T}}$, and at different redshifts up to $z\sim 3$. On the other hand, the NIR data from {\it Euclid} can be valuable for CSST for a better star-galaxy separation and thus to improve the precision of the PSF construction. 

The studies shown here assume the full availability of the data. The two missions have however different schedules. {\it Euclid} is expected to be launched in early 2023, and CSST targets late 2023 or early 2024 for launching. It is also noted that for CSST, different filters are distributed across the focal plane. Hence, the survey strategy will impact how quickly observations in all filters can be completed for a certain patch of the sky. With {\it Euclid} in orbit first, it might be worth thinking of a coordinated survey strategy for CSST, so that the synergies of the two missions can be realized in the early years of the surveys. The joint studies can benefit many more areas of astrophysical research beyond the WL analyses presented here. Given the great potential, possible survey coordination between the two missions, both for the wide and deep survey components, deserves further careful investigation.

\begin{acknowledgements}
    This research was initiated under the support of the ISSI/ISSI-BJ International Team Programs. All the Chinese coauthors acknowledge the science research grants from the China Manned Space Project with Nos.CMS-CSST-2021-A01 and CMS-CSST-2021-B01. The Bonn group acknowledges 
support from  the German Federal Ministry for Economic Affairs and Energy (BMWi) provided through DLR under projects 50QE1103 and 50QE2002, as well as funding provided  by the European Union's Horizon 2020 research and innovation programme under grant agreement No. 776247. DZL acknowledges the support from NSFC under the grant of 12103043. ZHF is supported in part by NSFC under 11933002, U1931210 and 11333001. RL acknowledges the NSFC grands of 11988101,11773032 and 12022306, and the support from K.C.Wong Education Foundation, CAS Project for Young Scientists in Basic Research (No. YSBR-062). DS is a member of the International Max Planck Research School (IMPRS) for Astronomy and Astrophysics at the Universities of Bonn and Cologne. HYS acknowledges the supports from NSFC under 11973070, the Shanghai Committee of Science and Technology under the grant No.19ZR1466600 and Key Research Program of Frontier Sciences, CAS, Grant No. ZDBS-LY-7013. YST acknowledges the financial support from the Australian Research Council through DECRA Fellowship DE220101520. LPF acknowledges the supports from NSFC grants 11933002, and the Dawn Program 19SG41 \& the Innovation Program 2019-01-07-00-02-E00032 of SMEC. HHi is supported by a Heisenberg grant of the Deutsche Forschungsgemeinschaft (Hi 1495/5-1). HHi and AHW acknowledge support by the ERC (Consolidator Grant No. 770935). XKL acknowledges the supports from NSFC of China under grant no. 11803028 and no. 12173033, and YNU grant no. C176220100008. 
    The COSMOS2015 sample is based on data products from observations made with ESO Telescopes at the La Silla Paranal Observatory under ESO programme ID 179.A-2005 and on data products produced by TERAPIX and the Cambridge Astronomy Survey Unit on behalf of the UltraVISTA consortium. The employed CANDELS images are based on observations taken by the CANDELS Multi-Cycle Treasury Program with the NASA/ESA HST, which is operated by the Association of Universities for Research in Astronomy, Inc., under NASA contract NAS5-26555.
\end{acknowledgements}
 \bibliographystyle{aa} 
\bibliography{euclid-cssos}

\begin{thebibliography}{101}
\expandafter\ifx\csname natexlab\endcsname\relax\def\natexlab#1{#1}\fi

\bibitem[{{Abbott} {et~al.}(2021){Abbott}, {Adam{\'o}w}, {Aguena}, {Allam},
  {Amon}, {Annis}, {Avila}, {Bacon}, {Banerji}, {Bechtol}, {Becker},
  {Bernstein}, {Bertin}, {Bhargava}, {Bridle}, {Brooks}, {Burke}, {Carnero
  Rosell}, {Carrasco Kind}, {Carretero}, {Castander}, {Cawthon}, {Chang},
  {Choi}, {Conselice}, {Costanzi}, {Crocce}, {da Costa}, {Davis}, {De Vicente},
  {DeRose}, {Desai}, {Diehl}, {Dietrich}, {Drlica-Wagner}, {Eckert},
  {Elvin-Poole}, {Everett}, {Evrard}, {Ferrero}, {Fert{\'e}}, {Flaugher},
  {Fosalba}, {Friedel}, {Frieman}, {Garc{\'\i}a-Bellido}, {Gaztanaga},
  {Gelman}, {Gerdes}, {Giannantonio}, {Gill}, {Gruen}, {Gruendl}, {Gschwend},
  {Gutierrez}, {Hartley}, {Hinton}, {Hollowood}, {Honscheid}, {Huterer},
  {James}, {Jeltema}, {Johnson}, {Kent}, {Kron}, {Kuehn}, {Kuropatkin},
  {Lahav}, {Li}, {Lidman}, {Lin}, {MacCrann}, {Maia}, {Manning}, {Maloney},
  {March}, {Marshall}, {Martini}, {Melchior}, {Menanteau}, {Miquel}, {Morgan},
  {Myles}, {Neilsen}, {Ogando}, {Palmese}, {Paz-Chinch{\'o}n}, {Petravick},
  {Pieres}, {Plazas}, {Pond}, {Rodriguez-Monroy}, {Romer}, {Roodman}, {Rykoff},
  {Sako}, {Sanchez}, {Santiago}, {Scarpine}, {Serrano}, {Sevilla-Noarbe},
  {Smith}, {Smith}, {Soares-Santos}, {Suchyta}, {Swanson}, {Tarle}, {Thomas},
  {To}, {Tremblay}, {Troxel}, {Tucker}, {Turner}, {Varga}, {Walker},
  {Wechsler}, {Weller}, {Wester}, {Wilkinson}, {Yanny}, {Zhang}, {Nikutta},
  {Fitzpatrick}, {Jacques}, {Scott}, {Olsen}, {Huang}, {Herrera}, {Juneau},
  {Nidever}, {Weaver}, {Adean}, {Correia}, {de Freitas}, {Freitas},
  {Singulani}, {Vila-Verde}, \& {Linea Science Server}}]{2021arXiv210105765D}
{Abbott}, T.~M.~C., {Adam{\'o}w}, M., {Aguena}, M., {et~al.} 2021, \apjs, 255,
  20

\bibitem[{{Abbott} {et~al.}(2022){Abbott}, {Aguena}, {Alarcon}, {Allam},
  {Alves}, {Amon}, {Andrade-Oliveira}, {Annis}, {Avila}, {Bacon}, {Baxter},
  {Bechtol}, {Becker}, {Bernstein}, {Bhargava}, {Birrer}, {Blazek},
  {Brandao-Souza}, {Bridle}, {Brooks}, {Buckley-Geer}, {Burke}, {Camacho},
  {Campos}, {Carnero Rosell}, {Carrasco Kind}, {Carretero}, {Castander},
  {Cawthon}, {Chang}, {Chen}, {Chen}, {Choi}, {Conselice}, {Cordero},
  {Costanzi}, {Crocce}, {da Costa}, {da Silva Pereira}, {Davis}, {Davis}, {De
  Vicente}, {DeRose}, {Desai}, {Di Valentino}, {Diehl}, {Dietrich}, {Dodelson},
  {Doel}, {Doux}, {Drlica-Wagner}, {Eckert}, {Eifler}, {Elsner}, {Elvin-Poole},
  {Everett}, {Evrard}, {Fang}, {Farahi}, {Fernandez}, {Ferrero}, {Fert{\'e}},
  {Fosalba}, {Friedrich}, {Frieman}, {Garc{\'\i}a-Bellido}, {Gatti},
  {Gaztanaga}, {Gerdes}, {Giannantonio}, {Giannini}, {Gruen}, {Gruendl},
  {Gschwend}, {Gutierrez}, {Harrison}, {Hartley}, {Herner}, {Hinton},
  {Hollowood}, {Honscheid}, {Hoyle}, {Huff}, {Huterer}, {Jain}, {James},
  {Jarvis}, {Jeffrey}, {Jeltema}, {Kovacs}, {Krause}, {Kron}, {Kuehn},
  {Kuropatkin}, {Lahav}, {Leget}, {Lemos}, {Liddle}, {Lidman}, {Lima}, {Lin},
  {MacCrann}, {Maia}, {Marshall}, {Martini}, {McCullough}, {Melchior},
  {Mena-Fern{\'a}ndez}, {Menanteau}, {Miquel}, {Mohr}, {Morgan}, {Muir},
  {Myles}, {Nadathur}, {Navarro-Alsina}, {Nichol}, {Ogando}, {Omori},
  {Palmese}, {Pandey}, {Park}, {Paz-Chinch{\'o}n}, {Petravick}, {Pieres},
  {Plazas Malag{\'o}n}, {Porredon}, {Prat}, {Raveri}, {Rodriguez-Monroy},
  {Rollins}, {Romer}, {Roodman}, {Rosenfeld}, {Ross}, {Rykoff}, {Samuroff},
  {S{\'a}nchez}, {Sanchez}, {Sanchez}, {Sanchez Cid}, {Scarpine}, {Schubnell},
  {Scolnic}, {Secco}, {Serrano}, {Sevilla-Noarbe}, {Sheldon}, {Shin}, {Smith},
  {Soares-Santos}, {Suchyta}, {Swanson}, {Tabbutt}, {Tarle}, {Thomas}, {To},
  {Troja}, {Troxel}, {Tucker}, {Tutusaus}, {Varga}, {Walker}, {Weaverdyck},
  {Wechsler}, {Weller}, {Yanny}, {Yin}, {Zhang}, {Zuntz}, \& {DES
  Collaboration}}]{DESY3}
{Abbott}, T.~M.~C., {Aguena}, M., {Alarcon}, A., {et~al.} 2022, \prd, 105,
  023520

\bibitem[{{Aihara} {et~al.}(2018){Aihara}, {Arimoto}, {Armstrong}, {Arnouts},
  {Bahcall}, {Bickerton}, {Bosch}, {Bundy}, {Capak}, {Chan}, {Chiba}, {Coupon},
  {Egami}, {Enoki}, {Finet}, {Fujimori}, {Fujimoto}, {Furusawa}, {Furusawa},
  {Goto}, {Goulding}, {Greco}, {Greene}, {Gunn}, {Hamana}, {Harikane},
  {Hashimoto}, {Hattori}, {Hayashi}, {Hayashi}, {He{\l}miniak}, {Higuchi},
  {Hikage}, {Ho}, {Hsieh}, {Huang}, {Huang}, {Ikeda}, {Imanishi}, {Inoue},
  {Iwasawa}, {Iwata}, {Jaelani}, {Jian}, {Kamata}, {Karoji}, {Kashikawa},
  {Katayama}, {Kawanomoto}, {Kayo}, {Koda}, {Koike}, {Kojima}, {Komiyama},
  {Konno}, {Koshida}, {Koyama}, {Kusakabe}, {Leauthaud}, {Lee}, {Lin}, {Lin},
  {Lupton}, {Mandelbaum}, {Matsuoka}, {Medezinski}, {Mineo}, {Miyama},
  {Miyatake}, {Miyazaki}, {Momose}, {More}, {More}, {Moritani}, {Moriya},
  {Morokuma}, {Mukae}, {Murata}, {Murayama}, {Nagao}, {Nakata}, {Niida},
  {Niikura}, {Nishizawa}, {Obuchi}, {Oguri}, {Oishi}, {Okabe}, {Okamoto},
  {Okura}, {Ono}, {Onodera}, {Onoue}, {Osato}, {Ouchi}, {Price}, {Pyo}, {Sako},
  {Sawicki}, {Shibuya}, {Shimasaku}, {Shimono}, {Shirasaki}, {Silverman},
  {Simet}, {Speagle}, {Spergel}, {Strauss}, {Sugahara}, {Sugiyama}, {Suto},
  {Suyu}, {Suzuki}, {Tait}, {Takada}, {Takata}, {Tamura}, {Tanaka}, {Tanaka},
  {Tanaka}, {Tanaka}, {Terai}, {Terashima}, {Toba}, {Tominaga}, {Toshikawa},
  {Turner}, {Uchida}, {Uchiyama}, {Umetsu}, {Uraguchi}, {Urata}, {Usuda},
  {Utsumi}, {Wang}, {Wang}, {Wong}, {Yabe}, {Yamada}, {Yamanoi}, {Yasuda},
  {Yeh}, {Yonehara}, \& {Yuma}}]{HSC}
{Aihara}, H., {Arimoto}, N., {Armstrong}, R., {et~al.} 2018, \pasj, 70, S4

\bibitem[{{Albrecht} {et~al.}(2006){Albrecht}, {Bernstein}, {Cahn}, {Freedman},
  {Hewitt}, {Hu}, {Huth}, {Kamionkowski}, {Kolb}, {Knox}, {Mather}, {Staggs},
  \& {Suntzeff}}]{albrecht06}
{Albrecht}, A., {Bernstein}, G., {Cahn}, R., {et~al.} 2006, astro-ph/0609591

\bibitem[{{Arcelin} {et~al.}(2021){Arcelin}, {Doux}, {Aubourg}, {Roucelle}, \&
  {LSST Dark Energy Science Collaboration}}]{Arcelin2021}
{Arcelin}, B., {Doux}, C., {Aubourg}, E., {Roucelle}, C., \& {LSST Dark Energy
  Science Collaboration}. 2021, \mnras, 500, 531

\bibitem[{{Astropy Collaboration} {et~al.}(2013){Astropy Collaboration},
  {Robitaille}, {Tollerud}, {Greenfield}, {Droettboom}, {Bray}, {Aldcroft},
  {Davis}, {Ginsburg}, {Price-Whelan}, {Kerzendorf}, {Conley}, {Crighton},
  {Barbary}, {Muna}, {Ferguson}, {Grollier}, {Parikh}, {Nair}, {Unther},
  {Deil}, {Woillez}, {Conseil}, {Kramer}, {Turner}, {Singer}, {Fox}, {Weaver},
  {Zabalza}, {Edwards}, {Azalee Bostroem}, {Burke}, {Casey}, {Crawford},
  {Dencheva}, {Ely}, {Jenness}, {Labrie}, {Lim}, {Pierfederici}, {Pontzen},
  {Ptak}, {Refsdal}, {Servillat}, \& {Streicher}}]{2013A&A...558A..33A}
{Astropy Collaboration}, {Robitaille}, T.~P., {Tollerud}, E.~J., {et~al.} 2013,
  \aap, 558, A33

\bibitem[{{Bartelmann} \& {Schneider}(2001)}]{BS2001}
{Bartelmann}, M. \& {Schneider}, P. 2001, Phys. Rept., 340, 291

\bibitem[{{Bertin} \& {Arnouts}(1996)}]{BA1996}
{Bertin}, E. \& {Arnouts}, S. 1996, A\&A Suppl., 117, 393

\bibitem[{{Boquien} {et~al.}(2019){Boquien}, {Burgarella}, {Roehlly}, {Buat},
  {Ciesla}, {Corre}, {Inoue}, \& {Salas}}]{2019A&A...622A.103B}
{Boquien}, M., {Burgarella}, D., {Roehlly}, Y., {et~al.} 2019, \aap, 622, A103

\bibitem[{{Bosch} {et~al.}(2018){Bosch}, {Armstrong}, {Bickerton}, {Furusawa},
  {Ikeda}, {Koike}, {Lupton}, {Mineo}, {Price}, {Takata}, {Tanaka}, {Yasuda},
  {AlSayyad}, {Becker}, {Coulton}, {Coupon}, {Garmilla}, {Huang}, {Krughoff},
  {Lang}, {Leauthaud}, {Lim}, {Lust}, {MacArthur}, {Mandelbaum}, {Miyatake},
  {Miyazaki}, {Murata}, {More}, {Okura}, {Owen}, {Swinbank}, {Strauss},
  {Yamada}, \& {Yamanoi}}]{Bosch2018}
{Bosch}, J., {Armstrong}, R., {Bickerton}, S., {et~al.} 2018, \pasj, 70, S5

\bibitem[{{Brammer} {et~al.}(2008){Brammer}, {van Dokkum}, \&
  {Coppi}}]{Brammer2008}
{Brammer}, G.~B., {van Dokkum}, P.~G., \& {Coppi}, P. 2008, \apj, 686, 1503

\bibitem[{{Cao} {et~al.}(2018){Cao}, {Gong}, {Meng}, {Xu}, {Chen}, {Guo}, {Li},
  {Liu}, {Xue}, {Cao}, {Fu}, {Zhang}, {Wang}, \& {Zhan}}]{cao18}
{Cao}, Y., {Gong}, Y., {Meng}, X.-M., {et~al.} 2018, \mnras, 480, 2178

\bibitem[{{Carlsten} {et~al.}(2018){Carlsten}, {Strauss}, {Lupton}, {Meyers},
  \& {Miyazaki}}]{2018MNRAS.479.1491C}
{Carlsten}, S.~G., {Strauss}, M.~A., {Lupton}, R.~H., {Meyers}, J.~E., \&
  {Miyazaki}, S. 2018, \mnras, 479, 1491

\bibitem[{{Casertano} {et~al.}(2000){Casertano}, {de Mello}, {Dickinson},
  {Ferguson}, {Fruchter}, {Gonzalez-Lopezlira}, {Heyer}, {Hook}, {Levay},
  {Lucas}, {Mack}, {Makidon}, {Mutchler}, {Smith}, {Stiavelli}, {Wiggs}, \&
  {Williams}}]{2000AJ....120.2747C}
{Casertano}, S., {de Mello}, D., {Dickinson}, M., {et~al.} 2000, \aj, 120, 2747

\bibitem[{{Chary} {et~al.}(2020){Chary}, {Helou}, {Brammer}, {Capak}, {Faisst},
  {Flynn}, {Groom}, {Ferguson}, {Grillmair}, {Hemmati}, {Koekemoer}, {Lee},
  {Malhotra}, {Miyatake}, {Melchior}, {Momcheva}, {Newman}, {Masiero},
  {Paladini}, {Prakash}, {Rusholme}, {Stickley}, {Smith}, {Wood-Vasey}, \&
  {Teplitz}}]{Chary2020}
{Chary}, R., {Helou}, G., {Brammer}, G., {et~al.} 2020, arXiv:2008.10663

\bibitem[{{Cypriano} {et~al.}(2010){Cypriano}, {Amara}, {Voigt}, {Bridle},
  {Abdalla}, {R{\'e}fr{\'e}gier}, {Seiffert}, \& {Rhodes}}]{cypriano10}
{Cypriano}, E.~S., {Amara}, A., {Voigt}, L.~M., {et~al.} 2010, \mnras, 405, 494

\bibitem[{{Dark Energy Survey Collaboration} {et~al.}(2016){Dark Energy Survey
  Collaboration}, {Abbott}, {Abdalla}, {Aleksic}, {Allam}, {Amara}, {Bacon},
  {Balbinot}, {Banerji}, {Bechtol}, {Benoit-Levy}, \& et~al.}]{DES}
{Dark Energy Survey Collaboration}, {Abbott}, T., {Abdalla}, F., {et~al.} 2016,
  \mnras, 460, 1270

\bibitem[{{Eifler} {et~al.}(2021){Eifler}, {Simet}, {Krause}, {Hirata},
  {Huang}, {Fang}, {Miranda}, {Mandelbaum}, {Doux}, {Heinrich}, \&
  et~al.}]{Eifler2021}
{Eifler}, T., {Simet}, M., {Krause}, E., {et~al.} 2021, \mnras, 507, 1514

\bibitem[{{Er} {et~al.}(2018){Er}, {Hoekstra}, {Schrabback}, {Cardone},
  {Scaramella}, {Maoli}, {Vicinanza}, {Gillis}, \& {Rhodes}}]{er18}
{Er}, X., {Hoekstra}, H., {Schrabback}, T., {et~al.} 2018, \mnras, 476, 5645

\bibitem[{{Eriksen} \& {Hoekstra}(2018)}]{eriksen18}
{Eriksen}, M. \& {Hoekstra}, H. 2018, \mnras, 477, 3433

\bibitem[{{Euclid Collaboration} {et~al.}(2020{\natexlab{a}}){Euclid
  Collaboration}, {Blanchard}, {Camera}, {Carbone}, {Cardone}, {Casas},
  {Clesse}, {Ili{\'c}}, {Kilbinger}, {Kitching}, {Kunz}, {Lacasa}, {Linder},
  {Majerotto}, {Markovi{\v{c}}}, {Martinelli}, {Pettorino}, {Pourtsidou},
  {Sakr}, {S{\'a}nchez}, {Sapone}, {Tutusaus}, {Yahia-Cherif}, {Yankelevich},
  {Andreon}, {Aussel}, {Balaguera-Antol{\'\i}nez}, {Baldi}, {Bardelli},
  {Bender}, {Biviano}, {Bonino}, {Boucaud}, {Bozzo}, {Branchini}, {Brau-Nogue},
  {Brescia}, {Brinchmann}, {Burigana}, {Cabanac}, {Capobianco}, {Cappi},
  {Carretero}, {Carvalho}, {Casas}, {Castander}, {Castellano}, {Cavuoti},
  {Cimatti}, {Cledassou}, {Colodro-Conde}, {Congedo}, {Conselice}, {Conversi},
  {Copin}, {Corcione}, {Coupon}, {Courtois}, {Cropper}, {Da Silva}, {de la
  Torre}, {Di Ferdinando}, {Dubath}, {Ducret}, {Duncan}, {Dupac}, {Dusini},
  {Fabbian}, {Fabricius}, {Farrens}, {Fosalba}, {Fotopoulou}, {Fourmanoit},
  {Frailis}, {Franceschi}, {Franzetti}, {Fumana}, {Galeotta}, {Gillard},
  {Gillis}, {Giocoli}, {G{\'o}mez-Alvarez}, {Graci{\'a}-Carpio}, {Grupp},
  {Guzzo}, {Hoekstra}, {Hormuth}, {Israel}, {Jahnke}, {Keihanen}, {Kermiche},
  {Kirkpatrick}, {Kohley}, {Kubik}, {Kurki-Suonio}, {Ligori}, {Lilje}, {Lloro},
  {Maino}, {Maiorano}, {Marggraf}, {Martinet}, {Marulli}, {Massey},
  {Medinaceli}, {Mei}, {Mellier}, {Metcalf}, {Metge}, {Meylan}, {Moresco},
  {Moscardini}, {Munari}, {Nichol}, {Niemi}, {Nucita}, {Padilla}, {Paltani},
  {Pasian}, {Percival}, {Pires}, {Polenta}, {Poncet}, {Pozzetti}, {Racca},
  {Raison}, {Renzi}, {Rhodes}, {Romelli}, {Roncarelli}, {Rossetti}, {Saglia},
  {Schneider}, {Scottez}, {Secroun}, {Sirri}, {Stanco}, {Starck}, {Sureau},
  {Tallada-Cresp{\'\i}}, {Tavagnacco}, {Taylor}, {Tenti}, {Tereno},
  {Toledo-Moreo}, {Torradeflot}, {Valenziano}, {Vassallo}, {Verdoes Kleijn},
  {Viel}, {Wang}, {Zacchei}, {Zoubian}, \& {Zucca}}]{Euclid-prep-VII_ISTF_20}
{Euclid Collaboration}, {Blanchard}, A., {Camera}, S., {et~al.}
  2020{\natexlab{a}}, \aap, 642, A191

\bibitem[{{Euclid Collaboration} {et~al.}(2020{\natexlab{b}}){Euclid
  Collaboration}, {Desprez}, {Paltani}, {Coupon}, {Almosallam},
  {Alvarez-Ayllon}, {Amaro}, {Brescia}, {Brodwin}, {Cavuoti}, \&
  et~al.}]{Euclidz2020}
{Euclid Collaboration}, {Desprez}, G., {Paltani}, S., {et~al.}
  2020{\natexlab{b}}, A\&A, 644, A31

\bibitem[{{Euclid Collaboration} {et~al.}(2022){Euclid Collaboration},
  {Scaramella}, {Amiaux}, {Mellier}, {Burigana}, {Carvalho}, {Cuillandre}, {Da
  Silva}, {Derosa}, {Dinis}, {Maiorano}, {Maris}, {Tereno}, {Laureijs},
  {Boenke}, {Buenadicha}, {Dupac}, {Gaspar Venancio}, {G{\'o}mez-{\'A}lvarez},
  {Hoar}, {Lorenzo Alvarez}, {Racca}, {Saavedra-Criado}, {Schwartz}, {Vavrek},
  {Schirmer}, {Aussel}, {Azzollini}, {Cardone}, {Cropper}, {Ealet}, {Garilli},
  {Gillard}, {Granett}, {Guzzo}, {Hoekstra}, {Jahnke}, {Kitching}, {Maciaszek},
  {Meneghetti}, {Miller}, {Nakajima}, {Niemi}, {Pasian}, {Percival},
  {Pottinger}, {Sauvage}, {Scodeggio}, {Wachter}, {Zacchei}, {Aghanim},
  {Amara}, {Auphan}, {Auricchio}, {Awan}, {Balestra}, {Bender}, {Bodendorf},
  {Bonino}, {Branchini}, {Brau-Nogue}, {Brescia}, {Candini}, {Capobianco},
  {Carbone}, {Carlberg}, {Carretero}, {Casas}, {Castander}, {Castellano},
  {Cavuoti}, {Cimatti}, {Cledassou}, {Congedo}, {Conselice}, {Conversi},
  {Copin}, {Corcione}, {Costille}, {Courbin}, {Degaudenzi}, {Douspis},
  {Dubath}, {Duncan}, {Dusini}, {Farrens}, {Ferriol}, {Fosalba}, {Fourmanoit},
  {Frailis}, {Franceschi}, {Franzetti}, {Fumana}, {Gillis}, {Giocoli},
  {Grazian}, {Grupp}, {Haugan}, {Holmes}, {Hormuth}, {Hudelot}, {Kermiche},
  {Kiessling}, {Kilbinger}, {Kohley}, {Kubik}, {K{\"u}mmel}, {Kunz},
  {Kurki-Suonio}, {Lahav}, {Ligori}, {Lilje}, {Lloro}, {Mansutti}, {Marggraf},
  {Markovic}, {Marulli}, {Massey}, {Maurogordato}, {Melchior}, {Merlin},
  {Meylan}, {Mohr}, {Moresco}, {Morin}, {Moscardini}, {Munari}, {Nichol},
  {Padilla}, {Paltani}, {Peacock}, {Pedersen}, {Pettorino}, {Pires}, {Poncet},
  {Popa}, {Pozzetti}, {Raison}, {Rebolo}, {Rhodes}, {Rix}, {Roncarelli},
  {Rossetti}, {Saglia}, {Schneider}, {Schrabback}, {Secroun}, {Seidel},
  {Serrano}, {Sirignano}, {Sirri}, {Skottfelt}, {Stanco}, {Starck},
  {Tallada-Cresp{\'\i}}, {Tavagnacco}, {Taylor}, {Teplitz}, {Toledo-Moreo},
  {Torradeflot}, {Trifoglio}, {Valentijn}, {Valenziano}, {Verdoes Kleijn},
  {Wang}, {Welikala}, {Weller}, {Wetzstein}, {Zamorani}, {Zoubian}, {Andreon},
  {Baldi}, {Bardelli}, {Boucaud}, {Camera}, {Di Ferdinando}, {Fabbian},
  {Farinelli}, {Galeotta}, {Graci{\'a}-Carpio}, {Maino}, {Medinaceli}, {Mei},
  {Neissner}, {Polenta}, {Renzi}, {Romelli}, {Rosset}, {Sureau}, {Tenti},
  {Vassallo}, {Zucca}, {Baccigalupi}, {Balaguera-Antol{\'\i}nez}, {Battaglia},
  {Biviano}, {Borgani}, {Bozzo}, {Cabanac}, {Cappi}, {Casas}, {Castignani},
  {Colodro-Conde}, {Coupon}, {Courtois}, {Cuby}, {de la Torre}, {Desai},
  {Dole}, {Fabricius}, {Farina}, {Ferreira}, {Finelli}, {Flose-Reimberg},
  {Fotopoulou}, {Ganga}, {Gozaliasl}, {Hook}, {Keihanen}, {Kirkpatrick},
  {Liebing}, {Lindholm}, {Mainetti}, {Martinelli}, {Martinet}, {Maturi},
  {McCracken}, {Metcalf}, {Morgante}, {Nightingale}, {Nucita}, {Patrizii},
  {Potter}, {Riccio}, {S{\'a}nchez}, {Sapone}, {Schewtschenko}, {Schultheis},
  {Scottez}, {Teyssier}, {Tutusaus}, {Valiviita}, {Viel}, {Vriend}, \&
  {Whittaker}}]{Scaramella2021}
{Euclid Collaboration}, {Scaramella}, R., {Amiaux}, J., {et~al.} 2022, \aap,
  662, A112

\bibitem[{{Fan}(2018)}]{Fan2018}
{Fan}, Z. 2018, 42nd COSPAR Scientific Assembly. Held 14-22 July 2018, in
  Pasadena, California, USA, Abstract id. E1.16

\bibitem[{{Farrens} {et~al.}(2022){Farrens}, {Lacan}, {Guinot}, \&
  {Vitorelli}}]{Farrens2022}
{Farrens}, S., {Lacan}, A., {Guinot}, A., \& {Vitorelli}, A.~Z. 2022, \aap,
  657, A98

\bibitem[{{Fu} \& {Fan}(2014)}]{FuFan2014}
{Fu}, L. \& {Fan}, Z. 2014, RAA, 14, 1061

\bibitem[{{Fu} {et~al.}(2008){Fu}, {Semboloni}, {Hoekstra}, {Kilbinger}, {van
  Waerbeke}, {Tereno}, {Mellier}, {Heymans}, {Coupon}, {Benabed}, {Benjamin},
  {Bertin}, {Dore}, {Hudson}, {Ilbert}, {Maoli}, {Marmo}, {McCracken}, \&
  {Menard}}]{Fu2008}
{Fu}, L., {Semboloni}, E., {Hoekstra}, H., {et~al.} 2008, A\&A, 479, 9

\bibitem[{{Gawiser} {et~al.}(2006){Gawiser}, {van Dokkum}, {Herrera}, {Maza},
  {Castander}, {Infante}, {Lira}, {Quadri}, {Toner}, {Treister}, {Urry},
  {Altmann}, {Assef}, {Christlein}, {Coppi}, {Dur{\'a}n}, {Franx}, {Galaz},
  {Huerta}, {Liu}, {L{\'o}pez}, {M{\'e}ndez}, {Moore}, {Rubio}, {Ruiz}, {Toft},
  \& {Yi}}]{2006ApJS..162....1G}
{Gawiser}, E., {van Dokkum}, P.~G., {Herrera}, D., {et~al.} 2006, \apjs, 162, 1

\bibitem[{{Giavalisco} {et~al.}(2004){Giavalisco}, {Ferguson}, {Koekemoer},
  {Dickinson}, {Alexander}, {Bauer}, {Bergeron}, {Biagetti}, {Brandt},
  {Casertano}, \& et~al.}]{Giavalisco2004}
{Giavalisco}, M., {Ferguson}, H., {Koekemoer}, A., {et~al.} 2004, \apj, 600,
  L93

\bibitem[{{Golob} {et~al.}(2021){Golob}, {Sawicki}, {Goulding}, \&
  {Coupon}}]{2021MNRAS.503.4136G}
{Golob}, A., {Sawicki}, M., {Goulding}, A.~D., \& {Coupon}, J. 2021, \mnras,
  503, 4136

\bibitem[{{Gong} {et~al.}(2019){Gong}, {Liu}, {Cao}, {Chen}, {Fan}, {Li}, {Li},
  {Li}, {Zhang}, \& {Zhan}}]{Gong2019}
{Gong}, Y., {Liu}, X., {Cao}, Y., {et~al.} 2019, \apj, 883, 203

\bibitem[{{Grogin} {et~al.}(2011){Grogin}, {Kocevski}, {Faber}, {Ferguson},
  {Koekemoer}, {Riess}, {Acquaviva}, {Alexander}, {Almaini}, {Ashby}, \&
  et~al.}]{Grogin2011}
{Grogin}, N., {Kocevski}, D., {Faber}, S., {et~al.} 2011, \apjs, 197, 35

\bibitem[{{Guinot} {et~al.}(2022){Guinot}, {Kilbinger}, {Farrens}, {Peel},
  {Pujol}, {Schmitz}, {Starck}, {Erben}, {Gavazzi}, {Gwyn}, {Hudson},
  {Hildebrandt}, {Tobias}, {Miller}, {Spitzer}, {Van Waerbeke}, {Cuillandre},
  {Fabbro}, {McConnachie}, \& {Mellier}}]{Guinot2022}
{Guinot}, A., {Kilbinger}, M., {Farrens}, S., {et~al.} 2022, \aap, 666, A162

\bibitem[{{Guo} {et~al.}(2013){Guo}, {Ferguson}, {Giavalisco}, {Barro},
  {Willner}, {Ashby}, {Dahlen}, {Donley}, {Faber}, {Fontana}, {Galametz},
  {Grazian}, {Huang}, {Kocevski}, {Koekemoer}, {Koo}, {McGrath}, {Peth},
  {Salvato}, {Wuyts}, {Castellano}, {Cooray}, {Dickinson}, {Dunlop}, {Fazio},
  {Gardner}, {Gawiser}, {Grogin}, {Hathi}, {Hsu}, {Lee}, {Lucas}, {Mobasher},
  {Nandra}, {Newman}, \& {van der Wel}}]{2013ApJS..207...24G}
{Guo}, Y., {Ferguson}, H.~C., {Giavalisco}, M., {et~al.} 2013, \apjs, 207, 24

\bibitem[{{Guy} {et~al.}(2022){Guy}, {Cuillandre}, {Bachelet}, {Banerji},
  {Bauer}, {Collett}, {Conselice}, {Eggl}, {Ferguson}, {Fontana}, {Heymans},
  {Hook}, {Aubourg}, {Aussel}, {Bosch}, {Carry}, {Hoekstra}, {Kuijken},
  {Lanusse}, {Melchior}, {Mohr}, {Moresco}, {Nakajima}, {Paltani}, {Troxel},
  {Allevato}, {Amara}, {Andreon}, {Anguita}, {Bardelli}, {Bechtol}, {Birrer},
  {Bisigello}, {Bolzonella}, {Botticella}, {Bouy}, {Brinchmann}, {Brough},
  {Camera}, {Cantiello}, {Cappellaro}, {Carlin}, {Castander}, {Castellano},
  {Chari}, {Chisari}, {Collins}, {Courbin}, {Cuby}, {Cucciati}, {Daylan},
  {Diego}, {Duc}, {Fotopoulou}, {Fouchez}, {Gavazzi}, {Gruen}, {Hatfield},
  {Hildebrandt}, {Landt}, {Hunt}, {Ibata}, {Ilbert}, {Jasche}, {Joachimi},
  {Joseph}, {Kotak}, {Laigle}, {Lan{\c{c}}on}, {Larsen}, {Lavaux}, {Leclercq},
  {Leonard}, {von der Linden}, {Liu}, {Longo}, {Magliocchetti}, {Maraston},
  {Marshall}, {Mart{\'\i}n}, {Mattila}, {Maturi}, {McCracken}, {Metcalf},
  {Montes}, {Mortlock}, {Moscardini}, {Narayan}, {Paolillo}, {Papaderos},
  {Pello}, {Pozzetti}, {Radovich}, {Rejkuba}, {Rom{\'a}n},
  {S{\'a}nchez-Janssen}, {Sarpa}, {Sartoris}, {Schrabback}, {Sluse}, {Smartt},
  {Smith}, {Snodgrass}, {Talia}, {Tao}, {Toft}, {Tortora}, {Tutusaus}, {Usher},
  {van Velzen}, {Verma}, {Vernardos}, {Voggel}, {Wandelt}, {Watkins}, {Weller},
  {Wright}, {Yoachim}, {Yoon}, \& {Zucca}}]{2022zndo...5836022G}
{Guy}, L.~P., {Cuillandre}, J.-C., {Bachelet}, E., {et~al.} 2022, in Zenodo id.
  5836022, Vol.~58, 5836022

\bibitem[{{Heymans} {et~al.}(2021){Heymans}, {Tr{\"o}ster}, {Asgari}, {Blake},
  {Hildebrandt}, {Joachimi}, {Kuijken}, {Lin}, {S{\'a}nchez}, {van den Busch},
  {Wright}, {Amon}, {Bilicki}, {de Jong}, {Crocce}, {Dvornik}, {Erben},
  {Fortuna}, {Getman}, {Giblin}, {Glazebrook}, {Hoekstra}, {Joudaki},
  {Kannawadi}, {K{\"o}hlinger}, {Lidman}, {Miller}, {Napolitano}, {Parkinson},
  {Schneider}, {Shan}, {Valentijn}, {Verdoes Kleijn}, \& {Wolf}}]{Heymans2021}
{Heymans}, C., {Tr{\"o}ster}, T., {Asgari}, M., {et~al.} 2021, \aap, 646, A140

\bibitem[{{Heymans} {et~al.}(2012){Heymans}, {van Waerbeke}, {Miller}, {Erben},
  {Hildebrandt}, {Hoekstra}, {Kitching}, {Mellier}, {Simon}, {Bonnet},
  {Coupon}, {Fu}, {Harnois Deraps}, {Hudson}, {Kilbinger}, {Kuijken}, {Rowe},
  {Schrabback}, {Semboloni}, {Edo Vafaei}, \& {Valander}}]{Heymans2012}
{Heymans}, C., {van Waerbeke}, L., {Miller}, L., {et~al.} 2012, \mnras, 427,
  146

\bibitem[{{Hikage} {et~al.}(2019){Hikage}, {Oguri}, {Hamana}, {More},
  {Mandelbaum}, {Takada}, {K{\"o}hlinger}, {Miyatake}, {Nishizawa}, {Aihara},
  {Armstrong}, {Bosch}, {Coupon}, {Ducout}, {Ho}, {Hsieh}, {Komiyama},
  {Lanusse}, {Leauthaud}, {Lupton}, {Medezinski}, {Mineo}, {Miyama},
  {Miyazaki}, {Murata}, {Murayama}, {Shirasaki}, {Sif{\'o}n}, {Simet},
  {Speagle}, {Spergel}, {Strauss}, {Sugiyama}, {Tanaka}, {Utsumi}, {Wang}, \&
  {Yamada}}]{HSC2019}
{Hikage}, C., {Oguri}, M., {Hamana}, T., {et~al.} 2019, \pasj, 71, 43

\bibitem[{{Hildebrandt} {et~al.}(2010){Hildebrandt}, {Arnouts}, {Capak},
  {Moustakas}, {Wolf}, {Abdalla}, {Assef}, {Banerji}, {Ben{\'\i}tez},
  {Brammer}, {Budav{\'a}ri}, {Carliles}, {Coe}, {Dahlen}, {Feldmann}, {Gerdes},
  {Gillis}, {Ilbert}, {Kotulla}, {Lahav}, {Li}, {Miralles}, {Purger},
  {Schmidt}, \& {Singal}}]{Hildebrandt2010}
{Hildebrandt}, H., {Arnouts}, S., {Capak}, P., {et~al.} 2010, \aap, 523, A31

\bibitem[{{Hildebrandt} {et~al.}(2012){Hildebrandt}, {Erben}, {Kuijken}, {van
  Waerbeke}, {Heymans}, {Coupon}, {Benjamin}, {Bonnett}, {Fu}, {Hoekstra},
  {Kitching}, {Mellier}, {Miller}, {Velander}, {Hudson}, {Rowe}, {Schrabback},
  {Semboloni}, \& {Ben{\'\i}tez}}]{2012MNRAS.421.2355H}
{Hildebrandt}, H., {Erben}, T., {Kuijken}, K., {et~al.} 2012, \mnras, 421, 2355

\bibitem[{{Hildebrandt} {et~al.}(2020){Hildebrandt}, {K{\"o}hlinger}, {van den
  Busch}, {Joachimi}, {Heymans}, {Kannawadi}, {Wright}, {Asgari}, {Blake},
  {Hoekstra}, {Joudaki}, {Kuijken}, {Miller}, {Morrison}, {Tr{\"o}ster},
  {Amon}, {Archidiacono}, {Brieden}, {Choi}, {de Jong}, {Erben}, {Giblin},
  {Mead}, {Peacock}, {Radovich}, {Schneider}, {Sif{\'o}n}, \&
  {Tewes}}]{Hildebrandt2020}
{Hildebrandt}, H., {K{\"o}hlinger}, F., {van den Busch}, J.~L., {et~al.} 2020,
  \aap, 633, A69

\bibitem[{{Hildebrandt} {et~al.}(2021){Hildebrandt}, {van den Busch}, {Wright},
  {Blake}, {Joachimi}, {Kuijken}, {Tr{\"o}ster}, {Asgari}, {Bilicki}, {de
  Jong}, {Dvornik}, {Erben}, {Getman}, {Giblin}, {Heymans}, {Kannawadi}, {Lin},
  \& {Shan}}]{Hildebrandt2021}
{Hildebrandt}, H., {van den Busch}, J.~L., {Wright}, A.~H., {et~al.} 2021,
  \aap, 647, A124

\bibitem[{{Hildebrandt} {et~al.}(2017){Hildebrandt}, {Viola}, {Heymans},
  {Joudaki}, {Kuijken}, {Blake}, {Erben}, {Joachimi}, {Klaes}, {Miller},
  {Morrison}, {Nakajima}, {Verdoes Kleijn}, {Amon}, {Choi}, {Covone}, {de
  Jong}, {Dvornik}, {Fenech Conti}, {Grado}, {Harnois-D{\'e}raps}, {Herbonnet},
  {Hoekstra}, {K{\"o}hlinger}, {McFarland}, {Mead}, {Merten}, {Napolitano},
  {Peacock}, {Radovich}, {Schneider}, {Simon}, {Valentijn}, {van den Busch},
  {van Uitert}, \& {Van Waerbeke}}]{Hildebrandt2017}
{Hildebrandt}, H., {Viola}, M., {Heymans}, C., {et~al.} 2017, \mnras, 465, 1454

\bibitem[{{Huang} {et~al.}(2018){Huang}, {Leauthaud}, {Murata}, {Bosch},
  {Price}, {Lupton}, {Mandelbaum}, {Lackner}, {Bickerton}, {Miyazaki},
  {Coupon}, \& {Tanaka}}]{Huang2018}
{Huang}, S., {Leauthaud}, A., {Murata}, R., {et~al.} 2018, \pasj, 70, S6

\bibitem[{{Ivezi{\'c}} {et~al.}(2019){Ivezi{\'c}}, {Kahn}, {Tyson}, {Abel},
  {Acosta}, {Allsman}, {Alonso}, {AlSayyad}, {Anderson}, {Andrew}, {Angel},
  {Angeli}, {Ansari}, {Antilogus}, {Araujo}, {Armstrong}, {Arndt}, {Astier},
  {Aubourg}, {Auza}, {Axelrod}, {Bard}, {Barr}, {Barrau}, {Bartlett}, {Bauer},
  {Bauman}, {Baumont}, {Bechtol}, {Bechtol}, {Becker}, {Becla}, {Beldica},
  {Bellavia}, {Bianco}, {Biswas}, {Blanc}, {Blazek}, {Blandford}, {Bloom},
  {Bogart}, {Bond}, {Booth}, {Borgland}, {Borne}, {Bosch}, {Boutigny},
  {Brackett}, {Bradshaw}, {Brandt}, {Brown}, {Bullock}, {Burchat}, {Burke},
  {Cagnoli}, {Calabrese}, {Callahan}, {Callen}, {Carlin}, {Carlson},
  {Chandrasekharan}, {Charles-Emerson}, {Chesley}, {Cheu}, {Chiang}, {Chiang},
  {Chirino}, {Chow}, {Ciardi}, {Claver}, {Cohen-Tanugi}, {Cockrum}, {Coles},
  {Connolly}, {Cook}, {Cooray}, {Covey}, {Cribbs}, {Cui}, {Cutri}, {Daly},
  {Daniel}, {Daruich}, {Daubard}, {Daues}, {Dawson}, {Delgado}, {Dellapenna},
  {de Peyster}, {de Val-Borro}, {Digel}, {Doherty}, {Dubois},
  {Dubois-Felsmann}, {Durech}, {Economou}, {Eifler}, {Eracleous}, {Emmons},
  {Fausti Neto}, {Ferguson}, {Figueroa}, {Fisher-Levine}, {Focke}, {Foss},
  {Frank}, {Freemon}, {Gangler}, {Gawiser}, {Geary}, {Gee}, {Geha}, {Gessner},
  {Gibson}, {Gilmore}, {Glanzman}, {Glick}, {Goldina}, {Goldstein}, {Goodenow},
  {Graham}, {Gressler}, {Gris}, {Guy}, {Guyonnet}, {Haller}, {Harris},
  {Hascall}, {Haupt}, {Hernandez}, {Herrmann}, {Hileman}, {Hoblitt}, {Hodgson},
  {Hogan}, {Howard}, {Huang}, {Huffer}, {Ingraham}, {Innes}, {Jacoby}, {Jain},
  {Jammes}, {Jee}, {Jenness}, {Jernigan}, {Jevremovi{\'c}}, {Johns}, {Johnson},
  {Johnson}, {Jones}, {Juramy-Gilles}, {Juri{\'c}}, {Kalirai}, {Kallivayalil},
  {Kalmbach}, {Kantor}, {Karst}, {Kasliwal}, {Kelly}, {Kessler}, {Kinnison},
  {Kirkby}, {Knox}, {Kotov}, {Krabbendam}, {Krughoff}, {Kub{\'a}nek},
  {Kuczewski}, {Kulkarni}, {Ku}, {Kurita}, {Lage}, {Lambert}, {Lange},
  {Langton}, {Le Guillou}, {Levine}, {Liang}, {Lim}, {Lintott}, {Long},
  {Lopez}, {Lotz}, {Lupton}, {Lust}, {MacArthur}, {Mahabal}, {Mandelbaum},
  {Markiewicz}, {Marsh}, {Marshall}, {Marshall}, {May}, {McKercher}, {McQueen},
  {Meyers}, {Migliore}, {Miller}, {Mills}, {Miraval}, {Moeyens}, {Moolekamp},
  {Monet}, {Moniez}, {Monkewitz}, {Montgomery}, {Morrison}, {Mueller},
  {Muller}, {Mu{\~n}oz Arancibia}, {Neill}, {Newbry}, {Nief}, {Nomerotski},
  {Nordby}, {O'Connor}, {Oliver}, {Olivier}, {Olsen}, {O'Mullane}, {Ortiz},
  {Osier}, {Owen}, {Pain}, {Palecek}, {Parejko}, {Parsons}, {Pease},
  {Peterson}, {Peterson}, {Petravick}, {Libby Petrick}, {Petry},
  {Pierfederici}, {Pietrowicz}, {Pike}, {Pinto}, {Plante}, {Plate}, {Plutchak},
  {Price}, {Prouza}, {Radeka}, {Rajagopal}, {Rasmussen}, {Regnault}, {Reil},
  {Reiss}, {Reuter}, {Ridgway}, {Riot}, {Ritz}, {Robinson}, {Roby}, {Roodman},
  {Rosing}, {Roucelle}, {Rumore}, {Russo}, {Saha}, {Sassolas}, {Schalk},
  {Schellart}, {Schindler}, {Schmidt}, {Schneider}, {Schneider}, {Schoening},
  {Schumacher}, {Schwamb}, {Sebag}, {Selvy}, {Sembroski}, {Seppala}, {Serio},
  {Serrano}, {Shaw}, {Shipsey}, {Sick}, {Silvestri}, {Slater}, {Smith},
  {Smith}, {Sobhani}, {Soldahl}, {Storrie-Lombardi}, {Stover}, {Strauss},
  {Street}, {Stubbs}, {Sullivan}, {Sweeney}, {Swinbank}, {Szalay}, {Takacs},
  {Tether}, {Thaler}, {Thayer}, {Thomas}, {Thornton}, {Thukral}, {Tice},
  {Trilling}, {Turri}, {Van Berg}, {Vanden Berk}, {Vetter}, {Virieux},
  {Vucina}, {Wahl}, {Walkowicz}, {Walsh}, {Walter}, {Wang}, {Wang}, {Warner},
  {Wiecha}, {Willman}, {Winters}, {Wittman}, {Wolff}, {Wood-Vasey}, {Wu},
  {Xin}, {Yoachim}, \& {Zhan}}]{LSST2019}
{Ivezi{\'c}}, {\v{Z}}., {Kahn}, S.~M., {Tyson}, J.~A., {et~al.} 2019, \apj,
  873, 111

\bibitem[{{Jain} {et~al.}(2015){Jain}, {Spergel}, {Bean}, {Connolly},
  {Dell'antonio}, {Frieman}, {Gawiser}, {Gehrels}, {Gladney}, {Heitmann},
  {Helou}, {Hirata}, {Ho}, {Ivezi{\'c}}, {Jarvis}, {Kahn}, {Kalirai}, {Kim},
  {Lupton}, {Mandelbaum}, {Marshall}, {Newman}, {Perlmutter}, {Postman},
  {Rhodes}, {Strauss}, {Tyson}, {Walkowicz}, \& {Wood-Vasey}}]{Jain2015}
{Jain}, B., {Spergel}, D., {Bean}, R., {et~al.} 2015, arXiv:1501.07897

\bibitem[{{Joseph} {et~al.}(2016){Joseph}, {Courbin}, \& {Starck}}]{MuSCADeT}
{Joseph}, R., {Courbin}, F., \& {Starck}, J. 2016, A\&A, 589, A2

\bibitem[{{Kilbinger}(2015)}]{Kilbinger2015}
{Kilbinger}, M. 2015, Rep. Prog. Phys., 78, 086901

\bibitem[{{Kilbinger} {et~al.}(2013){Kilbinger}, {Fu}, {Heymans}, {Simpson},
  {Benjamin}, {Erben}, {Harnois-Deraps}, {Hoekstra}, {Hildebrandt}, {Kitching},
  {Mellier}, {Miller}, {van Waerbeke}, {Benabed}, {Bonnet}, {Coupon}, {Hudson},
  {Kuijken}, {Rowe}, {Semboloni}, {Vafaei}, \& {Valander}}]{Kilbinger2013}
{Kilbinger}, M., {Fu}, L., {Heymans}, C., {et~al.} 2013, \mnras, 430, 2200

\bibitem[{{Koekemoer} {et~al.}(2011){Koekemoer}, {Faber}, {Ferguson}, {Grogin},
  {Kocevski}, {Koo}, {Lai}, {Lotz}, {Lucas}, {McGrath}, \&
  et~al.}]{Koekemoer2011}
{Koekemoer}, A., {Faber}, S., {Ferguson}, H., {et~al.} 2011, \apjs, 197, 36

\bibitem[{{Kron}(1980)}]{1980ApJS...43..305K}
{Kron}, R.~G. 1980, \apjs, 43, 305

\bibitem[{{Kuijken}(2008)}]{Kuijken2008}
{Kuijken}, K. 2008, \aap, 482, 1053

\bibitem[{{Kuijken} {et~al.}(2015){Kuijken}, {Heymans}, {Hildebrandt},
  {Nakajima}, {Erben}, {de Jong}, {Massimo}, {Hoekstra}, {Choi}, {Hoekstra},
  {Miller}, {van Uitert}, {Amon}, {Blake}, {Brouwer}, {Buddendiek}, {Conti},
  {Eriksen}, {Grado}, {Hlemich}, {Irisarri}, {Kitching}, {Klaes}, {La Barbera},
  {Napolitano}, {Radovich}, {Schneider}, {Sifon}, {Sikkema}, {simon},
  {Tudorica}, {Valentijn}, {Verdoes Kleijn}, \& {van Warebeke}}]{KiDS}
{Kuijken}, K., {Heymans}, C., {Hildebrandt}, H., {et~al.} 2015, \mnras, 454,
  3500

\bibitem[{{Laigle} {et~al.}(2016){Laigle}, {McCracken}, {Ilbert}, {Hsieh},
  {Davidzon}, {Capak}, {Hasinger}, {Silverman}, {Pichon}, {Coupon}, {Aussel},
  {Le Borgne}, {Caputi}, {Cassata}, {Chang}, {Civano}, {Dunlop}, {Fynbo},
  {Kartaltepe}, {Koekemoer}, {Le F{\`e}vre}, {Le Floc'h}, {Leauthaud}, {Lilly},
  {Lin}, {Marchesi}, {Milvang-Jensen}, {Salvato}, {Sanders}, {Scoville},
  {Smolcic}, {Stockmann}, {Taniguchi}, {Tasca}, {Toft}, {Vaccari}, \&
  {Zabl}}]{laigle16}
{Laigle}, C., {McCracken}, H.~J., {Ilbert}, O., {et~al.} 2016, \apjs, 224, 24

\bibitem[{{Laureijs} {et~al.}(2011){Laureijs}, {Amiaux}, {Arduini},
  {Augu{\`e}res}, {Brinchmann}, {Cole}, {Cropper}, {Dabin}, {Duvet}, {Ealet},
  {Garilli}, {Gondoin}, {Guzzo}, {Hoar}, {Hoekstra}, {Holmes}, {Kitching},
  {Maciaszek}, {Mellier}, {Pasian}, {Percival}, {Rhodes}, {Saavedra Criado},
  {Sauvage}, {Scaramella}, {Valenziano}, {Warren}, {Bender}, {Castander},
  {Cimatti}, {Le F{\`e}vre}, {Kurki-Suonio}, {Levi}, {Lilje}, {Meylan},
  {Nichol}, {Pedersen}, {Popa}, {Rebolo Lopez}, {Rix}, {Rottgering},
  {Zeilinger}, {Grupp}, {Hudelot}, {Massey}, {Meneghetti}, {Miller}, {Paltani},
  {Paulin-Henriksson}, {Pires}, {Saxton}, {Schrabback}, {Seidel}, {Walsh},
  {Aghanim}, {Amendola}, {Bartlett}, {Baccigalupi}, {Beaulieu}, {Benabed},
  {Cuby}, {Elbaz}, {Fosalba}, {Gavazzi}, {Helmi}, {Hook}, {Irwin}, {Kneib},
  {Kunz}, {Mannucci}, {Moscardini}, {Tao}, {Teyssier}, {Weller}, {Zamorani},
  {Zapatero Osorio}, {Boulade}, {Foumond}, {Di Giorgio}, {Guttridge}, {James},
  {Kemp}, {Martignac}, {Spencer}, {Walton}, {Bl{\"u}mchen}, {Bonoli},
  {Bortoletto}, {Cerna}, {Corcione}, {Fabron}, {Jahnke}, {Ligori}, {Madrid},
  {Martin}, {Morgante}, {Pamplona}, {Prieto}, {Riva}, {Toledo}, {Trifoglio},
  {Zerbi}, {Abdalla}, {Douspis}, {Grenet}, {Borgani}, {Bouwens}, {Courbin},
  {Delouis}, {Dubath}, {Fontana}, {Frailis}, {Grazian}, {Koppenh{\"o}fer},
  {Mansutti}, {Melchior}, {Mignoli}, {Mohr}, {Neissner}, {Noddle}, {Poncet},
  {Scodeggio}, {Serrano}, {Shane}, {Starck}, {Surace}, {Taylor},
  {Verdoes-Kleijn}, {Vuerli}, {Williams}, {Zacchei}, {Altieri}, {Escudero
  Sanz}, {Kohley}, {Oosterbroek}, {Astier}, {Bacon}, {Bardelli}, {Baugh},
  {Bellagamba}, {Benoist}, {Bianchi}, {Biviano}, {Branchini}, {Carbone},
  {Cardone}, {Clements}, {Colombi}, {Conselice}, {Cresci}, {Deacon}, {Dunlop},
  {Fedeli}, {Fontanot}, {Franzetti}, {Giocoli}, {Garcia-Bellido}, {Gow},
  {Heavens}, {Hewett}, {Heymans}, {Holland}, {Huang}, {Ilbert}, {Joachimi},
  {Jennins}, {Kerins}, {Kiessling}, {Kirk}, {Kotak}, {Krause}, {Lahav}, {van
  Leeuwen}, {Lesgourgues}, {Lombardi}, {Magliocchetti}, {Maguire}, {Majerotto},
  {Maoli}, {Marulli}, {Maurogordato}, {McCracken}, {McLure}, {Melchiorri},
  {Merson}, {Moresco}, {Nonino}, {Norberg}, {Peacock}, {Pello}, {Penny},
  {Pettorino}, {Di Porto}, {Pozzetti}, {Quercellini}, {Radovich}, {Rassat},
  {Roche}, {Ronayette}, {Rossetti}, {Sartoris}, {Schneider}, {Semboloni},
  {Serjeant}, {Simpson}, {Skordis}, {Smadja}, {Smartt}, {Spano}, {Spiro},
  {Sullivan}, {Tilquin}, {Trotta}, {Verde}, {Wang}, {Williger}, {Zhao},
  {Zoubian}, \& {Zucca}}]{laureijs11}
{Laureijs}, R., {Amiaux}, J., {Arduini}, S., {et~al.} 2011, arXiv:1110.3193

\bibitem[{{Leauthaud} {et~al.}(2007){Leauthaud}, {Massey}, {Kneib}, {Rhodes},
  {Johnston}, {Capak}, {Heymans}, {Ellis}, {Koekemoer}, {Le F{\`e}vre},
  {Mellier}, {R{\'e}fr{\'e}gier}, {Robin}, {Scoville}, {Tasca}, {Taylor}, \&
  {Van Waerbeke}}]{2007ApJS..172..219L}
{Leauthaud}, A., {Massey}, R., {Kneib}, J.-P., {et~al.} 2007, \apjs, 172, 219

\bibitem[{{Lemos} {et~al.}(2021){Lemos}, {Raveri}, {Campos}, {Park}, {Chang},
  {Weaverdyck}, {Huterer}, {Liddle}, {Blazek}, {Cawthon}, {Choi}, {DeRose},
  {Dodelson}, {Doux}, {Gatti}, {Gruen}, {Harrison}, {Krause}, {Lahav},
  {MacCrann}, {Muir}, {Prat}, {Rau}, {Rollins}, {Samuroff}, {Zuntz}, {Aguena},
  {Allam}, {Annis}, {Avila}, {Bacon}, {Bernstein}, {Bertin}, {Brooks}, {Burke},
  {Carnero Rosell}, {Carrasco Kind}, {Carretero}, {Castander}, {Conselice},
  {Costanzi}, {Crocce}, {Pereira}, {Davis}, {De Vicente}, {Desai}, {Diehl},
  {Doel}, {Eckert}, {Eifler}, {Elvin-Poole}, {Everett}, {Evrard}, {Ferrero},
  {Fert{\'e}}, {Flaugher}, {Fosalba}, {Frieman}, {Garc{\'\i}a-Bellido},
  {Gaztanaga}, {Gerdes}, {Giannantonio}, {Gruendl}, {Gschwend}, {Gutierrez},
  {Hartley}, {Hinton}, {Hollowood}, {Honscheid}, {Hoyle}, {Huff}, {James},
  {Jarvis}, {Lima}, {Maia}, {March}, {Marshall}, {Martini}, {Melchior},
  {Menanteau}, {Miquel}, {Mohr}, {Morgan}, {Myles}, {Ogando}, {Palmese},
  {Pandey}, {Paz-Chinch{\'o}n}, {Plazas Malag{\'o}n}, {Rodriguez-Monroy},
  {Roodman}, {Sanchez}, {Scarpine}, {Schubnell}, {Secco}, {Serrano},
  {Sevilla-Noarbe}, {Smith}, {Soares-Santos}, {Suchyta}, {Swanson}, {Tarle},
  {Thomas}, {To}, {Troxel}, {Varga}, {Weller}, {Wester}, \& {DES
  Collaboration}}]{Lemos2021}
{Lemos}, P., {Raveri}, M., {Campos}, A., {et~al.} 2021, \mnras, 505, 6179

\bibitem[{{Li} {et~al.}(2016){Li}, {Frenk}, {Cole}, {Gao}, {Bose}, \&
  {Hellwing}}]{Li2016}
{Li}, R., {Frenk}, C.~S., {Cole}, S., {et~al.} 2016, \mnras, 460, 363

\bibitem[{{Li} {et~al.}(2017){Li}, {Frenk}, {Cole}, {Wang}, \& {Gao}}]{Li2017}
{Li}, R., {Frenk}, C.~S., {Cole}, S., {Wang}, Q., \& {Gao}, L. 2017, \mnras,
  468, 1426

\bibitem[{{Liu} {et~al.}(2017){Liu}, {Yang}, {Yuan}, {Wu}, {Fan}, {Shan},
  {Yan}, \& {Zheng}}]{2017AJ....153...53L}
{Liu}, D., {Yang}, J., {Yuan}, S., {et~al.} 2017, \aj, 153, 53

\bibitem[{{MacDonald} \& {Bernstein}(2010)}]{2010PASP..122..485M}
{MacDonald}, C.~J. \& {Bernstein}, G. 2010, \pasp, 122, 485

\bibitem[{{Ma{\l}ek} {et~al.}(2013){Ma{\l}ek}, {Solarz}, {Pollo}, {Fritz},
  {Garilli}, {Scodeggio}, {Iovino}, {Granett}, {Abbas}, {Adami}, {Arnouts},
  {Bel}, {Bolzonella}, {Bottini}, {Branchini}, {Cappi}, {Coupon}, {Cucciati},
  {Davidzon}, {De Lucia}, {de la Torre}, {Franzetti}, {Fumana}, {Guzzo},
  {Ilbert}, {Krywult}, {Le Brun}, {Le Fevre}, {Maccagni}, {Marulli},
  {McCracken}, {Paioro}, {Polletta}, {Schlagenhaufer}, {Tasca}, {Tojeiro},
  {Vergani}, {Zanichelli}, {Burden}, {Di Porto}, {Marchetti}, {Marinoni},
  {Mellier}, {Moscardini}, {Nichol}, {Peacock}, {Percival}, {Phleps}, {Wolk},
  \& {Zamorani}}]{2013A&A...557A..16M}
{Ma{\l}ek}, K., {Solarz}, A., {Pollo}, A., {et~al.} 2013, \aap, 557, A16

\bibitem[{{Mandelbaum}(2018)}]{Mandelbaum2018}
{Mandelbaum}, R. 2018, \araa, 56, 393

\bibitem[{{McCracken} {et~al.}(2012){McCracken}, {Milvang-Jensen}, {Dunlop},
  {Franx}, {Fynbo}, {Le F{\`e}vre}, {Holt}, {Caputi}, {Goranova}, {Emerson},
  {Freudling}, {Hudelot}, {L{\'o}pez-Sanjuan}, {Magnard}, {Mellier}, {Moller},
  {Nilsson}, {Sutherland}, {Tasca}, \& {Zabl}}]{McCracken2012}
{McCracken}, H.~J., {Milvang-Jensen}, B., {Dunlop}, J., {et~al.} 2012, \aap,
  544, A156

\bibitem[{{Melchior} {et~al.}(2018){Melchior}, {Moolekamp}, {Jerdee},
  {Amstrong}, {Sun}, {Bosch}, \& {Lupton}}]{SCARLET}
{Melchior}, P., {Moolekamp}, F., {Jerdee}, M., {et~al.} 2018, Astronomy and
  Computing, 24, 129

\bibitem[{{Metcalf} {et~al.}(2019){Metcalf}, {Meneghetti}, {Avestruz},
  {Bellagamba}, {Bom}, {Bertin}, {Cabanac}, {Courbin}, {Davies},
  {Decenci{\`e}re}, {Flamary}, {Gavazzi}, {Geiger}, {Hartley},
  {Huertas-Company}, {Jackson}, {Jacobs}, {Jullo}, {Kneib}, {Koopmans},
  {Lanusse}, {Li}, {Ma}, {Makler}, {Li}, {Lightman}, {Petrillo}, {Serjeant},
  {Sch{\"a}fer}, {Sonnenfeld}, {Tagore}, {Tortora}, {Tuccillo},
  {Valent{\'\i}n}, {Velasco-Forero}, {Verdoes Kleijn}, \&
  {Vernardos}}]{Metcalf2019}
{Metcalf}, R.~B., {Meneghetti}, M., {Avestruz}, C., {et~al.} 2019, \aap, 625,
  A119

\bibitem[{{Meyers} \& {Burchat}(2015)}]{2015ApJ...807..182M}
{Meyers}, J.~E. \& {Burchat}, P.~R. 2015, \apj, 807, 182

\bibitem[{{Miller} {et~al.}(2013){Miller}, {Heymans}, {Kitching}, {van
  Waerbeke}, {Erben}, {Hildebrandt}, {Hoekstra}, {Mellier}, {Rowe}, {Coupon},
  \& et~al.}]{Miller2013}
{Miller}, L., {Heymans}, C., {Kitching}, T., {et~al.} 2013, \mnras, 429, 2858

\bibitem[{{Myles} {et~al.}(2021){Myles}, {Alarcon}, {Amon}, {S{\'a}nchez},
  {Everett}, {DeRose}, {McCullough}, {Gruen}, {Bernstein}, {Troxel},
  {Dodelson}, {Campos}, {MacCrann}, {Yin}, {Raveri}, {Amara}, {Becker}, {Choi},
  {Cordero}, {Eckert}, {Gatti}, {Giannini}, {Gschwend}, {Gruendl}, {Harrison},
  {Hartley}, {Huff}, {Kuropatkin}, {Lin}, {Masters}, {Miquel}, {Prat},
  {Roodman}, {Rykoff}, {Sevilla-Noarbe}, {Sheldon}, {Wechsler}, {Yanny},
  {Abbott}, {Aguena}, {Allam}, {Annis}, {Bacon}, {Bertin}, {Bhargava},
  {Bridle}, {Brooks}, {Burke}, {Carnero Rosell}, {Carrasco Kind}, {Carretero},
  {Castander}, {Conselice}, {Costanzi}, {Crocce}, {da Costa}, {Pereira},
  {Desai}, {Diehl}, {Eifler}, {Elvin-Poole}, {Evrard}, {Ferrero}, {Fert{\'e}},
  {Flaugher}, {Fosalba}, {Frieman}, {Garc{\'\i}a-Bellido}, {Gaztanaga},
  {Giannantonio}, {Hinton}, {Hollowood}, {Honscheid}, {Hoyle}, {Huterer},
  {James}, {Krause}, {Kuehn}, {Lahav}, {Lima}, {Maia}, {Marshall}, {Martini},
  {Melchior}, {Menanteau}, {Mohr}, {Morgan}, {Muir}, {Ogando}, {Palmese},
  {Paz-Chinch{\'o}n}, {Plazas}, {Rodriguez-Monroy}, {Samuroff}, {Sanchez},
  {Scarpine}, {Secco}, {Serrano}, {Smith}, {Soares-Santos}, {Suchyta},
  {Swanson}, {Tarle}, {Thomas}, {To}, {Varga}, {Weller}, \&
  {Wester}}]{Myles2021}
{Myles}, J., {Alarcon}, A., {Amon}, A., {et~al.} 2021, \mnras, 505, 4249

\bibitem[{{Newberry}(1991)}]{1991PASP..103..122N}
{Newberry}, M.~V. 1991, \pasp, 103, 122

\bibitem[{{Newman} {et~al.}(2015){Newman}, {Abate}, {Abdalla}, {Allam},
  {Allen}, {Ansari}, {Balley}, {Barkhouse}, {Beers}, {Blanton}, \&
  et~al.}]{Newman2015}
{Newman}, J., {Abate}, A., {Abdalla}, F., {et~al.} 2015, Astroparticle Physics,
  63, 81

\bibitem[{{Newman} \& {Gruen}(2022)}]{2022ARA&A..60..363N}
{Newman}, J.~A. \& {Gruen}, D. 2022, \araa, 60, 363

\bibitem[{{Oesch} {et~al.}(2018){Oesch}, {Montes}, {Reddy}, {Bouwens},
  {Illingworth}, {Magee}, {Atek}, {Carollo}, {Cibinel}, {Franx}, {Holden},
  {Labb{\'e}}, {Nelson}, {Steidel}, {van Dokkum}, {Morselli}, {Naidu}, \&
  {Wilkins}}]{Oesch2018}
{Oesch}, P.~A., {Montes}, M., {Reddy}, N., {et~al.} 2018, \apjs, 237, 12

\bibitem[{{Pearson} {et~al.}(2019){Pearson}, {Li}, \& {Dye}}]{Pearson2019}
{Pearson}, J., {Li}, N., \& {Dye}, S. 2019, \mnras, 488, 991

\bibitem[{{Petrillo} {et~al.}(2019){Petrillo}, {Tortora}, {Chatterjee},
  {Vernardos}, {Koopmans}, {Verdoes Kleijn}, {Napolitano}, {Covone}, {Kelvin},
  \& {Hopkins}}]{Petrillo2019}
{Petrillo}, C.~E., {Tortora}, C., {Chatterjee}, S., {et~al.} 2019, \mnras, 482,
  807

\bibitem[{{Planck Collaboration} {et~al.}(2020){Planck Collaboration},
  {Aghanim}, {Akrami}, {Ashdown}, {Aumont}, {Baccigalupi}, {Ballardini},
  {Banday}, {Barreiro}, {Bartolo}, {Basak}, {Battye}, {Benabed}, {Bernard},
  {Bersanelli}, {Bielewicz}, {Bock}, {Bond}, {Borrill}, {Bouchet}, {Boulanger},
  {Bucher}, {Burigana}, {Butler}, {Calabrese}, {Cardoso}, {Carron},
  {Challinor}, {Chiang}, {Chluba}, {Colombo}, {Combet}, {Contreras}, {Crill},
  {Cuttaia}, {de Bernardis}, {de Zotti}, {Delabrouille}, {Delouis}, {Di
  Valentino}, {Diego}, {Dor{\'e}}, {Douspis}, {Ducout}, {Dupac}, {Dusini},
  {Efstathiou}, {Elsner}, {En{\ss}lin}, {Eriksen}, {Fantaye}, {Farhang},
  {Fergusson}, {Fernandez-Cobos}, {Finelli}, {Forastieri}, {Frailis},
  {Fraisse}, {Franceschi}, {Frolov}, {Galeotta}, {Galli}, {Ganga},
  {G{\'e}nova-Santos}, {Gerbino}, {Ghosh}, {Gonz{\'a}lez-Nuevo}, {G{\'o}rski},
  {Gratton}, {Gruppuso}, {Gudmundsson}, {Hamann}, {Handley}, {Hansen},
  {Herranz}, {Hildebrandt}, {Hivon}, {Huang}, {Jaffe}, {Jones}, {Karakci},
  {Keih{\"a}nen}, {Keskitalo}, {Kiiveri}, {Kim}, {Kisner}, {Knox},
  {Krachmalnicoff}, {Kunz}, {Kurki-Suonio}, {Lagache}, {Lamarre}, {Lasenby},
  {Lattanzi}, {Lawrence}, {Le Jeune}, {Lemos}, {Lesgourgues}, {Levrier},
  {Lewis}, {Liguori}, {Lilje}, {Lilley}, {Lindholm}, {L{\'o}pez-Caniego},
  {Lubin}, {Ma}, {Mac{\'\i}as-P{\'e}rez}, {Maggio}, {Maino}, {Mandolesi},
  {Mangilli}, {Marcos-Caballero}, {Maris}, {Martin}, {Martinelli},
  {Mart{\'\i}nez-Gonz{\'a}lez}, {Matarrese}, {Mauri}, {McEwen}, {Meinhold},
  {Melchiorri}, {Mennella}, {Migliaccio}, {Millea}, {Mitra},
  {Miville-Desch{\^e}nes}, {Molinari}, {Montier}, {Morgante}, {Moss}, {Natoli},
  {N{\o}rgaard-Nielsen}, {Pagano}, {Paoletti}, {Partridge}, {Patanchon},
  {Peiris}, {Perrotta}, {Pettorino}, {Piacentini}, {Polastri}, {Polenta},
  {Puget}, {Rachen}, {Reinecke}, {Remazeilles}, {Renzi}, {Rocha}, {Rosset},
  {Roudier}, {Rubi{\~n}o-Mart{\'\i}n}, {Ruiz-Granados}, {Salvati}, {Sandri},
  {Savelainen}, {Scott}, {Shellard}, {Sirignano}, {Sirri}, {Spencer},
  {Sunyaev}, {Suur-Uski}, {Tauber}, {Tavagnacco}, {Tenti}, {Toffolatti},
  {Tomasi}, {Trombetti}, {Valenziano}, {Valiviita}, {Van Tent}, {Vibert},
  {Vielva}, {Villa}, {Vittorio}, {Wandelt}, {Wehus}, {White}, {White},
  {Zacchei}, \& {Zonca}}]{Planck2020}
{Planck Collaboration}, {Aghanim}, N., {Akrami}, Y., {et~al.} 2020, \aap, 641,
  A6

\bibitem[{{Rafelski} {et~al.}(2015){Rafelski}, {Teplitz}, {Gardner}, {Coe},
  {Bond}, {Koekemper}, {Grogin}, {Kurczynski}, {McGrath}, {Bourque}, \&
  et~al.}]{Rafelski2015}
{Rafelski}, M., {Teplitz}, H., {Gardner}, J., {et~al.} 2015, \aj, 150, 31

\bibitem[{{Reiman} \& {G{\"o}hre}(2019)}]{Reiman2019}
{Reiman}, D. \& {G{\"o}hre}, B. 2019, \mnras, 485, 2617

\bibitem[{{Rhodes} {et~al.}(2017){Rhodes}, {Nichol}, {Aubourg}, {Bean},
  {Boutigny}, {Bremer}, {Capak}, {Cardone}, {Carry}, {Conselice}, \&
  et~al.}]{Rhodes2017}
{Rhodes}, J., {Nichol}, R., {Aubourg}, E., {et~al.} 2017, \apjs, 233, 23

\bibitem[{{Rowe} {et~al.}(2015){Rowe}, {Jarvis}, {Mandelbaum}, {Bernstein},
  {Bosch}, {Simet}, {Meyers}, {Kacprzak}, {Nakajima}, {Zuntz}, {Miyatake},
  {Dietrich}, {Armstrong}, {Melchior}, \& {Gill}}]{2015A&C....10..121R}
{Rowe}, B.~T.~P., {Jarvis}, M., {Mandelbaum}, R., {et~al.} 2015, Astronomy and
  Computing, 10, 121

\bibitem[{{Salvato} {et~al.}(2019){Salvato}, {Ilbert}, \&
  {Hoyle}}]{Salvato2019}
{Salvato}, M., {Ilbert}, O., \& {Hoyle}, B. 2019, Nature Astronomy, 3, 212

\bibitem[{{Schade} {et~al.}(1996){Schade}, {Lilly}, {Le Fevre}, {Hammer}, \&
  {Crampton}}]{1996ApJ...464...79S}
{Schade}, D., {Lilly}, S.~J., {Le Fevre}, O., {Hammer}, F., \& {Crampton}, D.
  1996, \apj, 464, 79

\bibitem[{{Semboloni} {et~al.}(2013){Semboloni}, {Hoekstra}, {Huang},
  {Cardone}, {Cropper}, {Joachimi}, {Kitching}, {Kuijken}, {Lombardi}, {Maoli},
  {Mellier}, {Miller}, {Rhodes}, {Scaramella}, {Schrabback}, \&
  {Velander}}]{semboloni13}
{Semboloni}, E., {Hoekstra}, H., {Huang}, Z., {et~al.} 2013, \mnras, 432, 2385

\bibitem[{{Sevilla-Noarbe} {et~al.}(2021){Sevilla-Noarbe}, {Bechtol}, {Carrasco
  Kind}, {Carnero Rosell}, {Becker}, {Drlica-Wagner}, {Gruendl}, {Rykoff},
  {Sheldon}, {Yanny}, {Alarcon}, {Allam}, {Amon}, {Benoit-L{\'e}vy},
  {Bernstein}, {Bertin}, {Burke}, {Carretero}, {Choi}, {Diehl}, {Everett},
  {Flaugher}, {Gaztanaga}, {Gschwend}, {Harrison}, {Hartley}, {Hoyle},
  {Jarvis}, {Johnson}, {Kessler}, {Kron}, {Kuropatkin}, {Leistedt}, {Li},
  {Menanteau}, {Morganson}, {Ogando}, {Palmese}, {Paz-Chinch{\'o}n}, {Pieres},
  {Pond}, {Rodriguez-Monroy}, {Smith}, {Stringer}, {Troxel}, {Tucker}, {de
  Vicente}, {Wester}, {Zhang}, {Abbott}, {Aguena}, {Annis}, {Avila},
  {Bhargava}, {Bridle}, {Brooks}, {Brout}, {Castander}, {Cawthon}, {Chang},
  {Conselice}, {Costanzi}, {Crocce}, {da Costa}, {Pereira}, {Davis}, {Desai},
  {Dietrich}, {Doel}, {Eckert}, {Evrard}, {Ferrero}, {Fosalba},
  {Garc{\'\i}a-Bellido}, {Gerdes}, {Giannantonio}, {Gruen}, {Gutierrez},
  {Hinton}, {Hollowood}, {Honscheid}, {Huff}, {Huterer}, {James}, {Jeltema},
  {Kuehn}, {Lahav}, {Lidman}, {Lima}, {Lin}, {Maia}, {Marshall}, {Martini},
  {Melchior}, {Miquel}, {Mohr}, {Morgan}, {Neilsen}, {Plazas}, {Romer},
  {Roodman}, {Sanchez}, {Scarpine}, {Schubnell}, {Serrano}, {Smith}, {Suchyta},
  {Tarle}, {Thomas}, {To}, {Varga}, {Wechsler}, {Weller}, {Wilkinson}, \& {DES
  Collaboration}}]{Sevilla-Noarbe2021}
{Sevilla-Noarbe}, I., {Bechtol}, K., {Carrasco Kind}, M., {et~al.} 2021, \apjs,
  254, 24

\bibitem[{{Simard} {et~al.}(2002){Simard}, {Willmer}, {Vogt}, {Sarajedini},
  {Phillips}, {Weiner}, {Koo}, {Im}, {Illingworth}, \&
  {Faber}}]{2002ApJS..142....1S}
{Simard}, L., {Willmer}, C. N.~A., {Vogt}, N.~P., {et~al.} 2002, \apjs, 142, 1

\bibitem[{{Skelton} {et~al.}(2014){Skelton}, {Whitaker}, {Momcheva}, {Brammer},
  {van Dokkum}, {Labb{\'e}}, {Franx}, {van der Wel}, {Bezanson}, {Da Cunha},
  {Fumagalli}, {F{\"o}rster Schreiber}, {Kriek}, {Leja}, {Lundgren}, {Magee},
  {Marchesini}, {Maseda}, {Nelson}, {Oesch}, {Pacifici}, {Patel}, {Price},
  {Rix}, {Tal}, {Wake}, \& {Wuyts}}]{Skelton2014}
{Skelton}, R.~E., {Whitaker}, K.~E., {Momcheva}, I.~G., {et~al.} 2014, \apjs,
  214, 24

\bibitem[{{Spergel} {et~al.}(2015){Spergel}, {Gehrels}, {Baltay}, {Bennett},
  {Breckinridge}, {Donahue}, {Dressler}, {Gaudi}, {Greene}, {Guyon}, {Hirata},
  {Kalirai}, {Kasdin}, {Macintosh}, {Moos}, {Perlmutter}, {Postman},
  {Rauscher}, {Rhodes}, {Wang}, {Weinberg}, {Benford}, {Hudson}, {Jeong},
  {Mellier}, {Traub}, {Yamada}, {Capak}, {Colbert}, {Masters}, {Penny},
  {Savransky}, {Stern}, {Zimmerman}, {Barry}, {Bartusek}, {Carpenter}, {Cheng},
  {Content}, {Dekens}, {Demers}, {Grady}, {Jackson}, {Kuan}, {Kruk}, {Melton},
  {Nemati}, {Parvin}, {Poberezhskiy}, {Peddie}, {Ruffa}, {Wallace}, {Whipple},
  {Wollack}, \& {Zhao}}]{Roman2015}
{Spergel}, D., {Gehrels}, N., {Baltay}, C., {et~al.} 2015, arXiv:1503.03757

\bibitem[{{Stabenau} {et~al.}(2008){Stabenau}, {Connolly}, \&
  {Jain}}]{2008MNRAS.387.1215S}
{Stabenau}, H.~F., {Connolly}, A., \& {Jain}, B. 2008, \mnras, 387, 1215

\bibitem[{{Tanaka} {et~al.}(2018){Tanaka}, {Coupon}, {Hsieh}, {Mineo},
  {Nishizawa}, {Speagle}, {Furusawa}, {Miyazaki}, \& {Murayama}}]{Tanaka2018}
{Tanaka}, M., {Coupon}, J., {Hsieh}, B.-C., {et~al.} 2018, \pasj, 70, S9

\bibitem[{{Taniguchi} {et~al.}(2015){Taniguchi}, {Kajisawa}, {Kobayashi},
  {Shioya}, {Nagao}, {Capak}, {Aussel}, {Ichikawa}, {Murayama}, {Scoville}, \&
  et~al.}]{Taniguchi2015}
{Taniguchi}, Y., {Kajisawa}, M., {Kobayashi}, M., {et~al.} 2015, \pasj, 67, 104

\bibitem[{{Tr{\"o}ster} {et~al.}(2021){Tr{\"o}ster}, {Asgari}, {Blake},
  {Cataneo}, {Heymans}, {Hildebrandt}, {Joachimi}, {Lin}, {Sanchez}, {Wright},
  \& et~al.}]{KiDS1000}
{Tr{\"o}ster}, T., {Asgari}, M., {Blake}, C., {et~al.} 2021, A\&A, 649, id.A88

\bibitem[{{Voigt} {et~al.}(2012){Voigt}, {Bridle}, {Amara}, {Cropper},
  {Kitching}, {Massey}, {Rhodes}, \& {Schrabback}}]{voigt12}
{Voigt}, L.~M., {Bridle}, S.~L., {Amara}, A., {et~al.} 2012, \mnras, 421, 1385

\bibitem[{{Weinberg} {et~al.}(2013){Weinberg}, {Mortonson}, {Eisenstein},
  {Hirata}, {Riess}, \& {Rozo}}]{Weinberg2013}
{Weinberg}, D., {Mortonson}, M., {Eisenstein}, D., {et~al.} 2013, Phys Rept.,
  530, 87

\bibitem[{{Wen} \& {Han}(2021)}]{Wen2021}
{Wen}, Z. \& {Han}, J. 2021, \mnras, 500, 1003

\bibitem[{{Wright} {et~al.}(2019){Wright}, {Hildebrandt}, {Kuijken}, {Erben},
  {Blake}, {Buddelmeijer}, {Choi}, {Cross}, {de Jong}, {Edge},
  {Gonzalez-Fernandez}, {Gonz{\'a}lez Solares}, {Grado}, {Heymans}, {Irwin},
  {Kupcu Yoldas}, {Lewis}, {Mann}, {Napolitano}, {Radovich}, {Schneider},
  {Sif{\'o}n}, {Sutherland}, {Sutorius}, \& {Verdoes
  Kleijn}}]{2019A&A...632A..34W}
{Wright}, A.~H., {Hildebrandt}, H., {Kuijken}, K., {et~al.} 2019, \aap, 632,
  A34

\bibitem[{{Wright} {et~al.}(2016){Wright}, {Robotham}, {Bourne}, {Driver},
  {Dunne}, {Maddox}, {Alpaslan}, {Andrews}, {Bauer}, {Bland-Hawthorn},
  {Brough}, {Brown}, {Clarke}, {Cluver}, {Davies}, {Grootes}, {Holwerda},
  {Hopkins}, {Jarrett}, {Kafle}, {Lange}, {Liske}, {Loveday}, {Moffett},
  {Norberg}, {Popescu}, {Smith}, {Taylor}, {Tuffs}, {Wang}, \&
  {Wilkins}}]{Wright2016}
{Wright}, A.~H., {Robotham}, A.~S.~G., {Bourne}, N., {et~al.} 2016, \mnras,
  460, 765

\bibitem[{{Yang} {et~al.}(2020){Yang}, {Boquien}, {Buat}, {Burgarella},
  {Ciesla}, {Duras}, {Stalevski}, {Brandt}, \&
  {Papovich}}]{2020MNRAS.491..740Y}
{Yang}, G., {Boquien}, M., {Buat}, V., {et~al.} 2020, \mnras, 491, 740

\bibitem[{{Zhan}(2011)}]{Zhan2011}
{Zhan}, H. 2011, Scientia Sinica Physica, Mechanica \& Astronomica, 41, 1441

\bibitem[{{Zhan}(2018)}]{Zhan2018}
{Zhan}, H. 2018, 42nd COSPAR Scientific Assembly. Held 14-22 July 2018, in
  Pasadena, California, USA, Abstract id. E1.16

\bibitem[{{Zhan}(2021)}]{Zhan2021}
{Zhan}, H. 2021, Chinese Science Bulletin, 66, 1290

\bibitem[{{Zhou} {et~al.}(2021){Zhou}, {Gong}, {Meng}, {Zhang}, {Cao}, {Chen},
  {Amaro}, {Fan}, \& {Fu}}]{Zhou2021}
{Zhou}, X., {Gong}, Y., {Meng}, X., {et~al.} 2021, \apj, 909, 53

\end{thebibliography}

\end{document}